\let\orilabel\label % Fix for TexLive 2023
\documentclass[aps,prx,reprint,amsmath,amssymb,nofootinbib,floatfix]{revtex4-2}
\let\label\orilabel % Fix for TexLive 2023
\usepackage{figures/figures}        % Produces PGFplots
\usepackage{physics}
\usepackage{cleveref}
\usepackage{esint,tensor,isomath,mathtools}
\usepackage{multirow}
\usepackage{hhline}
\providecommand{\ii}{\text{i}}
\providecommand{\ee}{\text{e}}
\providecommand{\vc}{\vb*}
\providecommand{\uv}{\vu*}
\providecommand{\xhat}{\hat{\vc{x}}}
\providecommand{\yhat}{\hat{\vc{y}}}

\newcommand{\tens}[1]{\mathsfbfit{{#1}}}
\renewcommand{\Re}{\real}
\renewcommand{\Im}{\imaginary}

\begin{document}
\title{The electromagnetic symmetry sphere: a framework for energy, momentum, spin and other electromagnetic quantities}
\author{Sebastian Golat}
\email{sebastian.1.golat@kcl.ac.uk}
\affiliation{Department of Physics, King's College London, Strand, London WC2R 2LS, UK}
\affiliation{London Centre for Nanotechnology}

\author{Alex J. Vernon}
% \email{alexander.vernon@kcl.ac.uk}
\affiliation{Department of Physics, King's College London, Strand, London WC2R 2LS, UK}
\affiliation{London Centre for Nanotechnology}

\author{Francisco~J. Rodr\'iguez-Fortu\~no}
\email{francisco.rodriguez\_fortuno@kcl.ac.uk}
\affiliation{Department of Physics, King's College London, Strand, London WC2R 2LS, UK}
\affiliation{London Centre for Nanotechnology}

\date{\today}
\begin{abstract}
Electromagnetic quantities such as energy density, momentum, spin, and helicity bring meaning and intuition to electromagnetism and possess intricate interrelations, particularly prominent in complex non-paraxial near-fields. These quantities are conventionally expressed using electric and magnetic field vectors, yet the electric-magnetic basis is one among other often overlooked alternatives, including parallel-antiparallel and right-left-handed helicity bases, related to the parity and duality symmetries of electromagnetism. Projecting time-harmonic electromagnetic fields into a variety of bases allows re-interpreting established quantities and reveals underlying mathematical structures: a Bloch sphere which describes asymmetries in electromagnetic energy, a systematic path to unify and uncover relations between electromagnetic quantities, and the unlocking of symmetry-driven equations in light-matter interaction.
\end{abstract}

\maketitle%
% \section{Introduction}%
\paragraph*{Introduction}\!\!\!\!---\!\,\,\,%
Modern nanophotonics explores structured near fields in a non-paraxial regime, revealing phenomena like complex 3D polarizations, spin-orbit interactions, and oscillating energy flows, often leading to novel applications. Many physical quantities are defined and invoked in our attempts to gain intuition and give meaning to the field's behavior, such as the field's energy density, energy flow, linear momenta, spin and orbital angular momenta, helicity, and many others, which are often quadratic quantities in the fields. In-depth study of these quantities reveals intricate relations between them \cite{Cameron2012a,Bliokh2013,Aiello2015a,Barnett2012,Bliokh2011,Tang2010,Vernon2023,Vernon2023_2}. {The mathematical formulation of these quantities and their relations often seem convoluted, lacking discernible patterns or structure, complicating our efforts to understand them. In this work we present a theoretical framework, based on electromagnetic symmetries, to interpret, simplify and classify electromagnetic quadratic quantities, uncovering underlying mathematical structures, and leading to useful outcomes such as greatly simplified symmetry-based expressions for complex light-matter interactions.
}

\begin{figure}[b!]
    \centering
    \includegraphics{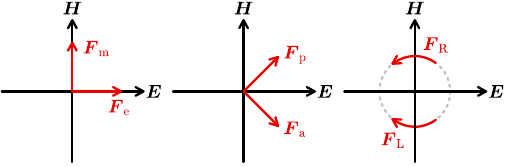}
    \caption{
    Components of the bispinor when represented in electric-magnetic (EM), parallel-antiparallel (PA) and right-left handed (RL) bases, drawn with respect to $\vc{E}$ and $\vc{H}$ vector-valued axes of the traditional EM basis.
    }
    \label{fig:C2basis}
\end{figure}%

\paragraph*{Bispinor formalism}\!\!\!\!---\!\,\,\,%
Bispinors, combining electric and magnetic fields into a single `wavefunction' $\vc{\psi}$, offer profound insights into electromagnetism's mathematical structure \cite{Alpeggiani2018,Bliokh2014,Bliokh2014a}. The foundation of this work is to embrace that, for monochromatic light, these bispinors live in a $\mathbb{C}^3\times\mathbb{C}^2\times\mathbb{L}^2(\mathbb{R}^3)$ vector space: $\mathbb{C}^3$ due to $\vc{E}$ and $\vc{H}$ being 3-dimensional complex phasors, $\mathbb{C}^2$ because $\vc{E}$ and $\vc{H}$ have a relative amplitude and phase with respect to each other, and $\mathbb{L}^2(\mathbb{R}^3)$ because each component is a square-integrable function of position (or momentum). In the literature, these bispinors are typically written in a basis in $\mathbb{C}^2$ space $\vc{\psi} =\tfrac{1}{2} (\sqrt{\varepsilon}\vc{E},\sqrt{\mu}\vc{H})^\intercal$ which separates electric and magnetic fields. But like any other $\mathbb{C}^2$ space, isomorphic with the Jones vector that describes 2D plane wave polarisation, different choices of basis can be made (see \cref{fig:C2basis}). Viewing an electromagnetic wave as composed of electric and magnetic fields (EM) is just as valid as considering it constituted by parallel and anti-parallel (PA), or right and left-handed (RL) fields. See {Supplemental Material (SM) section A} for a step-by-step justification {\cite{SM}}:%
\begin{equation}\label{eq:bases}
\begin{alignedat}{4}
    \vc{\psi}_\text{EM}(\vc{r})&=\pmqty{\vc{F}_\text{e}\\\vc{F}_\text{m}}&&=\frac{1}{2}\pmqty{\sqrt{\varepsilon}\vc{E}\\\sqrt{\mu}\vc{H}}\,,\\
    \vc{\psi}_\text{PA}(\vc{r})&=\pmqty{\vc{F}_\text{p}\\\vc{F}_\text{a}}&&=\frac{1}{2\sqrt{2}}\pmqty{\sqrt{\varepsilon}\vc{E}+\sqrt{\mu}\vc{H}\\\sqrt{\varepsilon}\vc{E}-\sqrt{\mu}\vc{H}}\,,\\
    \vc{\psi}_\text{RL}(\vc{r})&=\pmqty{\vc{F}_\text{R}\\\vc{F}_\text{L}}&&=\frac{1}{2\sqrt{2}}\pmqty{\sqrt{\varepsilon}\vc{E}+\ii\sqrt{\mu}\vc{H}\\\sqrt{\varepsilon}\vc{E}-\ii\sqrt{\mu}\vc{H}}\,.
\end{alignedat}    
\end{equation}%
The three bases EM, PA and RL relate to the fundamental electromagnetic symmetries: parity inversion $\hat{P}$: $\vc{r}\mapsto-\vc{r}$, time reversal $\hat{T}$: $t\mapsto-t$,
and discrete duality transformation $\hat{D}$. These act on the fields as
\begin{equation}
    \begin{split}
        \hat{D}:\;\qty(\sqrt{\varepsilon}\vc{E},\sqrt{\mu}\vc{H})^\intercal&\mapsto\qty(\sqrt{\mu}\vc{H},-\sqrt{\varepsilon}\vc{E})^\intercal,\\
        \hat{P}:\;\qty(\sqrt{\varepsilon}\vc{E},\sqrt{\mu}\vc{H})^\intercal&\mapsto\qty(-\sqrt{\varepsilon}\vc{E},\sqrt{\mu}\vc{H})^\intercal,\\
        \hat{T}:\;\qty(\sqrt{\varepsilon}\vc{E},\sqrt{\mu}\vc{H})^\intercal&\mapsto\qty(\sqrt{\varepsilon}\vc{E}^*,-\sqrt{\mu}\vc{H}^*)^\intercal,
    \end{split}
\end{equation}
so $\hat{D}$ is a 90 degree rotation in $\mathbb{C}^2$ space, while $\hat{P}$ is a mirror symmetry inverting the $\vc{E}$ axis. The components of the EM basis are eigenvectors of the $\hat{P}$ operator, RL basis of the $\hat{D}$ operator, and PA basis of the combined $\hat{P}\hat{D}$ operator.
Note that the fields $\vc{F}_\text{R}$ and $\vc{F}_\text{L}$ are similar to Riemann-Silberstein vectors \cite{BialynickiBirula_1994,BialynickiBirula_2003,Kaiser2004} but here defined using complex phasors $\vc{E}$ and $\vc{H}$.

\paragraph*{Field projections}\!\!\!\!---\!\,\,\,%
Any electromagnetic field $(\vc{E},\vc{H})${, including non-paraxial and structured near fields,} can be decomposed into a sum of two components, based on the different bases in $\mathbb{C}^2$ space:
\begin{equation}\label{eq:projectionsE}
\begin{alignedat}{4}
    \vc{E} &= \vc{E}_\text{e} &&+ \vc{E}_\text{m} &&= \quad\;\,\vc{E}&&+ \quad\;\,\vc{0} 
    \\
     &= \vc{E}_\text{p} &&+ \vc{E}_\text{a} &&= \frac{\vc{E} + \eta \vc{H}}{2} &&+ \frac{\vc{E} - \eta \vc{H}}{2} \\ &= \vc{E}_\text{R} &&+ \vc{E}_\text{L} &&= \frac{\vc{E} + \ii \eta \vc{H}}{2} &&+ \frac{\vc{E} - \ii \eta \vc{H}}{2}\,,
\end{alignedat}
\end{equation}
with the corresponding $\vc{H}_\text{e} = \vc{0}, \vc{H}_\text{m} = \vc{H}, \vc{H}_\text{p,a} = \pm \vc{E}_\text{p,a}/\eta$ and $\vc{H}_\text{R,L} = \mp \ii \vc{E}_\text{R,L}/\eta$, and $\eta=\sqrt{\mu/\varepsilon}$. These pairs $(\vc{E}_{i},\vc{H}_{i})$ represent the projection of the total electromagnetic field into each basis vector in $\mathbb{C}^2$ space, with $i=(\text{e},\text{m},\text{p},\text{a},\textsc{r},\textsc{l})$. For example, a linearly polarised field with $\vc{E} = E_0 \xhat$ and  $\vc{H} = E_0\yhat/\eta$ may be decomposed in the PA basis as the sum of parallel fields $(\vc{E}_\text{p},\vc{H}_\text{p}) = E_0 (\frac{\xhat+\yhat}{2},\frac{\xhat+\yhat}{2 \eta})$, which have $\vc{F}_\text{a} = 0$, plus antiparallel ones $(\vc{E}_\text{a},\vc{H}_\text{a}) = E_0 (\frac{\xhat-\yhat}{2},-\frac{\xhat-\yhat}{2 \eta})$, which have $\vc{F}_\text{p} = 0$. Each of these pairs of simultaneous electric and magnetic fields, given by $(\vc{E}_{i},\vc{H}_{i})$, are mathematical projections, not physically realizable in vacuum, except for the pairs $(\vc{E}_{\text{R}},\vc{H}_{\text{R}})$ and $(\vc{E}_{\text{L}},\vc{H}_{\text{L}})$ which correspond to realizable pure helicity fields \cite{Fernandez2014,Aiello2015a}. The total electromagnetic energy density $W$ is given by the norm $\norm{\vc{\psi}}^2$. Thanks to orthogonality and Parseval's identity, the energy density $W$ is the sum of the squared norm of each of the two $\vc{\psi}$ components in any of the three bases. Therefore, $W$, a quadratic quantity that is generally not linear with the fields, can be obtained \emph{by linear addition} in three different ways $W = W_\text{e} + W_\text{m} = W_\text{p} + W_\text{a} = W_\text{R} + W_\text{L}$, of which the first is well-known. A similar argument applies for the total spin angular momentum $\vc{S} = \vc{S}_\text{e} + \vc{S}_\text{m} = \vc{S}_\text{p} + \vc{S}_\text{a} = \vc{S}_\text{L} + \vc{S}_\text{R}$, and total stress tensor $\tens{T} = \tens{T}_\text{e} + \tens{T}_\text{m} = \tens{T}_\text{p} + \tens{T}_\text{a} = \tens{T}_\text{L} + \tens{T}_\text{R}$, by changing the dot product implicit in $W_i = |\vc{F}_i|^2 = \vc{F}_i^* \vdot \vc{F}_i$ into a cross or outer product. The individual components $W_i$, $\vc{S}_i$ and $\tens{T}_i$ (with a subscript $i=(\text{e},\text{m},\text{p},\text{a},\textsc{r},\textsc{l})$ corresponding to each of the six field projections) are the energy, spin, and stress tensor of each projection (using $\vc{E}_i$ and $\vc{H}_i$ from \cref{eq:projectionsE}), and can be defined laconically using the corresponding field coefficients $\vc{F}_i$ from \cref{eq:bases}:  
\begin{equation}\label{eq:projectionquadratics}
\begin{alignedat}{2}
W_i &= \vc{F}^*_i \vdot \vc{F}_i &&= \tfrac{1}{4}\left( \varepsilon \vc{E}_i^* \vdot \vc{E}_i + \mu \vc{H}_i^* \vdot \vc{H}_i \right)\,, \\
i \omega \vc{S}_i &= \vc{F}^*_i \!\cp\vc{F}_i &&= \tfrac{1}{4}\left( \varepsilon \vc{E}_i^* \!\cp\!\vc{E}_i + \mu \vc{H}_i^* \!\cp \vc{H}_i \right)\,, \\
\tens{T}_i + W_i\tens{I} &= \vc{F}^*_i \!\odot \vc{F}_i &&= \tfrac{1}{4}\left( \varepsilon \vc{E}_i^{\ast} \!\odot \vc{E}_i + \mu \vc{H}_i^{\ast} \!\odot \vc{H}_i\right) \,,
\end{alignedat}
\end{equation}
where $\!\odot$ represents the symmetric outer product defined as $\vc{A}\odot\vc{B}=\vc{A}\otimes\vc{B}+\vc{B}\otimes\vc{A}$, and $\otimes$ is the outer product. {Note $\vc{A}^{\ast}\odot\vc{A}=2\Re\lbrace \vc{A}^{\ast}\otimes\vc{A} \rbrace$ and $\vc{A}^{\ast}\times\vc{A} = \ii\Im\lbrace\vc{A}^{\ast}\times\vc{A}\rbrace$.}
\paragraph*{Table of quadratic quantities}\!\!\!\!---\!\,\,\,%
% QUADRATIC QUANTITIES TABLE
% --------------------------
\begin{table}
\centering
\renewcommand{\arraystretch}{1.3} % Default value: 1
\begin{tabular}{|c|c|c|c|c|c|}
\hline
& \text{Quantity} & \text{Definition} & $\hat{D}$ & $\hat{P}$ & $\hat{T}$ \\
\hline
\cellcolor{proper}\( W_0 \) & \( W \) & \( \frac{1}{4}({\varepsilon}\abs{\vc{E}}^2 + {\mu}\abs{\vc{H}}^2) \) & \cellcolor{proper}$+$ & \cellcolor{proper}$+$ & \cellcolor{proper}$+$ \\
\hline
\cellcolor{altroo}\( W_1 \) & \( W_\text{e} - W_\text{m} \) & \( \frac{1}{4}({\varepsilon}\abs{\vc{E}}^2 - {\mu}\abs{\vc{H}}^2) \) &\cellcolor{altroo}$-$ &\cellcolor{altroo}$+$ &\cellcolor{altroo}$+$ \\
\hline
\cellcolor{crlroo}\( W_2 \) & \(\mathrlap{W_\text{a}}{\phantom{W_\text{e}}} - \mathrlap{W_\text{p}}{\phantom{W_\text{m}}} \) & \(-\omega\mathfrak{S}_\text{reac}=- \frac{1}{2}\Re( \sqrt{\varepsilon\mu}{\vc{E}^*\!\vdot\vc{H}} ) \) &\cellcolor{crlroo}$-$ &\cellcolor{crlroo}$-$ &\cellcolor{crlroo}$-$ \\
\hline
\cellcolor{pseudo}\( W_3 \) & \(\mathrlap{W_\text{R}}{\phantom{W_\text{e}}} - \mathrlap{W_\text{L}}{\phantom{W_\text{m}}} \) & \( \omega\mathfrak{S}=-\frac{1}{2}\Im( \sqrt{\varepsilon\mu}{\vc{E}^*\!\vdot\vc{H}} ) \) & \cellcolor{pseudo}$+$ & \cellcolor{pseudo}$-$ & \cellcolor{pseudo}$+$ \\
\hline
\cellcolor{proper}\( \vc{p}_0 \) & \(\vc{p}\) & \( \frac{1}{4\omega}\Im\bqty{{\varepsilon}\vc{E}^\ast\!\!\vdot\!(\grad)\vc{E} + {\mu}\vc{H}^\ast\!\!\vdot\!(\grad)\vc{H}} \) &\cellcolor{proper}$+$ &$-$ &$-$ \\
\hline
\cellcolor{altroo}\( \vc{p}_1 \) & \(\vc{p}_\text{e} - \vc{p}_\text{m}\) & \( \frac{1}{4\omega}\Im\bqty{{\varepsilon}\vc{E}^\ast\!\!\vdot\!(\grad)\vc{E} - {\mu}\vc{H}^\ast\!\!\vdot\!(\grad)\vc{H}} \) &\cellcolor{altroo}$-$ &$-$ &$-$ \\
\hline
\cellcolor{crlroo}\( \vc{p}_2 \) & \(\mathrlap{\vc{p}_\text{a}}{\phantom{\vc{p}_\text{e}}} - \mathrlap{\vc{p}_\text{p}}{\phantom{\vc{p}_\text{m}}}  \) & \( \frac{1}{4\omega}\Im\{ \sqrt{\varepsilon\mu}\bqty{\vc{E}\!\vdot\!(\grad)\vc{H}^\ast\!-\vc{H}^*\!\!\vdot\!(\grad)\vc{E}} \}\! \) &\cellcolor{crlroo}$-$ &$+$ &$+$ \\
\hline
\cellcolor{pseudo}\( \vc{p}_3 \) & \(\mathrlap{\vc{p}_\text{R}}{\phantom{\vc{p}_\text{e}}} - \mathrlap{\vc{p}_\text{L}}{\phantom{\vc{p}_\text{m}}} \) & \( \frac{1}{4\omega}\Re\{  \sqrt{\varepsilon\mu}\bqty{\vc{E}\!\vdot\!(\grad)\vc{H}^\ast\!-\vc{H}^*\!\!\vdot\!(\grad)\vc{E}} \}\! \) &\cellcolor{pseudo}$+$ &$+$ &$-$ \\
\hline
\cellcolor{pseudo}\( \vc{S}_0 \) &  \(\vc{S}\) & \( \frac{1}{4\omega}\Im\left(\varepsilon{\vc{E}^*\!\cp\vc{E}} + \mu{\vc{H}^*\!\cp\vc{H}}) \right) \) &\cellcolor{pseudo}$+$ & $+$ &$-$ \\
\hline
\cellcolor{crlroo}\( \vc{S}_1 \) & \(\vc{S}_\text{e} - \vc{S}_\text{m}\) & \( \frac{1}{4\omega}\Im\left(\varepsilon{\vc{E}^*\!\cp\vc{E}} - \mu{\vc{H}^*\!\cp\vc{H}}) \right) \)  & \cellcolor{crlroo}$-$ & $+$ & \cellcolor{crlroo}$-$ \\
\hline
\cellcolor{altroo}\( \vc{S}_2 \) & \(\mathrlap{\vc{S}_\text{a}}{\phantom{\vc{S}_\text{e}}} - \mathrlap{\vc{S}_\text{p}}{\phantom{\vc{S}_\text{m}}} \) & \(  \frac{1}{\omega c}\Im\vc{\Pi}=\frac{1}{2\omega}\Im(\sqrt{\varepsilon\mu} \vc{E}\cp\vc{H}^* ) \) & \cellcolor{altroo}$-$ & $-$ & \cellcolor{altroo}$+$ \\
\hline
\cellcolor{proper}\( \vc{S}_3 \) & \(\mathrlap{\vc{S}_\text{R}}{\phantom{\vc{S}_\text{e}}} - \mathrlap{\vc{S}_\text{L}}{\phantom{\vc{S}_\text{m}}}\) & \( \frac{1}{\omega c}\Re\vc{\Pi}=\frac{1}{2\omega}\Re(\sqrt{\varepsilon\mu} \vc{E}\cp\vc{H}^* ) \) &\cellcolor{proper}$+$ &$-$ &$-$ \\
\hline
\cellcolor{proper}\( \tens{T}_0\) & \(\tens{T} \) & \( \frac{1}{4}(\varepsilon \vc{E}^{\ast} \!\odot \vc{E} + \mu \vc{H}^{\ast} \!\odot \vc{H})-W_0\tens{I} \) &\cellcolor{proper}$+$ &\cellcolor{proper}$+$ &\cellcolor{proper}$+$ \\
\hline
\cellcolor{altroo}\( \tens{T}_1 \) & \( \tens{T}_\text{e} - \tens{T}_\text{m} \) & \( \frac{1}{4}(\varepsilon \vc{E}^{\ast}\! \odot \vc{E} - \mu \vc{H}^{\ast}\! \odot \vc{H})-W_1\tens{I} \)&\cellcolor{altroo}$-$ &\cellcolor{altroo}$+$ &\cellcolor{altroo}$+$ \\
\hline
\cellcolor{crlroo}\( \tens{T}_2 \) & \(\mathrlap{\tens{T}_\text{a}}{\phantom{\tens{T}_\text{e}}} - \mathrlap{\tens{T}_\text{p}}{\phantom{\tens{T}_\text{m}}}\) & \(- \frac{1}{2}\Re(\sqrt{\varepsilon\mu}{\vc{E}^*\!\odot\vc{H}})- W_2\tens{I}\) &\cellcolor{crlroo}$-$ &\cellcolor{crlroo}$-$ &\cellcolor{crlroo}$-$ \\
\hline
\cellcolor{pseudo}\( \tens{T}_3 \) & \(\mathrlap{\tens{T}_\text{R}}{\phantom{\tens{T}_\text{e}}} - \mathrlap{\tens{T}_\text{L}}{\phantom{\tens{T}_\text{m}}}\) & \( -\frac{1}{2}\Im(\sqrt{\varepsilon\mu}{\vc{E}^*\!\odot\vc{H}})- W_3\tens{I} \) & \cellcolor{pseudo}$+$ & \cellcolor{pseudo}$-$ & \cellcolor{pseudo}$+$ \\
\hline
\end{tabular}
\caption{Quadratic quantities in the fields. We indicate symmetries under D, P, and T transformations. The symmetric outer product is to be understood as $\vc{A}\odot\vc{B}\equiv\vc{A}\otimes\vc{B}+\vc{B}\otimes\vc{A}$ where $\otimes$ is the outer product, while $\vc{A}\vdot(\grad)\vc{B} \equiv \sum_i ({A}_i \grad {B}_i)$.}
\label{tab:quadratics}
\end{table}%
% --------------------------
Even more interesting is the \emph{subtraction} of corresponding quantities $W_i$, $\vc{S}_i$ and $\tens{T}_i$, yielding a collection of physically meaningful quadratic quantities that are even or odd with respect to each of the symmetries. These are listed exhaustively in \cref{tab:quadratics}, {and are all purely real and defined for general non-paraxial fields}. A more rigorous mathematical motivation for performing these subtractions is given in the {SM section B \cite{SM}}. {Remarkably, most rows in \cref{tab:quadratics} correspond to well-known quadratic quantities, such as the complex Poynting vector $\vc{\Pi} = \frac{1}{2}\vc{E}\cp\vc{H}^*$ (whose real and imaginary parts appear in $\vc{S}_\text{R} - \vc{S}_\text{L}$ and $\vc{S}_\text{a} - \vc{S}_\text{p}$ respectively), the electromagnetic helicity $\mathfrak{S}$, known to be proportional to $W_\text{R} - W_\text{L}$ \cite{Trueba1996,Afanasiev1996,Berry2019}, and the reactive helicity \cite{NietoVesperinas2021}, also called magnetoelectric energy density \cite{Bliokh2014}, which appears in $W_\text{a} - W_\text{p}$. Chiral momentum corresponds to $\vc{p}_\text{R}-\vc{p}_\text{L}$ \cite{Bliokh2014,Vernon2023}.} We believe this provides a systematic and elegant symmetry-based way to interpret these quantities. In some cases, similar decompositions were made in different contexts, such as the helicity decomposition of the Poynting vector \cite{Aiello2015a} related to $\vc{S}_3$, and that of the local wavevector \cite{Berry2019} related to $\vc{p}_0$. \Cref{tab:quadratics} expresses all quantities in the familiar EM basis (i.e.\ in terms of electric and magnetic fields). But we can break free of such EM bias and express the same quantities in other bases, by using $\vc{F}_i$ from \cref{eq:bases}. This leads to simpler expressions as demonstrated for the energy densities in \cref{tab:energystokes} (the extension to the other quadratic quantities is given in the {SM section B4 \cite{SM}}). A clear pattern emerges in \cref{tab:energystokes}: the quadratic quantities, expressed in different bases, exhibit an \emph{exact} mathematical analogy with the Stokes plane polarisation parameters, motivating our subscript labelling $A=0,1,2,3$ in the first column of \cref{tab:quadratics,tab:energystokes}. 

% ENERGY STOKES TABLE
% ------------------------
\begin{table}
\centering
\renewcommand{\arraystretch}{1.3} % Default value: 1
\begin{tabular}{|c|c|c|c|}
\hline
&  \text{EM basis} & \text{PA basis} & \text{RL basis} \\
\hline
\cellcolor{proper}\( W_0 \) & \cellcolor{proper}\( \vc{F}_\text{e}^*\!\vdot\!\vc{F}_\text{e}^{\vphantom{*}} + \vc{F}_\text{m}^*\!\vdot\!\vc{F}_\text{m}^{\vphantom{*}} \) & \cellcolor{proper}\( \vc{F}_\text{a}^*\!\vdot\!\vc{F}_\text{a}^{\vphantom{*}} + \vc{F}_\text{p}^*\!\vdot\!\vc{F}_\text{p}^{\vphantom{*}} \) & \cellcolor{proper}\( \vc{F}_\text{R}^*\!\vdot\!\vc{F}_\text{R}^{\vphantom{*}} + \vc{F}_\text{L}^*\!\vdot\!\vc{F}_\text{L}^{\vphantom{*}} \) \\
\hline
\cellcolor{altroo}\( W_1 \) & \cellcolor{altroo}\( \vc{F}_\text{e}^*\!\vdot\!\vc{F}_\text{e} - \vc{F}_\text{m}^*\!\vdot\!\vc{F}_\text{m} \) & \( 2\Re(\vc{F}_\text{p}^*\vdot\vc{F}_\text{a}^{\vphantom{*}}) \) & \( 2\Re(\vc{F}_\text{R}^*\vdot\vc{F}_\text{L}^{\vphantom{*}}) \) \\
\hline
\cellcolor{crlroo}\( W_2 \) & \(- 2\Re( {\vc{F}_\text{e}^*\vdot\vc{F}_\text{m}} ) \) & \cellcolor{crlroo}\( \vc{F}_\text{a}^*\!\vdot\!\vc{F}_\text{a}^{\vphantom{*}} - \vc{F}_\text{p}^*\!\vdot\!\vc{F}_\text{p}^{\vphantom{*}} \) & \(2\Im(\vc{F}_\text{R}^* \vdot \vc{F}_\text{L}^{\vphantom{*}})\) \\
\hline
\cellcolor{pseudo}\( W_3 \) & \(- 2\Im( {\vc{F}_\text{e}^*\!\vdot\vc{F}_\text{m}} ) \) & \( 2\Im(\vc{F}_\text{p}^*\vdot\vc{F}_\text{a}^{\vphantom{*}}) \) & \cellcolor{pseudo}\( \vc{F}_\text{R}^*\!\vdot\!\vc{F}_\text{R}^{\vphantom{*}} - \vc{F}_\text{L}^*\!\vdot\!\vc{F}_\text{L}^{\vphantom{*}} \) \\
\hline
\end{tabular}
\caption{
Energy densities in different bases, revealing the similarity with Stokes vectors (for an extension to the other quadratic quantities see {SM section B4 \cite{SM}}).
}
\label{tab:energystokes}
\end{table}
% ------------------------

\begin{figure*}[!t]
    \centering
    \includegraphics{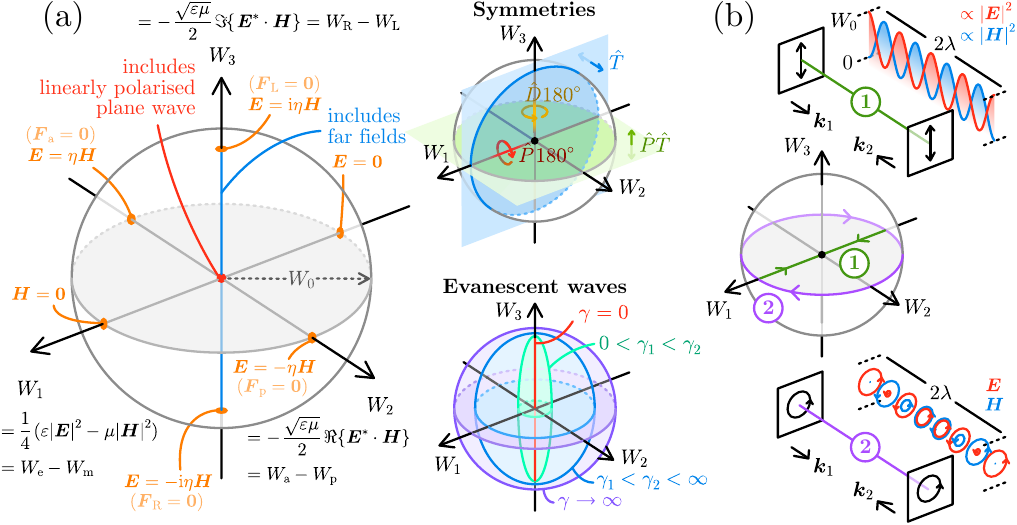}
    \caption{
    (a) The energy symmetry sphere (ESS), axes defined by the energy densities $W_{1-3}$, and its special points (orange).
    Far fields lie along the $W_3$ axis while a linearly polarised plane wave (preserving all symmetries) resides at the sphere centre.
    The upper inset depicts the effect of symmetry transformations on a field plotted in the sphere. The lower inset shows how evanescent waves can map to any point on an ellipsoid surface within the ESS, depending on the decay constant.
    (b) Traces on the ESS produced by two standing wave examples (top and bottom).
    The upper example (green line) maps the optical axis (a $2\lambda$ section) of two counter-propagating plane waves with identical linear polarisations to the sphere, giving a trace along the $W_1$ axis as energy is shared space-periodically among $\vc{E}$ and $\vc{H}$.
    In the lower example, a standing wave where both plane waves are cicularly polarised with the same spin direction (corresponding to the purple line) produces a trace around the equator of the ESS.
    Arrowheads (on the polarisation circles) indicate the phase of the $\vc{E}$ and $\vc{H}$ fields at various points along the standing wave, showing that the time-dependent electric and magnetic field vectors are always (anti)parallel (hence the ESS trace lies at $W_3=0$).
    \label{fig:symmetrysphere} %label inside caption, otherwise doesnt work! (very interesting)
    }
    
\end{figure*}
\paragraph*{Bloch sphere}\!\!\!\!---\!\,\,\,%
Given the mathematical analogy with Stokes parameters, it is natural to plot the real vector $\vc{W} = (W_1, W_2, W_3)$ in $W$-space. As with Stokes parameters, this vector is contained on a sphere of radius $W_0$, analogous to the Poincare or Bloch sphere. We term this the \emph{energy symmetry sphere}, depicted in \cref{fig:symmetrysphere}, as it reveals how the energy is distributed among the symmetries. Time-reversal $\hat{T}$ mirrors the sphere on the $(W_1,W_3)$ plane, parity $\hat{P}$ rotates it 180 degrees around the $W_1$ axis, and duality $\hat{D}$ rotates it around $W_3$. Unlike the Stokes parameters in the Poincare sphere, which only describe 2D polarisation, the symmetry sphere can describe \emph{any} electromagnetic field $(\vc{E},\vc{H})$, including non-paraxial, structured and near-fields, with arbitrary three-dimensional polarisation. The absolute spatial orientation of polarisation ellipses is not relevant to the energy symmetry sphere, but the relative orientation and phase between $\vc{E}$ and $\vc{H}$ plays a crucial role. Unlike the Poincare sphere, fully polarised electromagnetic fields can exist anywhere in the surface \emph{or inside} the sphere. The distance to the origin indicates how asymmetrically-distributed the energy is. Linearly polarised plane waves lie at the origin of the symmetry sphere, with maximum symmetry $(W_\text{e}=W_\text{m}=W_\text{p}=W_\text{a}=W_\text{R}=W_\text{L}=W/2)$. All possible homogeneous plane waves and paraxial far fields exist along the $W_3$ axis ($W_1=W_2=0$) with circular polarisation at the poles, while homogeneous evanescent waves can stray away from the $W_3$ axis, more so when more confined. The states at the very surface of the symmetry sphere (where $W_1^2+W_2^2+W_3^2 = W_0^2$) correspond to maximum-asymmetry between the energies in the two orthogonal states of some basis in $\mathbb{C}^2$ space. As shown in the {SM section B5 \cite{SM}}, this corresponds to $\vc{E} = \delta \eta \vc{H}$ with $\delta$ a complex scalar. This means that states in the surface have fewer degrees of freedom, the electric and magnetic fields are proportional and share the same polarisation structure. This was pointed out for pure helicity fields \cite{BialynickiBirula_2003,Kaiser2004}, which is a particular case of being at the surface (north and south poles). In contrast, any states \emph{inside} the sphere must have linearly independent fields. \Cref{fig:symmetrysphere} depicts special points on the surface where $\vc{F}_i=0$ and hence $W_i=0$ for $i=(\text{e},\text{m},\text{p},\text{a},\textsc{r},\textsc{l})$, corresponding to the field projections from \cref{eq:projectionsE}. The polar angle with respect to the axis joining the origin to any $\vc{F}_i=0$ point on the surface is related to the amplitude ratio between the two orthogonal components in that basis, while the azimuthal angle around the axis represents their phase difference (see {SM section B5 \cite{SM}}). 

Any spatial profile of electromagnetic fields given by ${(\vc{E}(\vc{r}),\vc{H}(\vc{r}))}$ can be mapped into the symmetry sphere, with every position $\vc{r}$ mapped to a point $\vc{W}(\vc{r})$ within the sphere. Doing this on simple standing waves leads to simple curves as in \cref{fig:symmetrysphere}(b), while doing it on more complicated fields, such as dipolar fields, leads to intriguing and sometimes beautiful patterns (see {SM section B6 \cite{SM}}).

An electromagnetic field $(\vc{E},\vc{H})$ cannot be superimposed onto its mirror image (accounting for the pseudovector nature of $\vc{H}$) if either $W_3 \neq 0$ or $ W_2 \neq 0$, so both quantities (helicity and reactive helicity) represent a form of time-averaged chirality. Indeed, chiral gradient forces are directed along $\grad W_3$ and $\grad W_2$ for chiral and non-reciprocal particles, respectively \cite{Golat2023}. While both $W_{2,3}$ change sign under parity, only $W_2$ changes sign under time reversal, making $W_2$ PT-symmetric \cite{Bliokh2014}.

The energy symmetry sphere requirement $W_1^2+W_2^2+W_3^2\leq W_0^2$ represents a fundamental limit. For instance, maximum helicity $W_3 = \pm W_0$ is attained when $W_1=W_2=0$, requiring equal electric and magnetic energies, while purely electric or magnetic fields $W_1 = \pm W_0$ (where $\vc{H}$ or $\vc{E}$ equal $0$) require $W_2=W_3=0$, and hence cannot be locally chiral.

% \section{Plane waves and evanescent waves}%
\paragraph*{Plane waves and evanescent waves}\!\!\!\!---\!\,\,\,%
A deep relationship between the quadratic quantities $W_A$ and the Stokes polarisation parameters $\mathcal{S}_A$ is revealed if we consider a plane wave, whose fields are:
\begin{equation}\label{eq:planewaveEfield}
\begin{split} \sqrt{\varepsilon}\vc{E}(\vc{r}) &= \pmqty{0,A_\text{v},-A_\text{h}}^\intercal \ee^{i k x}, \\\sqrt{\mu}\vc{H}(\vc{r}) &= \pmqty{0,A_\text{h},A_\text{v}}^\intercal \ee^{i k x}.
\end{split}    
\end{equation}
The Stokes parameters which describe the polarisation of this plane wave are defined as
\begin{alignat*}{4}
\mathcal{S}_0 &=\abs{A_\text{h}}^2+\abs{A_\text{v}}^2,%\\
&\quad
\mathcal{S}_2  &=2 \Re(A_\text{h}^*A_\text{v})\,, \\
\mathcal{S}_1&=\abs{A_\text{h}}^2-\abs{A_\text{v}}^2\,,%\\
&\quad
\mathcal{S}_3  &=2 \Im(A_\text{h}^*A_\text{v})\,.
\end{alignat*}
The quadratic quantities from \cref{tab:quadratics} evaluated for this simple plane wave field are given in \cref{tab:quadplanewave}, illuminating the relation to the Stokes parameters. As mentioned earlier, plane waves have {$W_1=W_2=0$}, but also ${\vc{S}_1=\vc{S}_2=0}$. Even more intricate relations appear for an evanescent wave with effective index $n=k_x/k$ (associated to $\kappa=\sqrt{n^2-1}$) such that $\vc{k}=k(n \hat{\vc{x}} + \ii \kappa \hat{\vc{z}})$, with arbitrary polarisation (but still ensuring that \( \grad \vdot \vc{E} = \grad \vdot \vc{H} = 0 \)):
\begin{equation}\label{EHevanescent}
\begin{split}
     \sqrt{\varepsilon}\vc{E}&=\pmqty{
        \ii \kappa A_\text{h},A_\text{v},-n A_\text{h} 
    }^\intercal \ee^{\ii \vc{k} \vdot \vc{r} }, \\
    \sqrt{\mu}\vc{H}&=\pmqty{
        -\ii \kappa A_\text{v},A_\text{h},n A_\text{v} 
    }^\intercal \ee^{\ii \vc{k} \vdot \vc{r} },
\end{split}
\end{equation}
where $\ee^{\ii \vc{k} \vdot \vc{r} } = \ee^{\ii k n x - k \kappa z}$ and \cref{eq:planewaveEfield} is recovered when $(n=1, \kappa=0)$. The quadratic quantities for this evanescent wave are given in \cref{tab:quadplanewave}, neatly collecting many recent findings related to evanescent waves, namely, the transverse spin ($\vc{S}_0\vdot\uv{y}$), associated with spin-momentum locking of guided modes  \cite{Bliokh2012,Bliokh2014a,Aiello2015,Bliokh2015,Bliokh2015a,Bekshaev2015,Eismann2020}, the transverse Poynting vector ($\vc{S}_3\vdot\uv{y}$), also known as Belinfante's momentum \cite{Bliokh2014,Liu2018,Antognozzi2016,Wei2020}, the imaginary Poynting vector structure of evanescent waves in the direction of decay ($\vc{S}_2\vdot\uv{z}$) \cite{Wei2018,Wei2018a,NietoVesperinas2021}, and in addition to these, new aspects of evanescent waves are revealed, such as the transverse components of $\vc{S}_1 = \vc{S}_\text{e} - \vc{S}_\text{m}$.
\begin{table*}[ht]
    \centering
    \renewcommand{\arraystretch}{1.4} % Default value: 1
    \begin{tabular}{|c|c||c|c||c|c|c|}
    \hhline{--::--::--}
    \multicolumn{2}{|c||}{Quantity} & \multicolumn{2}{c||}{Plane wave} & \multicolumn{2}{c|}{Evanescent wave}\\
    \hhline{==::==::==}
    $W_{0/3}$&$ W_{1/2}$ & $\frac{1}{2}\mathcal{S}_{0/3}$&$ 0$ 
    &$\frac{n^2}{2} \ee^{-2 k\kappa z}  \mathcal{S}_{0/3}$&$ \frac{\kappa^2}{2} \ee^{-2 k\kappa z}  \mathcal{S}_{1/2}$
    \\
    \hhline{--||--||--}
    $\vc{S}_{0/3}$&$\vc{S}_{1/2}$ & $ \frac{1}{2 \omega}({\mathcal{S}_{3/0} , 0 , 0})^\intercal\!\!$&$ ({ 0 , 0 , 0 })^\intercal\!\!$ 
    &$\frac{n}{2 \omega}\ee^{-2 k\kappa z}({\mathcal{S}_{3/0} , \pm \kappa \mathcal{S}_{0/3} , 0})^\intercal\!\!$&$ \frac{\kappa}{2 \omega}\ee^{-2 k\kappa z}({0,\pm n  \mathcal{S}_{1/2} ,  \mathcal{S}_{2/1}})^\intercal\!\!$
    \\
    \hhline{--||--||--}
     $\tens{T}_{0/3}$&$ \tens{T}_{1/2}$ & $ \frac{1}{2} \pmqty{ -\mathcal{S}_{0/3} & 0 & 0\\ 0 & 0 & 0\\ 0 & 0 & 0}\!\!$&$ \frac{1}{2} \pmqty{0 & 0 & 0\\ 0 & -\mathcal{S}_{1/2} & \mp\mathcal{S}_{2/1}\\ 0 & \mp\mathcal{S}_{2/1} & \mathcal{S}_{1/2}}$ 
     &
     $ \frac{\ee^{-2 k\kappa z}}{2} \pmqty{ -\mathcal{S}_{0/3} & \kappa \mathcal{S}_{3/0}  & 0\\  \kappa \mathcal{S}_{3/0} & -\kappa^2 \mathcal{S}_{0/3} & 0\\ 0 & 0 & 0}\!\!$&$ \frac{\ee^{-2 k\kappa z}}{2} \pmqty{0 & 0 & 0\\ 0 & -n^2 \mathcal{S}_{1/2} & \mp n \mathcal{S}_{2/1}\\ 0 & \mp n \mathcal{S}_{2/1} & \mathcal{S}_{1/2}}$\\
    \hhline{|--||--||--|}
    \end{tabular}
    \caption{Representation of quadratic quantities in a plane wave and an evanescent wave. The momenta-like quantities $\vc{p}_A$ can be obtained from $\vc{p}_A=W_A\Re(\vc{k})/\omega$. {Plane wave properties are recovered from evanescent wave ones when $\kappa=0$ and $n=1$.}}
    \label{tab:quadplanewave}
\end{table*}

\paragraph*{Relating quadratic quantities}\!\!\!\!---\!\,\,\,%
The quadratic quantities defined in \cref{tab:quadratics} are intimately related to one another via their curl and divergence. Below is a list of curls and divergences of $\vc{S}_A$, $\vc{p}_A$, and $\tens{T}_A$ valid in free space. Starting with the spin-like quantities, we have:
\begin{equation}
    \label{eq:spindiffs}
\begin{alignedat}{2}
\curl \vc{S}_0&=2(k\vc{S}_3-\vc{p}_0) \,,\quad & \div \vc{S}_0&=0 \,,\\
\curl \vc{S}_1&=-{2}\vc{p}_1  \,,\quad & \div \vc{S}_1&={+}\tfrac{2}{c}W_2 \,,\\
\curl \vc{S}_2&=-{2}\vc{p}_2  \,,\quad & \div \vc{S}_2&=-\tfrac{2}{c}W_1 \,,\\
\curl \vc{S}_3&=2(k\vc{S}_0-\vc{p}_3) \,,\quad & \div \vc{S}_3&=0\,.
\end{alignedat}
\end{equation}
These equations, with divergences seen as continuity conditions, elegantly incorporate many known relations and facts, such as: the Poynting theorem for time-harmonic fields in free space ($\div \vc{S}_3=0$) and the known fact that reactive power (imaginary Poynting vector $\propto \vc{S}_2$ \cite{Xu2019,Zhou2022}) {flows out of regions with high magnetic energy into regions of high electric energy.} The curl equations include the known Belinfante-Rosenfeld decomposition of the real Poynting vector ($\omega c \vc{S}_3$) into canonical momentum ($\vc{p}_0$) and spin momentum ($\curl\vc{S}_0$) 
\cite{Belinfante1940,Berry2009,Soper_book,Bliokh2013,Bliokh2014a}, and the lesser-known decomposition of total spin ($\vc{S}_0$) into canonical spin ($\propto\vc{p}_3$) and Poynting spin ($\curl \vc{S}_3$) \cite{Bliokh2014,Vernon2023}. {Note how \cref{eq:spindiffs} forms two pairs $A=(0,3)$ and $(1,2)$, linking quantities within each pair together, but not between the two pairs.}

Combining the previous results together with the identity $\curl(\curl\vc{A})=\grad(\div\vc{A})-\grad^2\vc{A}$ we can obtain the curls of $\vc{p}_A$, while all $\div\vc{p}_A=0$ from \cref{eq:spindiffs}. Finally, the tensor divergences are:
\begin{equation}\label{eq:tensorcontinuity}
    \begin{alignedat}{2}
    \div\tens{T}_{0/3}=0\,,\quad  & \div\tens{T}_{1/2}=\mp2k\omega\vc{S}_{2/1}\,.
    \end{alignedat}
\end{equation}
{Intriguingly,} \cref{eq:spindiffs} implies a Helmholtz/Poisson equation for the spin-like quantities with double and zero frequency as follows: 
\begin{equation}\label{eq:helmholtz}
    \begin{split}
       (\laplacian+4k^2)\vc{S}_{0/3}&=4k\vc{p}_{3/0}+2\curl\vc{p}_{0/3}\,, \\ 
       (\laplacian+0k^2)\vc{S}_{1/2}&=\pm\tfrac{2}{c}\grad W_{2/1}\pm2\curl\vc{p}_{1/2}\,.
    \end{split}
\end{equation}
{meaning that the total spin ($\vc{S}_0$) and the Poynting vector ($\omega c \vc{S}_3$) form propagating waves, while $\vc{S}_1$ and $\vc{S}_2$ are static-like, and quickly tend to zero in the far-field, like $W_1$ and $W_2$. In fact, all quantities with $A=1,2$ are associated with what the literature refers to as reactive quantities, like reactive power and reactive helicity, known to be only present in near fields.}
A partial version of the first line in \cref{eq:helmholtz} was presented in \cite{Shi2021}, with a claim of equivalence to Maxwell's equations and with calculations of extraordinary spin and momentum which were scrutinised in \cite{Bekshaev2022}. In fact, some signs that are crucial to Maxwell's equations do not match those in \cref{eq:spindiffs}. However, we can define new complex quantities $\vc{S}_A \pm \ii \vc{S}_B$, $W_A \pm \ii W_B$, and $\vc{p}_A \pm \ii \vc{p}_B$ which exactly match all signs in Maxwell's equations (see {SM section B7 \cite{SM}}). 
\Cref{eq:spindiffs,eq:tensorcontinuity,eq:helmholtz} laconically unify many known relations, while others are novel to our knowledge, and we highlight how elegantly and systematically they all arise from the quadratic quantities in \cref{tab:quadratics}.

\paragraph*{Relativistic interpretation}\!\!\!\!---\!\,\,\,%
In relativistic electrodynamics, the full energy-momentum-stress tensor $T^{\mu\nu}$ represents the flux of 4-momentum:
\begin{equation*}
    \spmqty{\text{4-momentum flux}}=\spmqty{\text{energy density}&\text{energy flux}/c\\c(\text{momentum density})&\text{momentum flux}}.
\end{equation*}
When energy and momentum are conserved, the continuity equation for this tensor becomes $\partial_\nu T^{\mu\nu}=0$, which can be written in general (no time average) as
\begin{equation}
    \label{eq:relativitycontinuityequations}
    \begin{split}
        \pmqty{\partial_\nu T^{0\nu}\\\partial_\nu T^{i\nu}\uv{e}_i}=\pmqty{\frac{1}{c}\qty(\pdv{t}W+\div\vc{\varPi})\\\pdv{t}\vc{p}+\div\tens{T}}=0\,,
    \end{split}
\end{equation}
where $W(t)$, $\vc{p}(t)$, $\vc{\varPi}(t)$ and $\tens{T}(t)$ are instantaneous energy and momentum densities, Poynting vector and Maxwell's stress tensor respectively. In the case of time-harmonic fields, the time derivatives of time-averaged quantities vanish. \Cref{eq:spindiffs,eq:tensorcontinuity}, when compared to \cref{eq:relativitycontinuityequations}, show that it is possible to define a tensor and pseudotensor that obey the continuity equation $\partial_\nu T^{\mu\nu}=0$ as:
\begin{alignat}{2}
    (T_0^{\mu\nu})&=\pmqty{W_0&\omega\vc{S}_3\\c\vc{p}_0&\tens{T}_0},
    \quad&(T_3^{\mu\nu})&=\pmqty{W_3&\omega\vc{S}_0\\c\vc{p}_3&\tens{T}_3}.
\end{alignat}
This illuminates the known fact that $W_0$ (energy) and $W_3$ (helicity) are densities of conserved quantities \cite{Cameron2012,Bliokh2013,Cameron2014a,Fernandez2014,NietoVesperinas2015,NietoVesperinas2015_2} whose flow is $\omega \vc{S}_3$ ($\propto$ Poynting vector) and $\omega \vc{S}_0$ ($\propto$ total spin), respectively (note the reversal $0 \leftrightarrow 3$), while $c \vc{p}_0$ (electromagnetic linear momentum density) and $c \vc{p}_3 = c (\vc{p}_R - \vc{p}_L)$ (chiral momentum) are conserved vector densities, with flow tensors $\tens{T}_0$ and $\tens{T}_3$. Notice that $T_0^{\mu\nu} = T_\text{R}^{\mu\nu} + T_\text{L}^{\mu\nu}$ and $T_3^{\mu\nu} = T_\text{R}^{\mu\nu} - T_\text{L}^{\mu\nu}$ can be added or subtracted to get the energy-momentum tensors of the pure helicity components $T_\text{R/L}^{\mu\nu}$, such that
\begin{equation}
    (T_{0/3}^{\mu\nu})=\pmqty{W_\text{R}\pm W_\text{L}&\omega(\vc{S}_\text{R}\mp\vc{S}_\text{L})\\c(\vc{p}_\text{R}\pm\vc{p}_\text{L})&\tens{T}_\text{R}\pm\tens{T}_\text{L}}.
\end{equation}
This provides a relativistic framework for the quadratic quantities related to $A=0,3$, but leaves out the $A=1,2$ quantities. These do not admit a 4-tensor grouping. However, $W_1 = W_\text{e} - W_\text{m}$ and $W_2 = W_\text{p} - W_\text{a}$ are the time-average of the well-known relativistic electromagnetic invariants \cite{Landau1994}:
\begin{equation}
    \begin{split}
        W_1&=\expval*{\varepsilon\norm{\vc{\mathcal{E}}(t)}^2-\tfrac{1}{\mu}\norm{\vc{\mathcal{B}}(t)}^2},\\
        W_2&=-\expval*{\tfrac{1}{\eta}\vc{\mathcal{B}}(t)\vdot\vc{\mathcal{E}}(t)},\\
    \end{split}
\end{equation}
where $\vc{\mathcal{E}}(t)$ and $\vc{\mathcal{H}}(t)$ are the time-varying electric and magnetic fields. Hence, Lorentz boosts change both $W_0$ and $W_3$, leaving $W_{1,2}$ unchanged. The normalised quantity $W_3/W_0$ is unchanged by Lorentz boosts, while $W_{1,2}/W_0$ change.

\paragraph*{Interaction with matter}\!\!\!\!---\!\,\,\,%
The different bases in the $\mathbb{C}^2$ space of $(\vc{E},\vc{H})$ are associated with different symmetries, giving them a fundamental role in the interaction with symmetry-breaking matter. For instance, matter at optical frequencies has a predominantly electric response with negligible magnetism, breaking the EM symmetry and hence experiencing forces and torques related to $W_1 = W_\text{e}-W_\text{m}$, $\vc{p}_1 = \vc{p}_\text{e} - \vc{p}_\text{m}$, $\vc{S}_1 = \vc{S}_\text{e} - \vc{S}_\text{m}$. Chiral matter breaks RL symmetry, hence experiencing forces and torques \cite{Golat2023} related to $W_3 = W_\text{R}-W_\text{L}$, $\vc{p}_3$, $\vc{S}_3$. 

For small particles that scatter like an electric ($\vc{p}$) and a magnetic ($\vc{m}$) dipole, one can define the dipole moment bispinor $\vc{\pi}_\text{EM} = \tfrac{1}{2}(\vc{p}/\sqrt{\varepsilon},\sqrt{\mu}\vc{m})^\intercal$ \cite{Bliokh2014}. In analogy to the electromagnetic field bispinor $\vc{\psi}$ from \cref{eq:bases}, this dipolar bispinor can be expressed in different bases such as $\vc{\pi}_\text{PA}$ and $\vc{\pi}_\text{RL}$. For linear response, dipole moments are proportional to applied fields, $\vc{\pi} = \tens{A} \vc{\psi}$, leading to a dipolar polarisabily matrix $\tens{A}$ which can too be expressed in the different bases (see {SM section C \cite{SM}}). This notation simplifies many expressions on the interaction between light and dipolar particles. For example, the extinction power by a dipole under illumination ${P_\text{ext} =\tfrac{\omega}{2}\Im(\vc{E}^*\!\vdot\vc{p} + \mu \vc{H}^*\!\vdot\vc{m})}$ can be written using bispinors as $2\omega\Im(\vc{\psi}^\dagger\vdot\vc{\pi})$ and subsequently using the polarisability matrix via $2\omega\Im(\vc{\psi}^\dagger\!\vdot\tens{A} \vc{\psi})$ which, assuming isotropy, reduces (see {SM section C \cite{SM}}) to the elegant $\mathbb{C}^2$-basis-independent expression $P_\text{ext}={2}{\omega}(\Im\alpha_0 W_0 + \Im\alpha_1 W_1 + \Im\alpha_2 W_2 + \Im\alpha_3 W_3)$, where $\alpha_0 = \tfrac{1}{2}(\alpha_\text{e} + \alpha_\text{m})$, $\alpha_1 = \tfrac{1}{2}(\alpha_\text{e} - \alpha_\text{m})$, $\alpha_2 = \tfrac{1}{2}(\alpha_\text{a} - \alpha_\text{p})$ (non-reciprocal polarisability) and $\alpha_3 = \tfrac{1}{2}(\alpha_\text{R} - \alpha_\text{L})$ (chiral polarisability). Another example is the interaction force on a dipole
${\vc{F}}=\frac{1}{2}\Re\lbrace\vc{p}^\ast\!\!\vdot\!(\grad)\vc{E}+\mu\vc{m}^\ast\!\!\vdot\!(\grad)\vc{H}\rbrace$ which can be written \cite{Golat2023} in a $\mathbb{C}^2$-basis-independent way as ${\vc{F}}=\sum_{A=0}^3\qty(\Re[{\alpha}_A]\grad W_A+2\omega\Im[{\alpha}_A]\vc{p}_A)$, revealing gradient and pressure forces associated to each symmetry breaking{, and greatly simplifying the --usually very long-- expressions for the dipolar force with chiral terms \cite{Golat2023}}. The fact that $W_1 = W_2 = 0$ in paraxial fields means that experiments with paraxial light will only access $\alpha_\text{e} + \alpha_\text{m}$ (total extinction spectra) and $\alpha_\text{R} - \alpha_\text{L}$ (optical activity and circular dichroism). Illumination with $W_1 \neq 0$ (such as an evanescent or standing wave) can be used to access $\alpha_1$ and hence discern $\alpha_\text{e}$ from $\alpha_\text{m}$. 

The use of different bases for electromagnetic bispinors also provides insights into material response. The constitutive relations that define an isotropic linear material (allowing chirality and non-reciprocity) are traditionally written in the EM basis \cite{Mun2020}:
\begin{equation}
\label{eq:constitutive}
    \underbrace{\frac{1}{2}\!\! \pmqty{\vc{D}/\sqrt{\varepsilon_0} \\ \vc{B}/\sqrt{\mu_0}}}_{\vc{\gamma}_{\text{EM}}} = \underbrace{\Bigg(\mqty{ \overbrace{1+\chi_\text{e}}^{\varepsilon_r} & \chi_t+\ii \chi_c \\ \chi_t - \ii \chi_c & \underbrace{1+\chi_\text{m}}_{\mu_r}}\Bigg)}_{\tens{I} + \tens{\chi}_\text{EM}} \underbrace{\frac{1}{2} \!\!\pmqty{\sqrt{\varepsilon_0}\vc{E}\\\sqrt{\mu_0}\vc{H}}}_{\vc{\psi}_{\text{EM}}},
\end{equation}
 % \noindent 
 where $\chi_\text{e}$ is the electric susceptibility, $\varepsilon_r = 1+\chi_\text{e}$ the relative electric permittivity, $\chi_\text{m}$ the magnetic susceptibility, $\mu_r = 1+\chi_\text{m}$ the relative magnetic permeability, $\chi_c$ the chiral susceptibility (typically written as $\kappa$ chiral material parameter in many works), and $\chi_t$ the non-reciprocal susceptibility. The constitutive relations motivate us to define a bispinor $\vc{\gamma}_{\text{EM}} = \tfrac{1}{2}(\vc{D}/\sqrt{\varepsilon_0}, \vc{B}/\sqrt{\mu_0})^\intercal$, which allows writing the constitutive relations in a basis-independent notation as $\vc{\gamma} = (\tens{I} + \tens{\chi}) \vc{\psi}$. Under the lens of our framework, we propose the susceptibility tensor $\tens{\chi}$ may be written in the EM, PA, or RL basis (see {SM section C \cite{SM}}) which motivates the use of four basis-independent material degrees of freedom, given by $\chi_0 = \tfrac{1}{2}(\varepsilon_r + \mu_r) - 1 = \tfrac{1}{2}(\chi_\text{e} + \chi_\text{m})$, and the differences $\chi_1 = \tfrac{1}{2}(\chi_\text{e} - \chi_\text{m})$, $\chi_2 = -\chi_t = -\tfrac{1}{2}(\chi_\text{p} - \chi_\text{a})$ and $\chi_3 = \chi_c = \tfrac{1}{2}(\chi_\text{R} - \chi_\text{L})$.
Additionally, as derived in the {SM section C \cite{SM}}, the time-harmonic macroscopic Maxwell equations (in absence of free sources) can be written as:
\begin{equation}
\label{eq:macroscopicmaxwell}
\curl \vc{\psi} =\ii k_0 \hat{D} \vc{\gamma},
\end{equation}
where the curl $\curl = \spmqty{ 1 & 0 \\ 0 & 1}\curl$
acts on the $\mathbb{C}^3 \times \mathbb{L}^3$ sub-space of bispinors, and $\hat{D}$ represents the duality transformation acting on the $\mathbb{C}^2$ sub-space, given by $\hat{D}_\text{EM} = \spmqty{ 0 & 1 \\ -1 & 0 }$ in the EM basis. Combining \cref{eq:constitutive,eq:macroscopicmaxwell} one arrives at $[ \ii k_0 \hat{D} (\tens{I} + \tens{\chi}) - \curl ]\vc{\psi} = 0$, which is of the form $\tens{M} \vc{\psi} = 0$. This has been done before in the EM basis, e.g.~\cite{muljarov2018resonant}, but we here stress that the representation of $\tens{M}$ depends on the $\mathbb{C}^2$ basis being used (see {SM section C \cite{SM}}). This brings interesting insight. In the EM basis, the tensor $\tens{M}$ and hence Maxwell's equations become diagonal (in $\mathbb{C}^2$) when $\varepsilon_r(\vc{r})=0$ and $\mu_r(\vc{r})=0$, which means that $\vc{F}_\text{e}(\vc{r})$ becomes uncoupled from $\vc{F}_\text{m}(\vc{r})$. Indeed, in such $\varepsilon$-and-$\mu$-near-zero materials, it is known that Maxwell's equations become static-like and the electric and magnetic fields uncouple from each other \cite{ziolkowski2004propagation,mahmoud2014wave}. In the RL basis, diagonalisation of $\tens{M}$ occurs under the condition $\varepsilon_r(\vc{r})=\mu_r(\vc{r})$ and $\chi_t(\vc{r})=0$. It is known that in these so-called dual materials (free space being a trivial example), scattered light preserves the incident helicity \cite{zambrana2013duality}, as $\vc{F}_\text{R}(\vc{r})$ is uncoupled from $\vc{F}_\text{L}(\vc{r})$. Finally, in the PA basis, diagonalisation occurs when $\varepsilon_r(\vc{r})=-\mu_r(\vc{r})$ and $\chi_t(\vc{r})=0$, meaning $\vc{F}_\text{p}(\vc{r})$ and $\vc{F}_\text{a}(\vc{r})$ become uncoupled. This condition occurs in opaque plasmonic materials. We believe that further interesting insights in light-matter interactions can arise from this symmetry-based formalism.

% \section{Conclusions}%
\paragraph*{Conclusions}\!\!\!\!---\!\,\,\,%
By freeing ourselves from the electric-magnetic basis, and expressing electromagnetic fields in the parallel-antiparallel and right-left basis, a flood of insights and analogies is uncovered. By subtracting the energy and spin between field projections, a re-interpretation and systematization of known electromagnetic quadratic quantities and their relations results, in some cases leading to novel insights such as the wave-like behavior of the spin-like quantities. The symmetry sphere is a valuable resource to gain intuition and understanding of symmetry in electromagnetism and its interaction with matter, and we feel there is much room for its exploration. Our current analysis is limited to time-harmonic fields and quadratic quantities only, though we strongly believe the framework can be adapted to more general situations.

\begin{acknowledgments}
\paragraph*{Acknowledgments}\!\!\!\!---\!\,\,\,%
SG and FJRF acknowledge support from EIC-Pathfinder-CHIRALFORCE (101046961) which is funded by Innovate UK Horizon Europe Guarantee (UKRI project 10045438). AJV is supported by EPSRC Grant EP/R513064/1. {We thank M Nieto-Vesperinas and N. Levy for useful discussions to improve clarity.}
\end{acknowledgments}

\hbadness 10000\relax
\bibliography{main}

%apsrev4-2.bst 2019-01-14 (MD) hand-edited version of apsrev4-1.bst
%Control: key (0)
%Control: author (72) initials jnrlst
%Control: editor formatted (1) identically to author
%Control: production of article title (0) allowed
%Control: page (1) range
%Control: year (1) truncated
%Control: production of eprint (0) enabled
\begin{thebibliography}{49}%
\makeatletter
\providecommand \@ifxundefined [1]{%
 \@ifx{#1\undefined}
}%
\providecommand \@ifnum [1]{%
 \ifnum #1\expandafter \@firstoftwo
 \else \expandafter \@secondoftwo
 \fi
}%
\providecommand \@ifx [1]{%
 \ifx #1\expandafter \@firstoftwo
 \else \expandafter \@secondoftwo
 \fi
}%
\providecommand \natexlab [1]{#1}%
\providecommand \enquote  [1]{``#1''}%
\providecommand \bibnamefont  [1]{#1}%
\providecommand \bibfnamefont [1]{#1}%
\providecommand \citenamefont [1]{#1}%
\providecommand \href@noop [0]{\@secondoftwo}%
\providecommand \href [0]{\begingroup \@sanitize@url \@href}%
\providecommand \@href[1]{\@@startlink{#1}\@@href}%
\providecommand \@@href[1]{\endgroup#1\@@endlink}%
\providecommand \@sanitize@url [0]{\catcode `\\12\catcode `\$12\catcode `\&12\catcode `\#12\catcode `\^12\catcode `\_12\catcode `\%12\relax}%
\providecommand \@@startlink[1]{}%
\providecommand \@@endlink[0]{}%
\providecommand \url  [0]{\begingroup\@sanitize@url \@url }%
\providecommand \@url [1]{\endgroup\@href {#1}{\urlprefix }}%
\providecommand \urlprefix  [0]{URL }%
\providecommand \Eprint [0]{\href }%
\providecommand \doibase [0]{https://doi.org/}%
\providecommand \selectlanguage [0]{\@gobble}%
\providecommand \bibinfo  [0]{\@secondoftwo}%
\providecommand \bibfield  [0]{\@secondoftwo}%
\providecommand \translation [1]{[#1]}%
\providecommand \BibitemOpen [0]{}%
\providecommand \bibitemStop [0]{}%
\providecommand \bibitemNoStop [0]{.\EOS\space}%
\providecommand \EOS [0]{\spacefactor3000\relax}%
\providecommand \BibitemShut  [1]{\csname bibitem#1\endcsname}%
\let\auto@bib@innerbib\@empty
%</preamble>
\bibitem [{\citenamefont {Cameron}\ \emph {et~al.}(2012)\citenamefont {Cameron}, \citenamefont {Barnett},\ and\ \citenamefont {Yao}}]{Cameron2012a}%
  \BibitemOpen
  \bibfield  {author} {\bibinfo {author} {\bibfnamefont {R.~P.}\ \bibnamefont {Cameron}}, \bibinfo {author} {\bibfnamefont {S.~M.}\ \bibnamefont {Barnett}},\ and\ \bibinfo {author} {\bibfnamefont {A.~M.}\ \bibnamefont {Yao}},\ }\bibfield  {title} {\bibinfo {title} {Optical helicity, optical spin and related quantities in electromagnetic theory},\ }\href {https://doi.org/10.1088/1367-2630/14/5/053050} {\bibfield  {journal} {\bibinfo  {journal} {New J. Phys.}\ }\textbf {\bibinfo {volume} {14}},\ \bibinfo {pages} {053050} (\bibinfo {year} {2012})}\BibitemShut {NoStop}%
\bibitem [{\citenamefont {Bliokh}\ \emph {et~al.}(2013)\citenamefont {Bliokh}, \citenamefont {Bekshaev},\ and\ \citenamefont {Nori}}]{Bliokh2013}%
  \BibitemOpen
  \bibfield  {author} {\bibinfo {author} {\bibfnamefont {K.~Y.}\ \bibnamefont {Bliokh}}, \bibinfo {author} {\bibfnamefont {A.~Y.}\ \bibnamefont {Bekshaev}},\ and\ \bibinfo {author} {\bibfnamefont {F.}~\bibnamefont {Nori}},\ }\bibfield  {title} {\bibinfo {title} {Dual electromagnetism: helicity, spin, momentum and angular momentum},\ }\href {https://doi.org/10.1088/1367-2630/15/3/033026} {\bibfield  {journal} {\bibinfo  {journal} {New J. Phys.}\ }\textbf {\bibinfo {volume} {15}},\ \bibinfo {pages} {033026} (\bibinfo {year} {2013})}\BibitemShut {NoStop}%
\bibitem [{\citenamefont {Aiello}\ and\ \citenamefont {Berry}(2015)}]{Aiello2015a}%
  \BibitemOpen
  \bibfield  {author} {\bibinfo {author} {\bibfnamefont {A.}~\bibnamefont {Aiello}}\ and\ \bibinfo {author} {\bibfnamefont {M.~V.}\ \bibnamefont {Berry}},\ }\bibfield  {title} {\bibinfo {title} {Note on the helicity decomposition of spin and orbital optical currents},\ }\href {https://doi.org/10.1088/2040-8978/17/6/062001} {\bibfield  {journal} {\bibinfo  {journal} {Journal of Optics}\ }\textbf {\bibinfo {volume} {17}},\ \bibinfo {pages} {062001} (\bibinfo {year} {2015})}\BibitemShut {NoStop}%
\bibitem [{\citenamefont {Barnett}\ \emph {et~al.}(2012)\citenamefont {Barnett}, \citenamefont {Cameron},\ and\ \citenamefont {Yao}}]{Barnett2012}%
  \BibitemOpen
  \bibfield  {author} {\bibinfo {author} {\bibfnamefont {S.~M.}\ \bibnamefont {Barnett}}, \bibinfo {author} {\bibfnamefont {R.~P.}\ \bibnamefont {Cameron}},\ and\ \bibinfo {author} {\bibfnamefont {A.~M.}\ \bibnamefont {Yao}},\ }\bibfield  {title} {\bibinfo {title} {Duplex symmetry and its relation to the conservation of optical helicity},\ }\href {https://doi.org/10.1103/PhysRevA.86.013845} {\bibfield  {journal} {\bibinfo  {journal} {Physical Review A}\ }\textbf {\bibinfo {volume} {86}},\ \bibinfo {pages} {013845} (\bibinfo {year} {2012})}\BibitemShut {NoStop}%
\bibitem [{\citenamefont {Bliokh}\ and\ \citenamefont {Nori}(2011)}]{Bliokh2011}%
  \BibitemOpen
  \bibfield  {author} {\bibinfo {author} {\bibfnamefont {K.~Y.}\ \bibnamefont {Bliokh}}\ and\ \bibinfo {author} {\bibfnamefont {F.}~\bibnamefont {Nori}},\ }\bibfield  {title} {\bibinfo {title} {Characterizing optical chirality},\ }\href {https://doi.org/10.1103/PhysRevA.83.021803} {\bibfield  {journal} {\bibinfo  {journal} {Physical Review A}\ }\textbf {\bibinfo {volume} {83}},\ \bibinfo {pages} {021803(R)} (\bibinfo {year} {2011})}\BibitemShut {NoStop}%
\bibitem [{\citenamefont {Tang}\ and\ \citenamefont {Cohen}(2010)}]{Tang2010}%
  \BibitemOpen
  \bibfield  {author} {\bibinfo {author} {\bibfnamefont {Y.}~\bibnamefont {Tang}}\ and\ \bibinfo {author} {\bibfnamefont {A.~E.}\ \bibnamefont {Cohen}},\ }\bibfield  {title} {\bibinfo {title} {Optical chirality and its interaction with matter},\ }\href {https://doi.org/10.1103/PhysRevLett.104.163901} {\bibfield  {journal} {\bibinfo  {journal} {Physical Review Letters}\ }\textbf {\bibinfo {volume} {104}},\ \bibinfo {pages} {163901} (\bibinfo {year} {2010})}\BibitemShut {NoStop}%
\bibitem [{\citenamefont {Vernon}\ \emph {et~al.}(2023{\natexlab{a}})\citenamefont {Vernon}, \citenamefont {Golat}, \citenamefont {Rigouzzo}, \citenamefont {Lim},\ and\ \citenamefont {Rodríguez-Fortuño}}]{Vernon2023}%
  \BibitemOpen
  \bibfield  {author} {\bibinfo {author} {\bibfnamefont {A.~J.}\ \bibnamefont {Vernon}}, \bibinfo {author} {\bibfnamefont {S.}~\bibnamefont {Golat}}, \bibinfo {author} {\bibfnamefont {C.}~\bibnamefont {Rigouzzo}}, \bibinfo {author} {\bibfnamefont {E.~A.}\ \bibnamefont {Lim}},\ and\ \bibinfo {author} {\bibfnamefont {F.~J.}\ \bibnamefont {Rodríguez-Fortuño}},\ }\href@noop {} {\bibinfo {title} {A decomposition of light's spin angular momentum density}} (\bibinfo {year} {2023}{\natexlab{a}}),\ \Eprint {https://arxiv.org/abs/2310.03804} {arXiv:2310.03804 [physics.optics]} \BibitemShut {NoStop}%
\bibitem [{\citenamefont {Vernon}\ \emph {et~al.}(2023{\natexlab{b}})\citenamefont {Vernon}, \citenamefont {Dennis},\ and\ \citenamefont {Rodríguez-Fortuño}}]{Vernon2023_2}%
  \BibitemOpen
  \bibfield  {author} {\bibinfo {author} {\bibfnamefont {A.~J.}\ \bibnamefont {Vernon}}, \bibinfo {author} {\bibfnamefont {M.~R.}\ \bibnamefont {Dennis}},\ and\ \bibinfo {author} {\bibfnamefont {F.~J.}\ \bibnamefont {Rodríguez-Fortuño}},\ }\bibfield  {title} {\bibinfo {title} {3d zeros in electromagnetic fields},\ }\href {https://doi.org/10.1364/OPTICA.487333} {\bibfield  {journal} {\bibinfo  {journal} {Optica}\ }\textbf {\bibinfo {volume} {10}},\ \bibinfo {pages} {1231} (\bibinfo {year} {2023}{\natexlab{b}})}\BibitemShut {NoStop}%
\bibitem [{\citenamefont {Alpeggiani}\ \emph {et~al.}(2018)\citenamefont {Alpeggiani}, \citenamefont {Bliokh}, \citenamefont {Nori},\ and\ \citenamefont {Kuipers}}]{Alpeggiani2018}%
  \BibitemOpen
  \bibfield  {author} {\bibinfo {author} {\bibfnamefont {F.}~\bibnamefont {Alpeggiani}}, \bibinfo {author} {\bibfnamefont {K.~Y.}\ \bibnamefont {Bliokh}}, \bibinfo {author} {\bibfnamefont {F.}~\bibnamefont {Nori}},\ and\ \bibinfo {author} {\bibfnamefont {L.}~\bibnamefont {Kuipers}},\ }\bibfield  {title} {\bibinfo {title} {Electromagnetic helicity in complex media},\ }\href {https://doi.org/10.1103/PhysRevLett.120.243605} {\bibfield  {journal} {\bibinfo  {journal} {Phys. Rev. Lett.}\ }\textbf {\bibinfo {volume} {120}},\ \bibinfo {pages} {243605} (\bibinfo {year} {2018})}\BibitemShut {NoStop}%
\bibitem [{\citenamefont {Bliokh}\ \emph {et~al.}(2014{\natexlab{a}})\citenamefont {Bliokh}, \citenamefont {Kivshar},\ and\ \citenamefont {Nori}}]{Bliokh2014}%
  \BibitemOpen
  \bibfield  {author} {\bibinfo {author} {\bibfnamefont {K.~Y.}\ \bibnamefont {Bliokh}}, \bibinfo {author} {\bibfnamefont {Y.~S.}\ \bibnamefont {Kivshar}},\ and\ \bibinfo {author} {\bibfnamefont {F.}~\bibnamefont {Nori}},\ }\bibfield  {title} {\bibinfo {title} {Magnetoelectric effects in local light-matter interactions},\ }\href {https://doi.org/10.1103/PhysRevLett.113.033601} {\bibfield  {journal} {\bibinfo  {journal} {Phys. Rev. Lett.}\ }\textbf {\bibinfo {volume} {113}},\ \bibinfo {pages} {033601} (\bibinfo {year} {2014}{\natexlab{a}})}\BibitemShut {NoStop}%
\bibitem [{\citenamefont {Bliokh}\ \emph {et~al.}(2014{\natexlab{b}})\citenamefont {Bliokh}, \citenamefont {Bekshaev},\ and\ \citenamefont {Nori}}]{Bliokh2014a}%
  \BibitemOpen
  \bibfield  {author} {\bibinfo {author} {\bibfnamefont {K.~Y.}\ \bibnamefont {Bliokh}}, \bibinfo {author} {\bibfnamefont {A.~Y.}\ \bibnamefont {Bekshaev}},\ and\ \bibinfo {author} {\bibfnamefont {F.}~\bibnamefont {Nori}},\ }\bibfield  {title} {\bibinfo {title} {Extraordinary momentum and spin in evanescent waves},\ }\bibfield  {journal} {\bibinfo  {journal} {Nature Communications}\ }\textbf {\bibinfo {volume} {5}},\ \href {https://doi.org/10.1038/ncomms4300} {10.1038/ncomms4300} (\bibinfo {year} {2014}{\natexlab{b}})\BibitemShut {NoStop}%
\bibitem [{SM()}]{SM}%
  \BibitemOpen
  \href@noop {} {}\bibinfo {note} {See Supplemental Material at [URL will be inserted by publisher] for more details.}\BibitemShut {Stop}%
\bibitem [{\citenamefont {Białynicki-Birula}(1994)}]{BialynickiBirula_1994}%
  \BibitemOpen
  \bibfield  {author} {\bibinfo {author} {\bibfnamefont {I.}~\bibnamefont {Białynicki-Birula}},\ }\bibfield  {title} {\bibinfo {title} {On the wave function of the photon},\ }\href@noop {} {\bibfield  {journal} {\bibinfo  {journal} {Acta Physica Polonica A}\ }\textbf {\bibinfo {volume} {86}},\ \bibinfo {pages} {97–116} (\bibinfo {year} {1994})}\BibitemShut {NoStop}%
\bibitem [{\citenamefont {Bialynicki-Birula}\ and\ \citenamefont {Bialynicka-Birula}(2003)}]{BialynickiBirula_2003}%
  \BibitemOpen
  \bibfield  {author} {\bibinfo {author} {\bibfnamefont {I.}~\bibnamefont {Bialynicki-Birula}}\ and\ \bibinfo {author} {\bibfnamefont {Z.}~\bibnamefont {Bialynicka-Birula}},\ }\bibfield  {title} {\bibinfo {title} {Vortex lines of the electromagnetic field},\ }\href {https://doi.org/10.1103/PhysRevA.67.062114} {\bibfield  {journal} {\bibinfo  {journal} {Physical Review A}\ }\textbf {\bibinfo {volume} {67}},\ \bibinfo {pages} {062114} (\bibinfo {year} {2003})}\BibitemShut {NoStop}%
\bibitem [{\citenamefont {Kaiser}(2004)}]{Kaiser2004}%
  \BibitemOpen
  \bibfield  {author} {\bibinfo {author} {\bibfnamefont {G.}~\bibnamefont {Kaiser}},\ }\bibfield  {title} {\bibinfo {title} {Helicity, polarization and riemann–silberstein vortices},\ }\href {https://doi.org/10.1088/1464-4258/6/5/018} {\bibfield  {journal} {\bibinfo  {journal} {Journal of Optics A: Pure and Applied Optics}\ }\textbf {\bibinfo {volume} {6}},\ \bibinfo {pages} {S243--S245} (\bibinfo {year} {2004})}\BibitemShut {NoStop}%
\bibitem [{\citenamefont {Fernandez-Corbaton}(2014)}]{Fernandez2014}%
  \BibitemOpen
  \bibfield  {author} {\bibinfo {author} {\bibfnamefont {I.}~\bibnamefont {Fernandez-Corbaton}},\ }\emph {\bibinfo {title} {Helicity and duality symmetry in light matter interactions: Theory and applications}},\ \href@noop {} {Ph.D. thesis},\ \bibinfo  {school} {Macquarie University} (\bibinfo {year} {2014})\BibitemShut {NoStop}%
\bibitem [{\citenamefont {Trueba}\ and\ \citenamefont {Rañada}(1996)}]{Trueba1996}%
  \BibitemOpen
  \bibfield  {author} {\bibinfo {author} {\bibfnamefont {J.~L.}\ \bibnamefont {Trueba}}\ and\ \bibinfo {author} {\bibfnamefont {A.~F.}\ \bibnamefont {Rañada}},\ }\bibfield  {title} {\bibinfo {title} {The electromagnetic helicity},\ }\href {https://doi.org/10.1088/0143-0807/17/3/008} {\bibfield  {journal} {\bibinfo  {journal} {European Journal of Physics}\ }\textbf {\bibinfo {volume} {17}},\ \bibinfo {pages} {141--144} (\bibinfo {year} {1996})}\BibitemShut {NoStop}%
\bibitem [{\citenamefont {Afanasiev}\ and\ \citenamefont {Stepanovsky}(1996)}]{Afanasiev1996}%
  \BibitemOpen
  \bibfield  {author} {\bibinfo {author} {\bibfnamefont {G.~N.}\ \bibnamefont {Afanasiev}}\ and\ \bibinfo {author} {\bibfnamefont {Y.~P.}\ \bibnamefont {Stepanovsky}},\ }\bibfield  {title} {\bibinfo {title} {The helicity of the free electromagnetic field and its physical meaning},\ }\href {https://doi.org/10.1007/BF02731014} {\bibfield  {journal} {\bibinfo  {journal} {Il Nuovo Cimento A}\ }\textbf {\bibinfo {volume} {109}},\ \bibinfo {pages} {271--279} (\bibinfo {year} {1996})}\BibitemShut {NoStop}%
\bibitem [{\citenamefont {Berry}\ and\ \citenamefont {Shukla}(2019)}]{Berry2019}%
  \BibitemOpen
  \bibfield  {author} {\bibinfo {author} {\bibfnamefont {M.~V.}\ \bibnamefont {Berry}}\ and\ \bibinfo {author} {\bibfnamefont {P.}~\bibnamefont {Shukla}},\ }\bibfield  {title} {\bibinfo {title} {Geometry of 3d monochromatic light: local wavevectors, phases, curl forces, and superoscillations},\ }\bibfield  {journal} {\bibinfo  {journal} {Journal of Optics}\ }\textbf {\bibinfo {volume} {21}},\ \href {https://doi.org/10.1088/2040-8986/ab14c4} {10.1088/2040-8986/ab14c4} (\bibinfo {year} {2019})\BibitemShut {NoStop}%
\bibitem [{\citenamefont {Nieto-Vesperinas}\ and\ \citenamefont {Xu}(2021)}]{NietoVesperinas2021}%
  \BibitemOpen
  \bibfield  {author} {\bibinfo {author} {\bibfnamefont {M.}~\bibnamefont {Nieto-Vesperinas}}\ and\ \bibinfo {author} {\bibfnamefont {X.}~\bibnamefont {Xu}},\ }\bibfield  {title} {\bibinfo {title} {Reactive helicity and reactive power in nanoscale optics: Evanescent waves. kerker conditions. optical theorems and reactive dichroism},\ }\href {https://doi.org/10.1103/PhysRevResearch.3.043080} {\bibfield  {journal} {\bibinfo  {journal} {Physical Review Research}\ }\textbf {\bibinfo {volume} {3}},\ \bibinfo {pages} {043080} (\bibinfo {year} {2021})}\BibitemShut {NoStop}%
\bibitem [{\citenamefont {Golat}\ \emph {et~al.}(2023)\citenamefont {Golat}, \citenamefont {Kingsley-Smith}, \citenamefont {Diez}, \citenamefont {Martinez-Romeu}, \citenamefont {Martínez},\ and\ \citenamefont {Rodríguez-Fortuño}}]{Golat2023}%
  \BibitemOpen
  \bibfield  {author} {\bibinfo {author} {\bibfnamefont {S.}~\bibnamefont {Golat}}, \bibinfo {author} {\bibfnamefont {J.~J.}\ \bibnamefont {Kingsley-Smith}}, \bibinfo {author} {\bibfnamefont {I.}~\bibnamefont {Diez}}, \bibinfo {author} {\bibfnamefont {J.}~\bibnamefont {Martinez-Romeu}}, \bibinfo {author} {\bibfnamefont {A.}~\bibnamefont {Martínez}},\ and\ \bibinfo {author} {\bibfnamefont {F.~J.}\ \bibnamefont {Rodríguez-Fortuño}},\ }\href@noop {} {\bibinfo {title} {Optical chiral sorting forces and their manifestation in evanescent waves and nanofibres}} (\bibinfo {year} {2023})\BibitemShut {NoStop}%
\bibitem [{\citenamefont {Bliokh}\ and\ \citenamefont {Nori}(2012)}]{Bliokh2012}%
  \BibitemOpen
  \bibfield  {author} {\bibinfo {author} {\bibfnamefont {K.~Y.}\ \bibnamefont {Bliokh}}\ and\ \bibinfo {author} {\bibfnamefont {F.}~\bibnamefont {Nori}},\ }\bibfield  {title} {\bibinfo {title} {Transverse spin of a surface polariton},\ }\href {https://doi.org/10.1103/physreva.85.061801} {\bibfield  {journal} {\bibinfo  {journal} {Physical Review A}\ }\textbf {\bibinfo {volume} {85}},\ \bibinfo {pages} {061801(R)} (\bibinfo {year} {2012})}\BibitemShut {NoStop}%
\bibitem [{\citenamefont {Aiello}\ \emph {et~al.}(2015)\citenamefont {Aiello}, \citenamefont {Banzer}, \citenamefont {Neugebauer},\ and\ \citenamefont {Leuchs}}]{Aiello2015}%
  \BibitemOpen
  \bibfield  {author} {\bibinfo {author} {\bibfnamefont {A.}~\bibnamefont {Aiello}}, \bibinfo {author} {\bibfnamefont {P.}~\bibnamefont {Banzer}}, \bibinfo {author} {\bibfnamefont {M.}~\bibnamefont {Neugebauer}},\ and\ \bibinfo {author} {\bibfnamefont {G.}~\bibnamefont {Leuchs}},\ }\bibfield  {title} {\bibinfo {title} {From transverse angular momentum to photonic wheels},\ }\href {https://doi.org/10.1038/nphoton.2015.203} {\bibfield  {journal} {\bibinfo  {journal} {Nature Photonics}\ }\textbf {\bibinfo {volume} {9}},\ \bibinfo {pages} {789--795} (\bibinfo {year} {2015})}\BibitemShut {NoStop}%
\bibitem [{\citenamefont {Bliokh}\ and\ \citenamefont {Nori}(2015)}]{Bliokh2015}%
  \BibitemOpen
  \bibfield  {author} {\bibinfo {author} {\bibfnamefont {K.~Y.}\ \bibnamefont {Bliokh}}\ and\ \bibinfo {author} {\bibfnamefont {F.}~\bibnamefont {Nori}},\ }\bibfield  {title} {\bibinfo {title} {Transverse and longitudinal angular momenta of light},\ }\href {https://doi.org/10.1016/j.physrep.2015.06.003} {\bibfield  {journal} {\bibinfo  {journal} {Physics Reports}\ }\textbf {\bibinfo {volume} {592}},\ \bibinfo {pages} {1--38} (\bibinfo {year} {2015})}\BibitemShut {NoStop}%
\bibitem [{\citenamefont {Bliokh}\ \emph {et~al.}(2015)\citenamefont {Bliokh}, \citenamefont {Rodr{\'{\i}}guez-Fortu{\~{n}}o}, \citenamefont {Nori},\ and\ \citenamefont {Zayats}}]{Bliokh2015a}%
  \BibitemOpen
  \bibfield  {author} {\bibinfo {author} {\bibfnamefont {K.~Y.}\ \bibnamefont {Bliokh}}, \bibinfo {author} {\bibfnamefont {F.~J.}\ \bibnamefont {Rodr{\'{\i}}guez-Fortu{\~{n}}o}}, \bibinfo {author} {\bibfnamefont {F.}~\bibnamefont {Nori}},\ and\ \bibinfo {author} {\bibfnamefont {A.~V.}\ \bibnamefont {Zayats}},\ }\bibfield  {title} {\bibinfo {title} {Spin{\textendash}orbit interactions of light},\ }\href {https://doi.org/10.1038/nphoton.2015.201} {\bibfield  {journal} {\bibinfo  {journal} {Nature Photonics}\ }\textbf {\bibinfo {volume} {9}},\ \bibinfo {pages} {796--808} (\bibinfo {year} {2015})}\BibitemShut {NoStop}%
\bibitem [{\citenamefont {Bekshaev}\ \emph {et~al.}(2015)\citenamefont {Bekshaev}, \citenamefont {Bliokh},\ and\ \citenamefont {Nori}}]{Bekshaev2015}%
  \BibitemOpen
  \bibfield  {author} {\bibinfo {author} {\bibfnamefont {A.~Y.}\ \bibnamefont {Bekshaev}}, \bibinfo {author} {\bibfnamefont {K.~Y.}\ \bibnamefont {Bliokh}},\ and\ \bibinfo {author} {\bibfnamefont {F.}~\bibnamefont {Nori}},\ }\bibfield  {title} {\bibinfo {title} {Transverse spin and momentum in two-wave interference},\ }\href {https://doi.org/10.1103/PhysRevX.5.011039} {\bibfield  {journal} {\bibinfo  {journal} {Physical Review X}\ }\textbf {\bibinfo {volume} {5}},\ \bibinfo {pages} {011039} (\bibinfo {year} {2015})}\BibitemShut {NoStop}%
\bibitem [{\citenamefont {Eismann}\ \emph {et~al.}(2020)\citenamefont {Eismann}, \citenamefont {Nicholls}, \citenamefont {Roth}, \citenamefont {Alonso}, \citenamefont {Banzer}, \citenamefont {Rodr{\'{\i}}guez-Fortu{\~{n}}o}, \citenamefont {Zayats}, \citenamefont {Nori},\ and\ \citenamefont {Bliokh}}]{Eismann2020}%
  \BibitemOpen
  \bibfield  {author} {\bibinfo {author} {\bibfnamefont {J.~S.}\ \bibnamefont {Eismann}}, \bibinfo {author} {\bibfnamefont {L.~H.}\ \bibnamefont {Nicholls}}, \bibinfo {author} {\bibfnamefont {D.~J.}\ \bibnamefont {Roth}}, \bibinfo {author} {\bibfnamefont {M.~A.}\ \bibnamefont {Alonso}}, \bibinfo {author} {\bibfnamefont {P.}~\bibnamefont {Banzer}}, \bibinfo {author} {\bibfnamefont {F.~J.}\ \bibnamefont {Rodr{\'{\i}}guez-Fortu{\~{n}}o}}, \bibinfo {author} {\bibfnamefont {A.~V.}\ \bibnamefont {Zayats}}, \bibinfo {author} {\bibfnamefont {F.}~\bibnamefont {Nori}},\ and\ \bibinfo {author} {\bibfnamefont {K.~Y.}\ \bibnamefont {Bliokh}},\ }\bibfield  {title} {\bibinfo {title} {Transverse spinning of unpolarized light},\ }\href {https://doi.org/10.1038/s41566-020-00733-3} {\bibfield  {journal} {\bibinfo  {journal} {Nature Photonics}\ }\textbf {\bibinfo {volume} {15}},\ \bibinfo {pages} {156--161} (\bibinfo {year} {2020})}\BibitemShut {NoStop}%
\bibitem [{\citenamefont {Liu}\ \emph {et~al.}(2018)\citenamefont {Liu}, \citenamefont {DiDonato}, \citenamefont {Ginis}, \citenamefont {Kheifets}, \citenamefont {Amirzhan},\ and\ \citenamefont {Capasso}}]{Liu2018}%
  \BibitemOpen
  \bibfield  {author} {\bibinfo {author} {\bibfnamefont {L.}~\bibnamefont {Liu}}, \bibinfo {author} {\bibfnamefont {A.}~\bibnamefont {DiDonato}}, \bibinfo {author} {\bibfnamefont {V.}~\bibnamefont {Ginis}}, \bibinfo {author} {\bibfnamefont {S.}~\bibnamefont {Kheifets}}, \bibinfo {author} {\bibfnamefont {A.}~\bibnamefont {Amirzhan}},\ and\ \bibinfo {author} {\bibfnamefont {F.}~\bibnamefont {Capasso}},\ }\bibfield  {title} {\bibinfo {title} {Three-dimensional measurement of the helicity-dependent forces on a mie particle},\ }\href {https://doi.org/10.1103/physrevlett.120.223901} {\bibfield  {journal} {\bibinfo  {journal} {Physical Review Letters}\ }\textbf {\bibinfo {volume} {120}},\ \bibinfo {pages} {223901} (\bibinfo {year} {2018})}\BibitemShut {NoStop}%
\bibitem [{\citenamefont {Antognozzi}\ \emph {et~al.}(2016)\citenamefont {Antognozzi}, \citenamefont {Bermingham}, \citenamefont {Harniman}, \citenamefont {Simpson}, \citenamefont {Senior}, \citenamefont {Hayward}, \citenamefont {Hoerber}, \citenamefont {Dennis}, \citenamefont {Bekshaev}, \citenamefont {Bliokh},\ and\ \citenamefont {Nori}}]{Antognozzi2016}%
  \BibitemOpen
  \bibfield  {author} {\bibinfo {author} {\bibfnamefont {M.}~\bibnamefont {Antognozzi}}, \bibinfo {author} {\bibfnamefont {C.~R.}\ \bibnamefont {Bermingham}}, \bibinfo {author} {\bibfnamefont {R.~L.}\ \bibnamefont {Harniman}}, \bibinfo {author} {\bibfnamefont {S.}~\bibnamefont {Simpson}}, \bibinfo {author} {\bibfnamefont {J.}~\bibnamefont {Senior}}, \bibinfo {author} {\bibfnamefont {R.}~\bibnamefont {Hayward}}, \bibinfo {author} {\bibfnamefont {H.}~\bibnamefont {Hoerber}}, \bibinfo {author} {\bibfnamefont {M.~R.}\ \bibnamefont {Dennis}}, \bibinfo {author} {\bibfnamefont {A.~Y.}\ \bibnamefont {Bekshaev}}, \bibinfo {author} {\bibfnamefont {K.~Y.}\ \bibnamefont {Bliokh}},\ and\ \bibinfo {author} {\bibfnamefont {F.}~\bibnamefont {Nori}},\ }\bibfield  {title} {\bibinfo {title} {Direct measurements of the extraordinary optical momentum and transverse spin-dependent force using a nano-cantilever},\ }\href {https://doi.org/10.1038/nphys3732} {\bibfield  {journal} {\bibinfo  {journal} {Nature Physics}\ }\textbf
  {\bibinfo {volume} {12}},\ \bibinfo {pages} {731--735} (\bibinfo {year} {2016})}\BibitemShut {NoStop}%
\bibitem [{\citenamefont {Wei}\ and\ \citenamefont {Rodr{\'{\i}}guez-Fortu{\~{n}}o}(2020)}]{Wei2020}%
  \BibitemOpen
  \bibfield  {author} {\bibinfo {author} {\bibfnamefont {L.}~\bibnamefont {Wei}}\ and\ \bibinfo {author} {\bibfnamefont {F.~J.}\ \bibnamefont {Rodr{\'{\i}}guez-Fortu{\~{n}}o}},\ }\bibfield  {title} {\bibinfo {title} {Momentum-space geometric structure of helical evanescent waves and its implications on near-field directionality},\ }\href {https://doi.org/10.1103/physrevapplied.13.014008} {\bibfield  {journal} {\bibinfo  {journal} {Physical Review Applied}\ }\textbf {\bibinfo {volume} {13}},\ \bibinfo {pages} {014008} (\bibinfo {year} {2020})}\BibitemShut {NoStop}%
\bibitem [{\citenamefont {Wei}\ \emph {et~al.}(2018{\natexlab{a}})\citenamefont {Wei}, \citenamefont {Zayats},\ and\ \citenamefont {Rodr{\'{\i}}guez-Fortu{\~{n}}o}}]{Wei2018}%
  \BibitemOpen
  \bibfield  {author} {\bibinfo {author} {\bibfnamefont {L.}~\bibnamefont {Wei}}, \bibinfo {author} {\bibfnamefont {A.~V.}\ \bibnamefont {Zayats}},\ and\ \bibinfo {author} {\bibfnamefont {F.~J.}\ \bibnamefont {Rodr{\'{\i}}guez-Fortu{\~{n}}o}},\ }\bibfield  {title} {\bibinfo {title} {Interferometric evanescent wave excitation of a nanoantenna for ultrasensitive displacement and phase metrology},\ }\href {https://doi.org/10.1103/physrevlett.121.193901} {\bibfield  {journal} {\bibinfo  {journal} {Physical Review Letters}\ }\textbf {\bibinfo {volume} {121}},\ \bibinfo {pages} {193901} (\bibinfo {year} {2018}{\natexlab{a}})}\BibitemShut {NoStop}%
\bibitem [{\citenamefont {Wei}\ \emph {et~al.}(2018{\natexlab{b}})\citenamefont {Wei}, \citenamefont {Picardi}, \citenamefont {Kingsley-Smith}, \citenamefont {Zayats},\ and\ \citenamefont {Rodr{\'{\i}}guez-Fortu{\~{n}}o}}]{Wei2018a}%
  \BibitemOpen
  \bibfield  {author} {\bibinfo {author} {\bibfnamefont {L.}~\bibnamefont {Wei}}, \bibinfo {author} {\bibfnamefont {M.~F.}\ \bibnamefont {Picardi}}, \bibinfo {author} {\bibfnamefont {J.~J.}\ \bibnamefont {Kingsley-Smith}}, \bibinfo {author} {\bibfnamefont {A.~V.}\ \bibnamefont {Zayats}},\ and\ \bibinfo {author} {\bibfnamefont {F.~J.}\ \bibnamefont {Rodr{\'{\i}}guez-Fortu{\~{n}}o}},\ }\bibfield  {title} {\bibinfo {title} {Directional scattering from particles under evanescent wave illumination: the role of reactive power},\ }\href {https://doi.org/10.1364/ol.43.003393} {\bibfield  {journal} {\bibinfo  {journal} {Optics Letters}\ }\textbf {\bibinfo {volume} {43}},\ \bibinfo {pages} {3393} (\bibinfo {year} {2018}{\natexlab{b}})}\BibitemShut {NoStop}%
\bibitem [{\citenamefont {Xu}\ and\ \citenamefont {Nieto-Vesperinas}(2019)}]{Xu2019}%
  \BibitemOpen
  \bibfield  {author} {\bibinfo {author} {\bibfnamefont {X.}~\bibnamefont {Xu}}\ and\ \bibinfo {author} {\bibfnamefont {M.}~\bibnamefont {Nieto-Vesperinas}},\ }\bibfield  {title} {\bibinfo {title} {Azimuthal imaginary poynting momentum density},\ }\href {https://doi.org/10.1103/PhysRevLett.123.233902} {\bibfield  {journal} {\bibinfo  {journal} {Physical Review Letters}\ }\textbf {\bibinfo {volume} {123}},\ \bibinfo {pages} {233902} (\bibinfo {year} {2019})}\BibitemShut {NoStop}%
\bibitem [{\citenamefont {Zhou}\ \emph {et~al.}(2022)\citenamefont {Zhou}, \citenamefont {Xu}, \citenamefont {Zhang}, \citenamefont {Li}, \citenamefont {Yan}, \citenamefont {Nieto-Vesperinas}, \citenamefont {Li}, \citenamefont {Qiu},\ and\ \citenamefont {Yao}}]{Zhou2022}%
  \BibitemOpen
  \bibfield  {author} {\bibinfo {author} {\bibfnamefont {Y.}~\bibnamefont {Zhou}}, \bibinfo {author} {\bibfnamefont {X.}~\bibnamefont {Xu}}, \bibinfo {author} {\bibfnamefont {Y.}~\bibnamefont {Zhang}}, \bibinfo {author} {\bibfnamefont {M.}~\bibnamefont {Li}}, \bibinfo {author} {\bibfnamefont {S.}~\bibnamefont {Yan}}, \bibinfo {author} {\bibfnamefont {M.}~\bibnamefont {Nieto-Vesperinas}}, \bibinfo {author} {\bibfnamefont {B.}~\bibnamefont {Li}}, \bibinfo {author} {\bibfnamefont {C.-W.}\ \bibnamefont {Qiu}},\ and\ \bibinfo {author} {\bibfnamefont {B.}~\bibnamefont {Yao}},\ }\bibfield  {title} {\bibinfo {title} {Observation of high-order imaginary poynting momentum optomechanics in structured light},\ }\bibfield  {journal} {\bibinfo  {journal} {Proceedings of the National Academy of Sciences}\ }\textbf {\bibinfo {volume} {119}},\ \href {https://doi.org/10.1073/pnas.2209721119} {10.1073/pnas.2209721119} (\bibinfo {year} {2022})\BibitemShut {NoStop}%
\bibitem [{\citenamefont {Belinfante}(1940)}]{Belinfante1940}%
  \BibitemOpen
  \bibfield  {author} {\bibinfo {author} {\bibfnamefont {F.~J.}\ \bibnamefont {Belinfante}},\ }\bibfield  {title} {\bibinfo {title} {{On the current and the density of the electric charge, the energy, the linear momentum and the angular momentum of arbitrary fields}},\ }\href {https://doi.org/10.1016/S0031-8914(40)90091-X} {\bibfield  {journal} {\bibinfo  {journal} {Physica}\ }\textbf {\bibinfo {volume} {7}},\ \bibinfo {pages} {449--474} (\bibinfo {year} {1940})}\BibitemShut {NoStop}%
\bibitem [{\citenamefont {Berry}(2009)}]{Berry2009}%
  \BibitemOpen
  \bibfield  {author} {\bibinfo {author} {\bibfnamefont {M.~V.}\ \bibnamefont {Berry}},\ }\bibfield  {title} {\bibinfo {title} {Optical currents},\ }\bibfield  {journal} {\bibinfo  {journal} {Journal of Optics A: Pure and Applied Optics}\ }\textbf {\bibinfo {volume} {11}},\ \href {https://doi.org/10.1088/1464-4258/11/9/094001} {10.1088/1464-4258/11/9/094001} (\bibinfo {year} {2009})\BibitemShut {NoStop}%
\bibitem [{\citenamefont {Soper}(1976)}]{Soper_book}%
  \BibitemOpen
  \bibfield  {author} {\bibinfo {author} {\bibfnamefont {D.~E.}\ \bibnamefont {Soper}},\ }\href@noop {} {\emph {\bibinfo {title} {Classical {{Field Theory}}}}}\ (\bibinfo  {publisher} {Wiley},\ \bibinfo {year} {1976})\BibitemShut {NoStop}%
\bibitem [{\citenamefont {Shi}\ \emph {et~al.}(2021)\citenamefont {Shi}, \citenamefont {Du}, \citenamefont {Li}, \citenamefont {Zayats},\ and\ \citenamefont {Yuan}}]{Shi2021}%
  \BibitemOpen
  \bibfield  {author} {\bibinfo {author} {\bibfnamefont {P.}~\bibnamefont {Shi}}, \bibinfo {author} {\bibfnamefont {L.}~\bibnamefont {Du}}, \bibinfo {author} {\bibfnamefont {C.}~\bibnamefont {Li}}, \bibinfo {author} {\bibfnamefont {A.~V.}\ \bibnamefont {Zayats}},\ and\ \bibinfo {author} {\bibfnamefont {X.}~\bibnamefont {Yuan}},\ }\bibfield  {title} {\bibinfo {title} {Transverse spin dynamics in structured electromagnetic guided waves},\ }\bibfield  {journal} {\bibinfo  {journal} {Proceedings of the National Academy of Sciences}\ }\textbf {\bibinfo {volume} {118}},\ \href {https://doi.org/10.1073/pnas.2018816118} {10.1073/pnas.2018816118} (\bibinfo {year} {2021})\BibitemShut {NoStop}%
\bibitem [{\citenamefont {Bekshaev}(2022)}]{Bekshaev2022}%
  \BibitemOpen
  \bibfield  {author} {\bibinfo {author} {\bibfnamefont {A.~Y.}\ \bibnamefont {Bekshaev}},\ }\bibfield  {title} {\bibinfo {title} {Transverse spin and the hidden vorticity of propagating light fields},\ }\href {https://doi.org/10.1364/JOSAA.466360} {\bibfield  {journal} {\bibinfo  {journal} {Journal of the Optical Society of America A}\ }\textbf {\bibinfo {volume} {39}},\ \bibinfo {pages} {1577} (\bibinfo {year} {2022})}\BibitemShut {NoStop}%
\bibitem [{\citenamefont {Cameron}\ and\ \citenamefont {Barnett}(2012)}]{Cameron2012}%
  \BibitemOpen
  \bibfield  {author} {\bibinfo {author} {\bibfnamefont {R.~P.}\ \bibnamefont {Cameron}}\ and\ \bibinfo {author} {\bibfnamefont {S.~M.}\ \bibnamefont {Barnett}},\ }\bibfield  {title} {\bibinfo {title} {Electric{\textendash}magnetic symmetry and noether{\textquotesingle}s theorem},\ }\href {https://doi.org/10.1088/1367-2630/14/12/123019} {\bibfield  {journal} {\bibinfo  {journal} {New J. Phys.}\ }\textbf {\bibinfo {volume} {14}},\ \bibinfo {pages} {123019} (\bibinfo {year} {2012})}\BibitemShut {NoStop}%
\bibitem [{\citenamefont {Cameron}\ \emph {et~al.}(2014)\citenamefont {Cameron}, \citenamefont {Barnett},\ and\ \citenamefont {Yao}}]{Cameron2014a}%
  \BibitemOpen
  \bibfield  {author} {\bibinfo {author} {\bibfnamefont {R.~P.}\ \bibnamefont {Cameron}}, \bibinfo {author} {\bibfnamefont {S.~M.}\ \bibnamefont {Barnett}},\ and\ \bibinfo {author} {\bibfnamefont {A.~M.}\ \bibnamefont {Yao}},\ }\bibfield  {title} {\bibinfo {title} {Optical helicity of interfering waves},\ }\href {https://doi.org/10.1080/09500340.2013.829874} {\bibfield  {journal} {\bibinfo  {journal} {Journal of Modern Optics}\ }\textbf {\bibinfo {volume} {61}},\ \bibinfo {pages} {25--31} (\bibinfo {year} {2014})}\BibitemShut {NoStop}%
\bibitem [{\citenamefont {Nieto-Vesperinas}(2015{\natexlab{a}})}]{NietoVesperinas2015}%
  \BibitemOpen
  \bibfield  {author} {\bibinfo {author} {\bibfnamefont {M.}~\bibnamefont {Nieto-Vesperinas}},\ }\bibfield  {title} {\bibinfo {title} {Optical theorem for the conservation of electromagnetic helicity: Significance for molecular energy transfer and enantiomeric discrimination by circular dichroism},\ }\href {https://doi.org/10.1103/PhysRevA.92.023813} {\bibfield  {journal} {\bibinfo  {journal} {Physical Review A}\ }\textbf {\bibinfo {volume} {92}},\ \bibinfo {pages} {023813} (\bibinfo {year} {2015}{\natexlab{a}})}\BibitemShut {NoStop}%
\bibitem [{\citenamefont {Nieto-Vesperinas}(2015{\natexlab{b}})}]{NietoVesperinas2015_2}%
  \BibitemOpen
  \bibfield  {author} {\bibinfo {author} {\bibfnamefont {M.}~\bibnamefont {Nieto-Vesperinas}},\ }\bibfield  {title} {\bibinfo {title} {Optical torque: Electromagnetic spin and orbital-angular-momentum conservation laws and their significance},\ }\href {https://doi.org/10.1103/PhysRevA.92.043843} {\bibfield  {journal} {\bibinfo  {journal} {Physical Review A}\ }\textbf {\bibinfo {volume} {92}},\ \bibinfo {pages} {043843} (\bibinfo {year} {2015}{\natexlab{b}})}\BibitemShut {NoStop}%
\bibitem [{\citenamefont {Landau}\ and\ \citenamefont {Lifšic}(1994)}]{Landau1994}%
  \BibitemOpen
  \bibfield  {author} {\bibinfo {author} {\bibfnamefont {L.~D.}\ \bibnamefont {Landau}}\ and\ \bibinfo {author} {\bibfnamefont {E.~M.}\ \bibnamefont {Lifšic}},\ }\href@noop {} {\emph {\bibinfo {title} {Course of theoretical physics}}},\ \bibinfo {edition} {4th}\ ed.,\ Vol.\ \bibinfo {volume} {Vol. 2}\ (\bibinfo  {publisher} {Pergamon Press},\ \bibinfo {address} {Oxford [u.a.]},\ \bibinfo {year} {1994})\BibitemShut {NoStop}%
\bibitem [{\citenamefont {Mun}\ \emph {et~al.}(2020)\citenamefont {Mun}, \citenamefont {Kim}, \citenamefont {Yang}, \citenamefont {Badloe}, \citenamefont {Ni}, \citenamefont {Chen}, \citenamefont {Qiu},\ and\ \citenamefont {Rho}}]{Mun2020}%
  \BibitemOpen
  \bibfield  {author} {\bibinfo {author} {\bibfnamefont {J.}~\bibnamefont {Mun}}, \bibinfo {author} {\bibfnamefont {M.}~\bibnamefont {Kim}}, \bibinfo {author} {\bibfnamefont {Y.}~\bibnamefont {Yang}}, \bibinfo {author} {\bibfnamefont {T.}~\bibnamefont {Badloe}}, \bibinfo {author} {\bibfnamefont {J.}~\bibnamefont {Ni}}, \bibinfo {author} {\bibfnamefont {Y.}~\bibnamefont {Chen}}, \bibinfo {author} {\bibfnamefont {C.-W.}\ \bibnamefont {Qiu}},\ and\ \bibinfo {author} {\bibfnamefont {J.}~\bibnamefont {Rho}},\ }\bibfield  {title} {\bibinfo {title} {Electromagnetic chirality: from fundamentals to nontraditional chiroptical phenomena},\ }\bibfield  {journal} {\bibinfo  {journal} {Light Sci. Appl.}\ }\textbf {\bibinfo {volume} {9}},\ \href {https://doi.org/10.1038/s41377-020-00367-8} {10.1038/s41377-020-00367-8} (\bibinfo {year} {2020})\BibitemShut {NoStop}%
\bibitem [{\citenamefont {Muljarov}\ and\ \citenamefont {Weiss}(2018)}]{muljarov2018resonant}%
  \BibitemOpen
  \bibfield  {author} {\bibinfo {author} {\bibfnamefont {E.~A.}\ \bibnamefont {Muljarov}}\ and\ \bibinfo {author} {\bibfnamefont {T.}~\bibnamefont {Weiss}},\ }\bibfield  {title} {\bibinfo {title} {Resonant-state expansion for open optical systems: generalization to magnetic, chiral, and bi-anisotropic materials},\ }\href@noop {} {\bibfield  {journal} {\bibinfo  {journal} {Optics letters}\ }\textbf {\bibinfo {volume} {43}},\ \bibinfo {pages} {1978--1981} (\bibinfo {year} {2018})}\BibitemShut {NoStop}%
\bibitem [{\citenamefont {Ziolkowski}(2004)}]{ziolkowski2004propagation}%
  \BibitemOpen
  \bibfield  {author} {\bibinfo {author} {\bibfnamefont {R.~W.}\ \bibnamefont {Ziolkowski}},\ }\bibfield  {title} {\bibinfo {title} {Propagation in and scattering from a matched metamaterial having a zero index of refraction},\ }\href@noop {} {\bibfield  {journal} {\bibinfo  {journal} {Physical Review E}\ }\textbf {\bibinfo {volume} {70}},\ \bibinfo {pages} {046608} (\bibinfo {year} {2004})}\BibitemShut {NoStop}%
\bibitem [{\citenamefont {Mahmoud}\ and\ \citenamefont {Engheta}(2014)}]{mahmoud2014wave}%
  \BibitemOpen
  \bibfield  {author} {\bibinfo {author} {\bibfnamefont {A.~M.}\ \bibnamefont {Mahmoud}}\ and\ \bibinfo {author} {\bibfnamefont {N.}~\bibnamefont {Engheta}},\ }\bibfield  {title} {\bibinfo {title} {Wave--matter interactions in epsilon-and-mu-near-zero structures},\ }\href@noop {} {\bibfield  {journal} {\bibinfo  {journal} {Nature communications}\ }\textbf {\bibinfo {volume} {5}},\ \bibinfo {pages} {5638} (\bibinfo {year} {2014})}\BibitemShut {NoStop}%
\bibitem [{\citenamefont {Zambrana-Puyalto}\ \emph {et~al.}(2013)\citenamefont {Zambrana-Puyalto}, \citenamefont {Fernandez-Corbaton}, \citenamefont {Juan}, \citenamefont {Vidal},\ and\ \citenamefont {Molina-Terriza}}]{zambrana2013duality}%
  \BibitemOpen
  \bibfield  {author} {\bibinfo {author} {\bibfnamefont {X.}~\bibnamefont {Zambrana-Puyalto}}, \bibinfo {author} {\bibfnamefont {I.}~\bibnamefont {Fernandez-Corbaton}}, \bibinfo {author} {\bibfnamefont {M.}~\bibnamefont {Juan}}, \bibinfo {author} {\bibfnamefont {X.}~\bibnamefont {Vidal}},\ and\ \bibinfo {author} {\bibfnamefont {G.}~\bibnamefont {Molina-Terriza}},\ }\bibfield  {title} {\bibinfo {title} {Duality symmetry and kerker conditions},\ }\href@noop {} {\bibfield  {journal} {\bibinfo  {journal} {Optics letters}\ }\textbf {\bibinfo {volume} {38}},\ \bibinfo {pages} {1857--1859} (\bibinfo {year} {2013})}\BibitemShut {NoStop}%
\end{thebibliography}%


%apsrev4-2.bst 2019-01-14 (MD) hand-edited version of apsrev4-1.bst
%Control: key (0)
%Control: author (72) initials jnrlst
%Control: editor formatted (1) identically to author
%Control: production of article title (0) allowed
%Control: page (1) range
%Control: year (1) truncated
%Control: production of eprint (0) enabled
\begin{thebibliography}{16}%
\makeatletter
\providecommand \@ifxundefined [1]{%
 \@ifx{#1\undefined}
}%
\providecommand \@ifnum [1]{%
 \ifnum #1\expandafter \@firstoftwo
 \else \expandafter \@secondoftwo
 \fi
}%
\providecommand \@ifx [1]{%
 \ifx #1\expandafter \@firstoftwo
 \else \expandafter \@secondoftwo
 \fi
}%
\providecommand \natexlab [1]{#1}%
\providecommand \enquote  [1]{``#1''}%
\providecommand \bibnamefont  [1]{#1}%
\providecommand \bibfnamefont [1]{#1}%
\providecommand \citenamefont [1]{#1}%
\providecommand \href@noop [0]{\@secondoftwo}%
\providecommand \href [0]{\begingroup \@sanitize@url \@href}%
\providecommand \@href[1]{\@@startlink{#1}\@@href}%
\providecommand \@@href[1]{\endgroup#1\@@endlink}%
\providecommand \@sanitize@url [0]{\catcode `\\12\catcode `\$12\catcode `\&12\catcode `\#12\catcode `\^12\catcode `\_12\catcode `\%12\relax}%
\providecommand \@@startlink[1]{}%
\providecommand \@@endlink[0]{}%
\providecommand \url  [0]{\begingroup\@sanitize@url \@url }%
\providecommand \@url [1]{\endgroup\@href {#1}{\urlprefix }}%
\providecommand \urlprefix  [0]{URL }%
\providecommand \Eprint [0]{\href }%
\providecommand \doibase [0]{https://doi.org/}%
\providecommand \selectlanguage [0]{\@gobble}%
\providecommand \bibinfo  [0]{\@secondoftwo}%
\providecommand \bibfield  [0]{\@secondoftwo}%
\providecommand \translation [1]{[#1]}%
\providecommand \BibitemOpen [0]{}%
\providecommand \bibitemStop [0]{}%
\providecommand \bibitemNoStop [0]{.\EOS\space}%
\providecommand \EOS [0]{\spacefactor3000\relax}%
\providecommand \BibitemShut  [1]{\csname bibitem#1\endcsname}%
\let\auto@bib@innerbib\@empty
%</preamble>
\bibitem [{\citenamefont {Lounesto}(2001)}]{Lounesto2001}%
  \BibitemOpen
  \bibfield  {author} {\bibinfo {author} {\bibfnamefont {P.}~\bibnamefont {Lounesto}},\ }\bibfield  {title} {\bibinfo {title} {Clifford algebras and spinors},\ }in\ \href@noop {} {\emph {\bibinfo {booktitle} {Clifford Algebras and Their Applications in Mathematical Physics}}}\ (\bibinfo  {publisher} {Springer},\ \bibinfo {year} {2001})\ pp.\ \bibinfo {pages} {37--39}\BibitemShut {NoStop}%
\bibitem [{\citenamefont {Nieto-Vesperinas}\ and\ \citenamefont {Xu}(2022)}]{NietoVesperinas2022}%
  \BibitemOpen
  \bibfield  {author} {\bibinfo {author} {\bibfnamefont {M.}~\bibnamefont {Nieto-Vesperinas}}\ and\ \bibinfo {author} {\bibfnamefont {X.}~\bibnamefont {Xu}},\ }\bibfield  {title} {\bibinfo {title} {The complex maxwell stress tensor theorem: The imaginary stress tensor and the reactive strength of orbital momentum. a novel scenery underlying electromagnetic optical forces},\ }\href {https://doi.org/10.1038/s41377-022-00979-2} {\bibfield  {journal} {\bibinfo  {journal} {Light: Science \& Applications}\ }\textbf {\bibinfo {volume} {11}},\ \bibinfo {pages} {297} (\bibinfo {year} {2022})}\BibitemShut {NoStop}%
\bibitem [{\citenamefont {Jackson}(1998)}]{Jackson1998}%
  \BibitemOpen
  \bibfield  {author} {\bibinfo {author} {\bibfnamefont {J.~D.}\ \bibnamefont {Jackson}},\ }\href@noop {} {\emph {\bibinfo {title} {Classical Electrodynamics}}}\ (\bibinfo  {publisher} {John Wiley and Sons Ltd},\ \bibinfo {year} {1998})\BibitemShut {NoStop}%
\bibitem [{\citenamefont {Bergman}\ and\ \citenamefont {Carozzi}(2009)}]{Bergman2009}%
  \BibitemOpen
  \bibfield  {author} {\bibinfo {author} {\bibfnamefont {J.~E.~S.}\ \bibnamefont {Bergman}}\ and\ \bibinfo {author} {\bibfnamefont {T.~D.}\ \bibnamefont {Carozzi}},\ }\href@noop {} {\bibinfo {title} {Canonical electromagnetic observables for systematic characterization of electric and magnetic wave field data on board spacecraft}} (\bibinfo {year} {2009}),\ \Eprint {https://arxiv.org/abs/0804.2092} {arXiv:0804.2092 [physics.geo-ph]} \BibitemShut {NoStop}%
\bibitem [{\citenamefont {Bliokh}\ \emph {et~al.}(2014)\citenamefont {Bliokh}, \citenamefont {Kivshar},\ and\ \citenamefont {Nori}}]{Bliokh2014}%
  \BibitemOpen
  \bibfield  {author} {\bibinfo {author} {\bibfnamefont {K.~Y.}\ \bibnamefont {Bliokh}}, \bibinfo {author} {\bibfnamefont {Y.~S.}\ \bibnamefont {Kivshar}},\ and\ \bibinfo {author} {\bibfnamefont {F.}~\bibnamefont {Nori}},\ }\bibfield  {title} {\bibinfo {title} {Magnetoelectric effects in local light-matter interactions},\ }\href {https://doi.org/10.1103/PhysRevLett.113.033601} {\bibfield  {journal} {\bibinfo  {journal} {Phys. Rev. Lett.}\ }\textbf {\bibinfo {volume} {113}},\ \bibinfo {pages} {033601} (\bibinfo {year} {2014})}\BibitemShut {NoStop}%
\bibitem [{\citenamefont {Nieto-Vesperinas}\ and\ \citenamefont {Xu}(2021)}]{NietoVesperinas2021}%
  \BibitemOpen
  \bibfield  {author} {\bibinfo {author} {\bibfnamefont {M.}~\bibnamefont {Nieto-Vesperinas}}\ and\ \bibinfo {author} {\bibfnamefont {X.}~\bibnamefont {Xu}},\ }\bibfield  {title} {\bibinfo {title} {Reactive helicity and reactive power in nanoscale optics: Evanescent waves. kerker conditions. optical theorems and reactive dichroism},\ }\href {https://doi.org/10.1103/PhysRevResearch.3.043080} {\bibfield  {journal} {\bibinfo  {journal} {Physical Review Research}\ }\textbf {\bibinfo {volume} {3}},\ \bibinfo {pages} {043080} (\bibinfo {year} {2021})}\BibitemShut {NoStop}%
\bibitem [{\citenamefont {Kamenetskii}\ \emph {et~al.}(2015)\citenamefont {Kamenetskii}, \citenamefont {Berezin},\ and\ \citenamefont {Shavit}}]{kamenetskii2015microwave}%
  \BibitemOpen
  \bibfield  {author} {\bibinfo {author} {\bibfnamefont {E.}~\bibnamefont {Kamenetskii}}, \bibinfo {author} {\bibfnamefont {M.}~\bibnamefont {Berezin}},\ and\ \bibinfo {author} {\bibfnamefont {R.}~\bibnamefont {Shavit}},\ }\bibfield  {title} {\bibinfo {title} {Microwave magnetoelectric fields: helicities and reactive power flows},\ }\href@noop {} {\bibfield  {journal} {\bibinfo  {journal} {Applied Physics B}\ }\textbf {\bibinfo {volume} {121}},\ \bibinfo {pages} {31--47} (\bibinfo {year} {2015})}\BibitemShut {NoStop}%
\bibitem [{\citenamefont {Vernon}\ \emph {et~al.}(2023)\citenamefont {Vernon}, \citenamefont {Golat}, \citenamefont {Rigouzzo}, \citenamefont {Lim},\ and\ \citenamefont {Rodríguez-Fortuño}}]{Vernon2023}%
  \BibitemOpen
  \bibfield  {author} {\bibinfo {author} {\bibfnamefont {A.~J.}\ \bibnamefont {Vernon}}, \bibinfo {author} {\bibfnamefont {S.}~\bibnamefont {Golat}}, \bibinfo {author} {\bibfnamefont {C.}~\bibnamefont {Rigouzzo}}, \bibinfo {author} {\bibfnamefont {E.~A.}\ \bibnamefont {Lim}},\ and\ \bibinfo {author} {\bibfnamefont {F.~J.}\ \bibnamefont {Rodríguez-Fortuño}},\ }\href@noop {} {\bibinfo {title} {A decomposition of light's spin angular momentum density}} (\bibinfo {year} {2023}),\ \Eprint {https://arxiv.org/abs/2310.03804} {arXiv:2310.03804 [physics.optics]} \BibitemShut {NoStop}%
\bibitem [{\citenamefont {Picardi}\ \emph {et~al.}(2018)\citenamefont {Picardi}, \citenamefont {Zayats},\ and\ \citenamefont {Rodríguez-Fortuño}}]{Picardi2018_2}%
  \BibitemOpen
  \bibfield  {author} {\bibinfo {author} {\bibfnamefont {M.~F.}\ \bibnamefont {Picardi}}, \bibinfo {author} {\bibfnamefont {A.~V.}\ \bibnamefont {Zayats}},\ and\ \bibinfo {author} {\bibfnamefont {F.~J.}\ \bibnamefont {Rodríguez-Fortuño}},\ }\bibfield  {title} {\bibinfo {title} {Janus and huygens dipoles: Near-field directionality beyond spin-momentum locking},\ }\href {https://doi.org/10.1103/PhysRevLett.120.117402} {\bibfield  {journal} {\bibinfo  {journal} {Physical Review Letters}\ }\textbf {\bibinfo {volume} {120}},\ \bibinfo {pages} {117402} (\bibinfo {year} {2018})}\BibitemShut {NoStop}%
\bibitem [{\citenamefont {Vernon}\ \emph {et~al.}(2024)\citenamefont {Vernon}, \citenamefont {Kille}, \citenamefont {Rodríguez-Fortuño},\ and\ \citenamefont {Afanasev}}]{Vernon2024}%
  \BibitemOpen
  \bibfield  {author} {\bibinfo {author} {\bibfnamefont {A.~J.}\ \bibnamefont {Vernon}}, \bibinfo {author} {\bibfnamefont {A.}~\bibnamefont {Kille}}, \bibinfo {author} {\bibfnamefont {F.~J.}\ \bibnamefont {Rodríguez-Fortuño}},\ and\ \bibinfo {author} {\bibfnamefont {A.}~\bibnamefont {Afanasev}},\ }\bibfield  {title} {\bibinfo {title} {Non-diffracting polarization features around far-field zeros of electromagnetic radiation},\ }\href {https://doi.org/10.1364/OPTICA.502020} {\bibfield  {journal} {\bibinfo  {journal} {Optica}\ }\textbf {\bibinfo {volume} {11}},\ \bibinfo {pages} {120} (\bibinfo {year} {2024})}\BibitemShut {NoStop}%
\bibitem [{\citenamefont {Muljarov}\ and\ \citenamefont {Weiss}(2018)}]{muljarov2018resonant}%
  \BibitemOpen
  \bibfield  {author} {\bibinfo {author} {\bibfnamefont {E.~A.}\ \bibnamefont {Muljarov}}\ and\ \bibinfo {author} {\bibfnamefont {T.}~\bibnamefont {Weiss}},\ }\bibfield  {title} {\bibinfo {title} {Resonant-state expansion for open optical systems: generalization to magnetic, chiral, and bi-anisotropic materials},\ }\href@noop {} {\bibfield  {journal} {\bibinfo  {journal} {Optics letters}\ }\textbf {\bibinfo {volume} {43}},\ \bibinfo {pages} {1978--1981} (\bibinfo {year} {2018})}\BibitemShut {NoStop}%
\bibitem [{\citenamefont {Ziolkowski}(2004)}]{ziolkowski2004propagation}%
  \BibitemOpen
  \bibfield  {author} {\bibinfo {author} {\bibfnamefont {R.~W.}\ \bibnamefont {Ziolkowski}},\ }\bibfield  {title} {\bibinfo {title} {Propagation in and scattering from a matched metamaterial having a zero index of refraction},\ }\href@noop {} {\bibfield  {journal} {\bibinfo  {journal} {Physical Review E}\ }\textbf {\bibinfo {volume} {70}},\ \bibinfo {pages} {046608} (\bibinfo {year} {2004})}\BibitemShut {NoStop}%
\bibitem [{\citenamefont {Mahmoud}\ and\ \citenamefont {Engheta}(2014)}]{mahmoud2014wave}%
  \BibitemOpen
  \bibfield  {author} {\bibinfo {author} {\bibfnamefont {A.~M.}\ \bibnamefont {Mahmoud}}\ and\ \bibinfo {author} {\bibfnamefont {N.}~\bibnamefont {Engheta}},\ }\bibfield  {title} {\bibinfo {title} {Wave--matter interactions in epsilon-and-mu-near-zero structures},\ }\href@noop {} {\bibfield  {journal} {\bibinfo  {journal} {Nature communications}\ }\textbf {\bibinfo {volume} {5}},\ \bibinfo {pages} {5638} (\bibinfo {year} {2014})}\BibitemShut {NoStop}%
\bibitem [{\citenamefont {Zambrana-Puyalto}\ \emph {et~al.}(2013)\citenamefont {Zambrana-Puyalto}, \citenamefont {Fernandez-Corbaton}, \citenamefont {Juan}, \citenamefont {Vidal},\ and\ \citenamefont {Molina-Terriza}}]{zambrana2013duality}%
  \BibitemOpen
  \bibfield  {author} {\bibinfo {author} {\bibfnamefont {X.}~\bibnamefont {Zambrana-Puyalto}}, \bibinfo {author} {\bibfnamefont {I.}~\bibnamefont {Fernandez-Corbaton}}, \bibinfo {author} {\bibfnamefont {M.}~\bibnamefont {Juan}}, \bibinfo {author} {\bibfnamefont {X.}~\bibnamefont {Vidal}},\ and\ \bibinfo {author} {\bibfnamefont {G.}~\bibnamefont {Molina-Terriza}},\ }\bibfield  {title} {\bibinfo {title} {Duality symmetry and kerker conditions},\ }\href@noop {} {\bibfield  {journal} {\bibinfo  {journal} {Optics letters}\ }\textbf {\bibinfo {volume} {38}},\ \bibinfo {pages} {1857--1859} (\bibinfo {year} {2013})}\BibitemShut {NoStop}%
\bibitem [{\citenamefont {Carozzi}\ \emph {et~al.}(2000)\citenamefont {Carozzi}, \citenamefont {Karlsson},\ and\ \citenamefont {Bergman}}]{Carozzi2000}%
  \BibitemOpen
  \bibfield  {author} {\bibinfo {author} {\bibfnamefont {T.}~\bibnamefont {Carozzi}}, \bibinfo {author} {\bibfnamefont {R.}~\bibnamefont {Karlsson}},\ and\ \bibinfo {author} {\bibfnamefont {J.}~\bibnamefont {Bergman}},\ }\bibfield  {title} {\bibinfo {title} {Parameters characterizing electromagnetic wave polarization},\ }\href@noop {} {\bibfield  {journal} {\bibinfo  {journal} {Physical Review E}\ }\textbf {\bibinfo {volume} {61}},\ \bibinfo {pages} {2024} (\bibinfo {year} {2000})}\BibitemShut {NoStop}%
\bibitem [{\citenamefont {Set{\"a}l{\"a}}\ \emph {et~al.}(2002)\citenamefont {Set{\"a}l{\"a}}, \citenamefont {Shevchenko}, \citenamefont {Kaivola},\ and\ \citenamefont {Friberg}}]{Setala2002}%
  \BibitemOpen
  \bibfield  {author} {\bibinfo {author} {\bibfnamefont {T.}~\bibnamefont {Set{\"a}l{\"a}}}, \bibinfo {author} {\bibfnamefont {A.}~\bibnamefont {Shevchenko}}, \bibinfo {author} {\bibfnamefont {M.}~\bibnamefont {Kaivola}},\ and\ \bibinfo {author} {\bibfnamefont {A.~T.}\ \bibnamefont {Friberg}},\ }\bibfield  {title} {\bibinfo {title} {Degree of polarization for optical near fields},\ }\href@noop {} {\bibfield  {journal} {\bibinfo  {journal} {Physical Review E}\ }\textbf {\bibinfo {volume} {66}},\ \bibinfo {pages} {016615} (\bibinfo {year} {2002})}\BibitemShut {NoStop}%
\end{thebibliography}%

\end{document}

\appendix

\title{Supplemental Material for ``The electromagnetic symmetry sphere: a framework for energy, momentum, spin and other electromagnetic quantities''}

\author{Sebastian Golat}
\email{sebastian.1.golat@kcl.ac.uk}
\affiliation{Department of Physics, King's College London, Strand, London WC2R 2LS, UK}
\affiliation{London Centre for Nanotechnology}

\author{Alex J. Vernon}
% \email{alexander.vernon@kcl.ac.uk}
\affiliation{Department of Physics, King's College London, Strand, London WC2R 2LS, UK}
\affiliation{London Centre for Nanotechnology}

\author{Francisco~J. Rodr\'iguez-Fortu\~no}
\email{francisco.rodriguez\_fortuno@kcl.ac.uk}
\affiliation{Department of Physics, King's College London, Strand, London WC2R 2LS, UK}
\affiliation{London Centre for Nanotechnology}

\date{\today}
% \begin{abstract}

% \end{abstract}

\maketitle%
\section{Bispinor formalism}
\subsection{Bases in the electromagnetic bispinor vector space}

Electromagnetic bispinors are a mathematical description of electromagnetic fields. For monochromatic light, these bispinors live in a $\mathbb{C}^3\times\mathbb{C}^2\times\mathbb{L}^2(\mathbb{R}^3)$ vector space. The bispinor $\vc{\psi}(\vc{r})$ or $\vc{\psi}(\vc{k})$ contains all the information about the electromagnetic field at any point in real space $\vc{r}$ or momentum space $\vc{k}$. Indeed it has the same information as the two complex phasor vectors $\vc{E}$ and $\vc{H}$ defined at a given point, hence most works write it as $\vc{\psi} = (\vc{E},\vc{H})$, but this representation as a two-dimensional vector implicitly assumes a specific basis representation in the $\mathbb{C}^2$ space of electric and magnetic fields. Because we are going to choose different bases in that space, it is first convenient to re-scale our electric and magnetic field vectors in order to make their dimensions match and for them to be normalised in terms of energy. For this, consider the electromagnetic energy density in a linear, homogeneous, isotropic medium with permittivity $\varepsilon$ and permeability $\mu$:
\begin{equation}\label{eq:introEMenergydensitywithEH}
    W = \frac{1}{4}({\varepsilon}\abs{\vc{E}}^2 + {\mu}\abs{\vc{H}}^2). 
\end{equation}%
\noindent This expression suggests defining re-scaled vectors to describe electric and magnetic fields:
\begin{equation}\label{eq:rescaledFeFmintermsofEH}
\begin{alignedat}{1}
    \vc{F}_\text{e} &\equiv \frac{1}{2} \sqrt{\varepsilon} \vc{E}, \\
    \vc{F}_\text{m} &\equiv \frac{1}{2} \sqrt{\mu} \vc{H}, \\
\end{alignedat}   
\end{equation}%
\noindent such that they have the same dimensions. The electromagnetic energy density can then be written as:
\begin{equation}\label{eq:introEMenergydensitywithFeFm}
    W = \abs{\vc{F}_\text{e}}^2 +\abs{\vc{F}_\text{m}}^2. 
\end{equation}%

We can describe an electromagnetic field either with $(\vc{E},\vc{H})$ vectors, or with $(\vc{F}_\text{e},\vc{F}_\text{m})$, knowing these two descriptions to be equivalent and differing only by fixed constants in a specific medium. Now let's focus our attention on the $\mathbb{C}^3\times\mathbb{C}^2$ nature of the electromagnetic bi-spinor (the remaining $\mathbb{L}^2(\mathbb{R}^3)$ nature describes the spatial or momentum space dependence of each field). The most common basis for this bi-spinor is to use $(\uv{x},\uv{y},\uv{z})$ basis in the $\mathbb{C}^3$ space, and the electric-magnetic basis (let's introduce unit vectors $(\uv{e},\uv{m})$ to be defined later) in the $\mathbb{C}^2$ space, hence one can understand the bi-spinor as a six-dimensional entity composed of six scalar fields $\vc{\psi} = (F_{\text{e},x},F_{\text{e},y},F_{\text{e},z},F_{\text{m},x},F_{\text{m},y},F_{\text{m},z})$ where $F_{\text{e},x}$ is the x-component of $\vc{F}_\text{e}$, and so on. This can be understood as a linear combination of basis vectors as follows:
\begin{equation}\label{eq:bispinorfullexpansion}
\begin{alignedat}{4}
    \vc{\psi} &= F_{\text{e},x}\uv{x}\uv{e} + F_{\text{e},y}\uv{y}\uv{e}+F_{\text{e},z}\uv{z}\uv{e}+F_{\text{m},x}\uv{x}\uv{m} + F_{\text{m},y}\uv{y}\uv{m}+F_{\text{m},z}\uv{z}\uv{m}
\end{alignedat}    
\end{equation}%

\noindent where the product of unit vectors is an outer product between \emph{different} vector spaces. We can factor out this expression in either of two ways:
\begin{equation}\label{eq:basisexplanationxyz}
\begin{alignedat}{4}
    \vc{\psi} &= \vc{\psi}_x \uv{x} + \vc{\psi}_y \uv{y} + \vc{\psi}_z \uv{z} \quad \to \quad \vc{\psi}_{[x,y,z]} = \underbrace{\vc{\psi}_x}_{\mathbb{C}^2}  \underbrace{\pmqty{1\\0\\0}}_{\mathbb{C}^3} + \vc{\psi}_y \pmqty{0\\1\\0} + \vc{\psi}_z \pmqty{0\\0\\1} = \pmqty{\vc{\psi}_x\\\vc{\psi}_y\\\vc{\psi}_z},
\end{alignedat}    
\end{equation}%

\noindent where $\vc{\psi}_x = F_{\text{e},x}\uv{e} + F_{\text{m},x}\uv{m} $ is a $\mathbb{C}^2$ vector coefficient containing all information of electric and magnetic field's x-components, and similarly for $\vc{\psi}_y$ and $\vc{\psi}_z$. Alternatively, we can factor it as:

\begin{equation}\label{eq:basisexplanationEM}
\begin{alignedat}{4}
    \vc{\psi} &= \vc{F}_\text{e} \uv{e} + \vc{F}_\text{m} \uv{m}  \quad \to \quad \vc{\psi}_{\text{[EM]}} = \underbrace{\vc{F}_\text{e}}_{\mathbb{C}^3} \underbrace{\pmqty{1\\0}}_{\mathbb{C}^2}   + \vc{F}_\text{m} \pmqty{0\\1} = \pmqty{\vc{F}_\text{e}\\\vc{F}_\text{m}}
\end{alignedat}    
\end{equation}%

\noindent where $\vc{F}_\text{e} = F_{\text{e},x}\uv{x} + F_{\text{e},y}\uv{y} + F_{\text{e},z}\uv{z}$ is a vector in $\mathbb{C}^3$ space that contains the information of the electric field in all spatial directions, and similarly for $\vc{F}_\text{m}$ in the magnetic field.

The most unusual aspect of this notation for readers with an optics and electromagnetism background is that we are writing a linear combination of basis vectors in a given space, but the \emph{coefficients} of this linear combination are themselves vectors in a different vector space. That is why $\vc{\psi}_{\text{[EM]}} = (\vc{F}_\text{e},\vc{F}_\text{m})$ is a vector (in $\mathbb{C}^2$) whose elements are themselves vectors (in $\mathbb{C}^3$). Also, when we write down vectors as an array of numbers, we are inherently defining a basis, as we just did for $\vc{\psi}_{\text{[EM]}}$, and we use square bracket subscript to explicitly indicate this. Alternatively, we may use linear combinations of basis vectors, such as $\vc{\psi} = \vc{F}_\text{e} \uv{e} + \vc{F}_\text{m} \uv{m}$ to keep a basis independent notation.

It is well known that the $(\uv{x},\uv{y},\uv{z})$ basis is not the only basis to express three-dimensional vectors such as $\vc{E}$. Similarly, and this being the key point of this manuscript, the electric-magnetic $(\uv{e},\uv{m})$ basis used in $\mathbb{C}^2$ subspace in \cref{eq:basisexplanationEM} is not the only basis in that space. In fact, being a $\mathbb{C}^2$ space isomorphic with the Jones vectors of the field in a plane, we take inspiration on the horizontal-vertical, diagonal-antidiagonal, and right-left-handed polarisation bases, to define the following unit vectors:
\begin{equation}\label{eq:C2basisvectors}
\begin{alignedat}{2}
    \uv{p} &= (\uv{e}+\uv{m})/\sqrt{2}, \quad \quad \quad
    \uv{r} &= (\uv{e}+\ii\uv{m})/\sqrt{2}, \\
    \uv{a} &= (\uv{e}-\uv{m})/\sqrt{2},\quad \quad \quad
    \uv{l} &= (\uv{e}-\ii\uv{m})/\sqrt{2}.
\end{alignedat}    
\end{equation}%

These basis vectors form three orthonormal bases in the $\mathbb{C}^2$ subspace of $\vc{\psi}$: the EM basis $(\uv{e},\uv{m})$ distinguishing electric and magnetic fields, the PA basis $(\uv{p},\uv{a})$ distinguishing between states where electric and magnetic fields are parallel or antiparallel, and the RL basis $(\uv{r},\uv{l})$ which corresponds to distinguishing fields with different helicities. These three bases are depicted in the main text Fig.~1. Any electromagnetic field $\vc{\psi}$ can be written as a linear combination in any of these bases:
\begin{equation}\label{eq:phiexpressedinthreebases}
    \boxed{
        \quad \vc{\psi} = \vc{F}_\text{e} \uv{e} + \vc{F}_\text{m} \uv{m} = \vc{F}_\text{p} \uv{p} + \vc{F}_\text{a} \uv{a} = \vc{F}_\text{R} \uv{r} + \vc{F}_\text{L} \uv{l} \quad
    }
\end{equation}
where the coefficients $\vc{F}_i$ are vectors in $\mathbb{C}^3$. This implies that we may write this bispinor as an array of two elements by choosing different bases, resulting in the notation:
\begin{equation}\label{eq:phiexpressedinthreebasesarrayform}
    \vc{\psi}_{[\text{EM}]} =  \pmqty{\vc{F}_\text{e}\\\vc{F}_\text{m}} \quad \quad \vc{\psi}_{[\text{PA}]} =  \pmqty{\vc{F}_\text{p}\\\vc{F}_\text{a}} \quad \quad \vc{\psi}_{[\text{RL}]} =  \pmqty{\vc{F}_\text{R}\\\vc{F}_\text{L}} ,
\end{equation}%
as used in Eq.~(1) in the main text. The unit vectors are normalised such that they each correspond to electromagnetic fields with unit energy density. We can conclude that electromagnetic energy density from \cref{eq:introEMenergydensitywithFeFm} is the norm of the electromagnetic bispinor, and thus thanks to Parseval's theorem can be written in terms of the coefficients in any basis: 
\begin{equation}\label{eq:EMenergydensitywithFeFmFpFaFRFL}
    W = \norm{\vc{\psi}}^2= \abs{\vc{F}_\text{e}}^2 +\abs{\vc{F}_\text{m}}^2 = \abs{\vc{F}_\text{p}}^2 +\abs{\vc{F}_\text{a}}^2 = \abs{\vc{F}_\text{R}}^2 +\abs{\vc{F}_\text{L}}^2. 
\end{equation}%

For completeness and clarity, the simultaneous \cref{eq:phiexpressedinthreebases} can be written explicitly in each basis:
\begin{equation}\label{eq:expandphisuperpositioninthreebases}
\begin{alignedat}{6}
    \vc{\psi}_{[\text{EM}]} =  \pmqty{\vc{F}_\text{e}\\\vc{F}_\text{m}} &= \vc{F}_\text{e}  \pmqty{1\\0} &&+ \vc{F}_\text{m}  \pmqty{0\\1} &&= \vc{F}_\text{p} \pmqty{\frac{1}{\sqrt{2}}\\\frac{1}{\sqrt{2}}} &&+ \vc{F}_\text{a} \pmqty{\frac{1}{\sqrt{2}}\\\frac{-1}{\sqrt{2}}} &&= \vc{F}_\text{R} \pmqty{\frac{1}{\sqrt{2}}\\\frac{\ii}{\sqrt{2}}} &&+ \vc{F}_\text{L} \pmqty{\frac{1}{\sqrt{2}}\\\frac{-\ii}{\sqrt{2}}}\\
    \vc{\psi}_{[\text{PA}]} =  \pmqty{\vc{F}_\text{p}\\\vc{F}_\text{a}} &=  \vc{F}_\text{e}  \pmqty{\frac{1}{\sqrt{2}}\\\frac{1}{\sqrt{2}}} &&+ \vc{F}_\text{m}  \pmqty{\frac{1}{\sqrt{2}}\\\frac{-1}{\sqrt{2}}} &&= \vc{F}_\text{p} \pmqty{1\\0} &&+ \vc{F}_\text{a} \pmqty{0\\1} &&= \vc{F}_\text{R} \pmqty{\frac{1+i}{2}\\\frac{1-i}{2}} &&+ \vc{F}_\text{L} \pmqty{\frac{1-i}{2}\\\frac{1+i}{2}}\\
    \vc{\psi}_{[\text{RL}]} =  \pmqty{\vc{F}_\text{R}\\\vc{F}_\text{L}} &=  \vc{F}_\text{e}  \underbrace{\pmqty{\frac{1}{\sqrt{2}}\\\frac{1}{\sqrt{2}}}}_{\uv{e}} &&+ \vc{F}_\text{m}  \underbrace{\pmqty{\frac{-\ii}{\sqrt{2}}\\\frac{\ii}{\sqrt{2}}}}_{\uv{m}} &&= \vc{F}_\text{p} \underbrace{\pmqty{\frac{1-i}{2}\\\frac{1+i}{2}}}_{\uv{p}} &&+ \vc{F}_\text{a} \underbrace{\pmqty{\frac{1+i}{2}\\\frac{1-i}{2}}}_{\uv{a}} &&= \vc{F}_\text{R} \underbrace{\pmqty{1\\0}\ \ }_{\uv{r}} &&+ \vc{F}_\text{L} \underbrace{\pmqty{0\\1}\ \ }_{\uv{l}}\\
\end{alignedat}
\end{equation}
where we have expressed each of the six basis vectors using each of the three orthonormal bases, by re-arranging \cref{eq:C2basisvectors}, revealing the strict relations between the different $\vc{F}_i$.

This completes the formal description of our approach to the different bases for electromagnetic bi-spinors. It is however interesting to also consider how this approach translates into the usual language of $\vc{E}$ and $\vc{H}$ as we do in the next section.

\subsection{Bispinor basis formalism in terms of \texorpdfstring{$\vc{E}$}{E} and \texorpdfstring{$\vc{H}$}{H}}

\begin{table}[!ht]
\centering
\begin{tabular}{|c|c|c|}
\hline
\textbf{Basis name} & \text{$\vc{F}[\mathbb{C}^3] \otimes \vc{\hat{u}}[\mathbb{C}^2]$} & \text{Meaning in terms of $\vc{E}$ and $\vc{H}$} \\
\hline
EM & \( \vc{F}_\text{e} \uv{e} \) & \( \vc{E} = 2 \vc{F}_\text{e}/\sqrt{\varepsilon} \quad \text{and} \quad \vc{H} = \vc{0} \) \\
basis & \( \vc{F}_\text{m} \uv{m} \) & \( \vc{E} = \vc{0} \quad \text{and} \quad \vc{H} = 2 \vc{F}_\text{m}/\sqrt{\mu} \) \\
\hline
PA & \( \vc{F}_\text{p} \uv{p} \) & \( \vc{E} = \sqrt{2} \vc{F}_\text{p}/\sqrt{\varepsilon} \quad \text{and} \quad \vc{H} = \sqrt{2} \vc{F}_\text{p}/\sqrt{\mu} \) \\
basis & \( \vc{F}_\text{a} \uv{a} \) & \( \vc{E} = \sqrt{2} \vc{F}_\text{a}/\sqrt{\varepsilon} \quad \text{and} \quad \vc{H} = -\sqrt{2} \vc{F}_\text{a}/\sqrt{\mu} \) \\
\hline
RL & \( \vc{F}_\text{R} \uv{r} \) & \( \vc{E} = \sqrt{2} \vc{F}_\text{R}/\sqrt{\varepsilon} \quad \text{and} \quad \vc{H} = \ii \sqrt{2} \vc{F}_\text{R}/\sqrt{\mu} \) \\
basis & \( \vc{F}_\text{L} \uv{l} \) & \( \vc{E} = \sqrt{2} \vc{F}_\text{L}/\sqrt{\varepsilon} \quad \text{and} \quad \vc{H} = -\ii \sqrt{2} \vc{F}_\text{L}/\sqrt{\mu} \) \\
\hline
\end{tabular}
\caption{The precise meaning of the $\mathbb{C}^2$ basis vectors in terms of $\vc{E}$ and $\vc{H}$. To derive this table, start by the definitions in \cref{eq:rescaledFeFmintermsofEH} to find the first row, and then use the definitions in \cref{eq:C2basisvectors} to evaluate the other rows.}
\label{tab:definebasisvectorsC2space}
\end{table}

In \cref{tab:definebasisvectorsC2space} we write down what each $\mathbb{C}^2$ basis vector, when multiplied by a $\mathbb{C}^3$ vector $\vc{F}$, implies in terms of electric and magnetic fields. Using \cref{tab:definebasisvectorsC2space} we can expand the simultaneous linear combinations in \cref{eq:phiexpressedinthreebases} into its electric and magnetic counterparts:

\begin{equation}\label{eq:EandHexpressedinthreebases}
\begin{alignedat}{3}
     \vc{E} &= \frac{2}{\sqrt{\varepsilon}} (\vc{F}_\text{e}+ \vc{0}) &= \frac{\sqrt{2}}{\sqrt{\varepsilon}}(\vc{F}_\text{p}+\vc{F}_\text{a}) &= \frac{\sqrt{2}}{\sqrt{\varepsilon}}(\vc{F}_\text{R}+\vc{F}_\text{L}), \\
    \vc{H} &= \frac{2}{\sqrt{\mu}}(\vc{0} + \vc{F}_\text{m}) &= \frac{\sqrt{2}}{\sqrt{\mu}}(\vc{F}_\text{p}-\vc{F}_\text{a}) &= \frac{\sqrt{2}}{\sqrt{\mu}}(\ii \vc{F}_\text{R}-\ii \vc{F}_\text{L}). \\
\end{alignedat}
\end{equation}%
which we may rewrite by definining vector fields $\vc{E}_i$ and $\vc{H}_i$ with $i=\{\text{e},\text{m},\text{p},\text{a},\textsc{r},\textsc{l}\}$ in a way that highlights the decomposition of electromagnetic fields into the sum of the projections on each basis:
\begin{equation}\label{eq:EandHexpressedasprojections}
\begin{alignedat}{3}
     \vc{E} &= \vc{E}_\text{e} + \vc{E}_\text{m} &&= \vc{E}_\text{p} + \vc{E}_\text{a} &&= \vc{E}_\text{R} + \vc{E}_\text{L}, \\
     \vc{H} &= \vc{H}_\text{e} + \vc{H}_\text{m} &&= \vc{H}_\text{p} + \vc{H}_\text{a} &&= \vc{H}_\text{R} + \vc{H}_\text{L}. \\
\end{alignedat}
\end{equation}

Now, \cref{eq:EandHexpressedinthreebases} can be inverted to write the coefficients $\vc{F}_i$ in terms of $\vc{E}$ and $\vc{H}$:
\begin{equation}\label{eq:Fi_definitions}
\begin{alignedat}{3}
    \vc{F}_\text{e} &= \frac{1}{2} \sqrt{\varepsilon} \vc{E}, \quad \quad \vc{F}_\text{p} &= \frac{1}{2\sqrt{2}} (\sqrt{\varepsilon} \vc{E} + \sqrt{\mu} \vc{H}), \quad \quad \vc{F}_\text{R} &= \frac{1}{2\sqrt{2}} (\sqrt{\varepsilon} \vc{E} + \ii\sqrt{\mu} \vc{H}) \\
    \vc{F}_\text{m} &= \frac{1}{2} \sqrt{\mu} \vc{H}, \quad \quad \vc{F}_\text{a} &= \frac{1}{2\sqrt{2}} (\sqrt{\varepsilon} \vc{E} - \sqrt{\mu} \vc{H}), \quad \quad \vc{F}_\text{L} &= \frac{1}{2\sqrt{2}} (\sqrt{\varepsilon} \vc{E} - \ii\sqrt{\mu} \vc{H}).
\end{alignedat}   
\end{equation}%

Combining \cref{eq:phiexpressedinthreebasesarrayform,eq:Fi_definitions} we arrive at Eq. 1 in the main text, reproduced here:
\begin{equation}\label{eq:maintextbases}
\begin{alignedat}{4}
    \vc{\psi}_\text{[EM]}(\vc{r})&=\pmqty{\vc{F}_\text{e}\\\vc{F}_\text{m}}&&=\frac{1}{2}\pmqty{\sqrt{\varepsilon}\vc{E}\\\sqrt{\mu}\vc{H}}\,,\\
    \vc{\psi}_\text{[PA]}(\vc{r})&=\pmqty{\vc{F}_\text{p}\\\vc{F}_\text{a}}&&=\frac{1}{2\sqrt{2}}\pmqty{\sqrt{\varepsilon}\vc{E}+\sqrt{\mu}\vc{H}\\\sqrt{\varepsilon}\vc{E}-\sqrt{\mu}\vc{H}}\,,\\
    \vc{\psi}_\text{[RL]}(\vc{r})&=\pmqty{\vc{F}_\text{R}\\\vc{F}_\text{L}}&&=\frac{1}{2\sqrt{2}}\pmqty{\sqrt{\varepsilon}\vc{E}+\ii\sqrt{\mu}\vc{H}\\\sqrt{\varepsilon}\vc{E}-\ii\sqrt{\mu}\vc{H}}\,.
\end{alignedat}    
\end{equation}%
And combining \cref{eq:EandHexpressedinthreebases,eq:EandHexpressedasprojections,eq:Fi_definitions} we arrive at Eq. 3 from the main text, reproduced here:
\begin{equation}\label{eq:maintextprojectionsE}
\begin{alignedat}{4}
    \vc{E} &= \vc{E}_\text{e} &&+ \vc{E}_\text{m} &&= \vc{E}+ \vc{0} &&
    \\
     &= \vc{E}_\text{p} &&+ \vc{E}_\text{a} &&= \frac{\vc{E} + \eta \vc{H}}{2} &&+ \frac{\vc{E} - \eta \vc{H}}{2} \\ &= \vc{E}_\text{R} &&+ \vc{E}_\text{L} &&= \frac{\vc{E} + \ii \eta \vc{H}}{2} &&+ \frac{\vc{E} - \ii \eta \vc{H}}{2}
\end{alignedat}
\end{equation}
with the corresponding $\vc{H}_\text{e} = \vc{0}, \vc{H}_\text{m} = \vc{H}, \vc{H}_\text{p,a} = \pm \vc{E}_\text{p,a}/\eta$ and $\vc{H}_\text{R,L} = \mp \ii \vc{E}_\text{R,L}/\eta$, and $\eta=\sqrt{\mu/\varepsilon}$. \Cref{eq:maintextprojectionsE} provides a recipe for directly splitting any electromagnetic field $(\vc{E},\vc{H})$ into a mathematical sum of two projections into the two basis vectors of any of the basis EM, PA or RL.

\subsection{Microscopic Maxwell's equations in different bases}
The use of different bases to describe electromagnetic fields is theoretically on the same footing as the traditional $\vc{E}$ and $\vc{H}$ approach. In fact, one can rewrite Maxwell's equations using them, and the well-known duality symmetry of Maxwell's equations makes them particularly pleasant. Following standard notation with $\rho$ being charge density and $\vc{J}$ being current density, the dual symmetric (including magnetic charges and currents) microscopic Maxwell's equations for time-harmonic fields are
\begin{equation}
    \label{eq:maxwelldualnormalised}
    \begin{split}
        \div\sqrt{\varepsilon}\vc{E}&=\rho_\text{e}/\sqrt{\varepsilon}\,,\\
        \div\sqrt{\mu}\vc{H}&=\rho_\text{m}/\sqrt{\mu}\,,\\
        \curl\sqrt{\varepsilon}\vc{E}&=+\ii k\sqrt{\mu}\vc{H}-\sqrt{\varepsilon}\vc{J}_\text{m}\,,\\
        \curl\sqrt{\mu}\vc{H}&=-\ii k\sqrt{\varepsilon}\vc{E}+\sqrt{\mu}\vc{J}_\text{e}\,.
    \end{split}
\end{equation}
These can be rewritten in terms of the basis coefficients $\vc{F}_{i}$ where $i\in\{\text{e,m,p,a,\textsc{r},\textsc{l}}\}$. In order to do that we need to find what charge and current densities look in those different bases. We may define the following scaled versions of charge and current densities:
\begin{equation}
    \begin{alignedat}{2}
        f_{\text{e}} &= \frac{\rho_\text{e}}{2\sqrt{\varepsilon}}, \quad \quad \vc{g}_{\text{e}} &&= \frac{\vc{J}_\text{e}}{2 c \sqrt{\varepsilon}},\\
        f_{\text{m}} &= \frac{\rho_\text{m}}{2\sqrt{\mu}}, \quad \quad \vc{g}_{\text{m}} &&= \frac{\vc{J}_\text{m}}{2 c \sqrt{\mu}},
    \end{alignedat}
\end{equation}
such that, $\vc{F}_\text{e} = \sqrt{\varepsilon} \vc{E}/2$ and $\vc{F}_\text{m} = \sqrt{\mu} \vc{H}/2$ constitute a simple re-scaling of Maxwell's equations,
\begin{equation}
    \begin{split}
    \label{eq:definedchargesbasis}
        \div\vc{F}_\text{e/m}&=f_\text{e/m}\,,\\
        \curl\vc{F}_\text{e/m}&=\pm [\ii k\vc{F}_\text{m/e}-\vc{g}_\text{m/e}]\,.
    \end{split}
\end{equation}

Next, to get the equations in the right-left basis we need to combine them to get fields $\vc{F}_\text{R/L}=(\vc{F}_\text{e}\pm\ii\vc{F}_\text{m})/\sqrt{2}$, which will lead to charge and current density combinations: $f_\text{R/L}=(f_\text{e}\pm\ii f_\text{m})/\sqrt{2}$ and $\vc{g}_\text{R/L}=(\vc{g}_\text{e}\pm\ii\vc{g}_\text{m})/\sqrt{2}$, leading to
\begin{equation}
    \begin{split}
        \div\vc{F}_\text{R/L}&=f_\text{R/L}\,,\\
        \curl\vc{F}_\text{R/L}&=\pm [k\vc{F}_\text{R/L}+\ii\vc{g}_\text{R/L}]\,.
    \end{split}
\end{equation}
Note that in this basis the right and left fields don't mix, unlike in the usual electromagnetic basis.
To obtain Maxwell's equations in the parallel-antiparallel basis, we need the fields to be $\vc{F}_{p/a}=(\vc{F}_\text{e}\pm\vc{F}_\text{m})/\sqrt{2}$, charge densities to be $f_\text{p/a}=(f_\text{e}\pm f_\text{m})/\sqrt{2}$, and current densities $\vc{g}_\text{p/a}=(\vc{g}_\text{e}\pm\vc{g}_\text{m})/\sqrt{2}$. These combinations yield 
\begin{equation}
    \begin{split}
        \div\vc{F}_\text{p/a}&={f_\text{p/a}}\,,\\
        \curl\vc{F}_\text{p/a}&=\mp [\ii k\vc{F}_\text{a/p}-{\vc{g}_\text{a/p}}]\,.
    \end{split}
\end{equation}
Of course, these are just some arbitrary choices of bases in $\mathbb{C}^2$ and we could make a different choice (take a different linear combination of Maxwell's equations), however, one can define a charge density in the $\mathbb{C}^2\times\mathbb{L}^2(\mathbb{R}^3)$ vector space as
\begin{equation}
    %\boxed{
        \quad \vc{f} = f_\text{e} \uv{e} + f_\text{m} \uv{m} = f_\text{p} \uv{p} + f_\text{a} \uv{a} = f_\text{R} \uv{r} + f_\text{L} \uv{l} \quad
    %}
\end{equation}
and a current bispinor in the $\mathbb{C}^3\times\mathbb{C}^2\times\mathbb{L}^2(\mathbb{R}^3)$ vector space as
\begin{equation}
    %\boxed{
        \quad \vc{g} = \vc{g}_\text{e} \uv{e} + \vc{g}_\text{m} \uv{m} = \vc{g}_\text{p} \uv{p} + \vc{g}_\text{a} \uv{a} = \vc{g}_\text{R} \uv{r} + \vc{g}_\text{L} \uv{l} \quad
    %}
\end{equation}
to write the time-harmonic microscopic Maxwell's equations in a basis-independent manner as 
\begin{equation}
\label{eq:microscopicmaxwell_basisindependent}
    \begin{split}
        \div\vc{\psi}&=\vc{f}\,,\\
        \curl\vc{\psi}&=\hat{D}(\ii k\vc{\psi}-\vc{g})\,,
    \end{split}
\end{equation}
where $\hat{D}$ is the electromagnetic duality transformation
$\hat{D}\uv{e}=\uv{m}$ and $\hat{D}\uv{m}=-\uv{e}$.

\section{Quadratic quantities and symmetry sphere}
\subsection{Linear transformations in \texorpdfstring{$\mathbb{C}^2$}{C²} sub-space written in different bases}

Any linear transformation $\tens{A}$ in the $\mathbb{C}^2$ subspace can be represented, in any given basis, as a $2\times2$ matrix:
\begin{equation}
\tens{A}=\begin{pmatrix}
a_{11} & a_{12} \\
a_{21} & a_{22}
\end{pmatrix}.
\end{equation}
Such a transformation has four complex degrees of freedom embodied by the four scalars $a_{ij}$. It is well-known that any such transformation can be written as a linear combination of four matrices that form a complete basis in this space: the identity matrix $\tens{I}$, and the three Pauli matrices $\tens{\sigma}_{1,2,3}$, with complex coefficients $a_{0,1,2,3}$ being the four equivalent degrees of freedom:
\begin{equation}
\label{eq:general2x2decomposition}
\begin{alignedat}{3}
\begin{pmatrix}
a_{11} & a_{12} \\
a_{21} & a_{22}
\end{pmatrix} = a_0\underbrace{\mqty(\pmat{0})}_{\tens{I}}+a_1\underbrace{\mqty(\pmat{1})}_{\tens{\sigma}_1}+a_2\underbrace{\mqty(\pmat{2})}_{\tens{\sigma}_2}+a_3\underbrace{\mqty(\pmat{3})}_{\tens{\sigma}_3} = \begin{pmatrix}
a_0 + a_1 & a_3 - \ii a_2 \\
a_3 + \ii a_2 & a_0 - a_1
\end{pmatrix}\,,
\end{alignedat}
\end{equation}%
from where we can obtain the determinant $\det(\tens{A})=a_{11} a_{22} - a_{12} a_{21} = a_0^2 - a_1^2 - a_2^2 - a_3^2$. \Cref{eq:general2x2decomposition} enables writing any transformation in a basis-independent way, via the four coefficients, as:
\begin{equation}
\label{eq:2x2decomposition_basisindependent}
\tens{A}=a_0\tens{W}_0 + a_1 \tens{W}_1 + a_2 \tens{W}_2 + a_3 \tens{W}_3.
\end{equation}

This is extremely useful. Because the Pauli matrices swap into one another when changing between the $\mathbb{C}^2$ bases (using change of basis matrices and similarity transformations, which we won't review here), this notation allows us to easily change the basis for any arbitrary linear transformation,  preserving the same set of four coefficients $a_{0,1,2,3}$, only requiring using the appropriate Pauli matrix, and with the correct sign, in place of each $\tens{W}_i$:
\begin{table}[h]
\centering
\begin{tabular}{|c|c|c|c|c|}
\hline
\ & $\tens{W}_0$ & $\tens{W}_1$ & $\tens{W}_2$ & $\tens{W}_3$ \\ \hline
RL basis & $\tens{I}$ & $\tens{\sigma}_1$ & $\tens{\sigma}_2$ & $\tens{\sigma}_3$ \\ \hline
EM basis & $\tens{I}$ & $\tens{\sigma}_3$ & $-\tens{\sigma}_1$ & $-\tens{\sigma}_2$ \\ \hline
PA basis & $\tens{I}$ & $\tens{\sigma}_1$ & $-\tens{\sigma}_3$ & $\tens{\sigma}_2$ \\ \hline
transformation & 1 & $-\hat{P}$ & $\hat{P}\hat{D}$ & $\ii\hat{D}$ \\ \hline
\end{tabular}
\caption{Changing the basis of linear transformations in $\mathbb{C}^2$.}
\label{tab:transposed_matrix_basis}
\end{table}

For completeness and clarity, we write here how any arbitrary linear transformation would be written in each of the three bases:
\begin{equation}
\label{eq:2by2_in_different_bases}
\begin{alignedat}{3}
    \tens{A}_\text{[RL]}&=\pmqty{a_\text{R}&a_\text{RL}\\a_\text{LR}&a_\text{L}}&&=a_0\mqty(\pmat{0})+a_1\underbrace{\mqty(\pmat{1})}_{\tens{\sigma}_1}\mathop{+}a_2\underbrace{\mqty(\pmat{2})}_{\tens{\sigma}_2}\mathop{+}a_3\underbrace{\mqty(\pmat{3})}_{\tens{\sigma}_3}\,,\\
    \tens{A}_\text{[EM]}&=\pmqty{a_\text{e}&a_\text{em}\\a_\text{me}&a_\text{m}} &&=a_0\mqty(\pmat{0})+a_1\underbrace{\mqty(\pmat{3})}_{\tens{\sigma}_3}\mathop{-}a_2\underbrace{\mqty(\pmat{1})}_{\tens{\sigma}_1}\mathop{-}a_3\underbrace{\mqty(\pmat{2})}_{\tens{\sigma}_2}\,,\\
    \tens{A}_\text{[PA]}&=\pmqty{a_\text{p}&a_\text{pa}\\a_\text{ap}&a_\text{a}}&&=a_0\mqty(\pmat{0})+a_1\underbrace{\mqty(\pmat{1})}_{\tens{\sigma}_1}\mathop{-}a_2\underbrace{\mqty(\pmat{3})}_{\tens{\sigma}_3}\mathop{+}a_3\underbrace{\mqty(\pmat{2})}_{\tens{\sigma}_2}\,,
\end{alignedat}
\end{equation}%

\noindent which provides a nice interpretation for the meaning of the coefficients: 
\begin{equation}
\label{eq:definition_of_ai}
\begin{alignedat}{2}
a_0 = \tfrac{1}{2}(a_\text{R} + a_\text{L}) &= \tfrac{1}{2}(a_\text{e} + a_\text{m}) = \tfrac{1}{2}(a_\text{p} + a_\text{a})\,, \\
a_1 &= \tfrac{1}{2}(a_\text{e} - a_\text{m})\,, \\
a_2 &= -\tfrac{1}{2}(a_\text{p} - a_\text{a})\,, \\
a_3 &= \tfrac{1}{2}(a_\text{R} - a_\text{L})\,.
\end{alignedat}
\end{equation}

The three matrix representations $\tens{A}_{[\text{RL}]}$, $\tens{A}_{[\text{EM}]}$ and $\tens{A}_{[\text{PA}]}$ are \emph{similar matrices} in the mathematical sense, as they represent the same linear transformation in different bases, and therefore they all have the same trace and determinant:

\begin{equation}
\begin{alignedat}{1}
\label{eq:traceanddeterminant}
\tr\tens{A} &= 2 a_0, \\
\det\tens{A} &= a_0^2 - a_1^2 - a_2^2 - a_3^2.
\end{alignedat}
\end{equation}

We made a choice of matching the subscript $i$ of the basis-independent tensor $\tens{W}_i$ and its coefficient $a_i$ to the number of the corresponding Pauli matrix $\tens{\sigma}_i$ for the RL basis. This arbirary choice was taken after some consideration, such that the numbering $0, 1,2,3$ matches the Stokes parameters when calculating the different energies in the main text. This arbitrary choice, which we carefully considered, has some consequences in other definitions. For example, $W_2=-(W_\text{p}-W_\text{a})$ with a negative sign, because \cref{eq:2by2_in_different_bases} has a negative sign in the coefficient of $\spmqty{ 1 & 0 \\ 0 & -1}$ for the PA basis. Similarly, in a later definition of material susceptibility, \cref{eq:constitutiverelation_basisindependent}, we have $\chi_2 = -\chi_\text{t} = -(\chi_\text{p}-\chi_\text{a})/2$, with again a negative sign, due to this choice.

\subsection{Stokes parameters and the Poincaré sphere}
As mentioned before the $\mathbb{C}^2$ subspace of electromagnetic bispinors is analogous to the Jones vectors.
Any polarised paraxial light can be represented by a $\mathbb{C}^2$ vector $\vc{J}$ such that the electric field phasor can be written as
\begin{equation}
    \sqrt{\varepsilon}\vc{E}=(\vc{J}^\intercal\uv{e})\ee^{\ii kz}=\pmqty{A_\text{h} & A_\text{v}}  \pmqty{\uv{e}_\text{h}\\ \uv{e}_\text{v}}\ee^{\ii kz}=\pmqty{A_\text{d} & A_\text{a}}  \pmqty{\uv{e}_\text{d}\\ \uv{e}_\text{a}}\ee^{\ii kz}=\pmqty{A_\text{R} & A_\text{L}}  \pmqty{\uv{e}_\text{R}\\ \uv{e}_\text{L}}\ee^{\ii kz}\,,
\end{equation}
where ($A_\text{h}$, $A_\text{v}$) are linear horizontal-vertical, ($A_\text{d}$, $A_\text{a}$) diagonal-antidiagonal and ($A_\text{R}$, $A_\text{L}$) right-left circular polarisation complex amplitudes. The outer product of the Jones vector with itself produces a tensor which, following \cref{eq:general2x2decomposition}, may be written as a linear combination of Pauli matrices, whose coefficients are one half of the Stokes parameters as they are conventionally defined
\begin{equation}
\begin{alignedat}{2}
\vc{J} \vc{J}^\dagger _{[\text{ps}]}&= \pmqty{A_\text{h} \\ A_\text{v}} \pmqty{A^*_\text{h} & A^*_\text{v}} = \begin{pmatrix}
\abs{A_\text{h}}^2  & A_\text{h}  A_\text{v}^* \\
A_\text{v}  A_\text{h}^* & \abs{A_\text{v}}^2
\end{pmatrix} = \frac{1}{2}\begin{pmatrix}
\mathcal{S}_0+\mathcal{S}_1  & \mathcal{S}_2-\ii\mathcal{S}_3 \\
\mathcal{S}_2+\ii\mathcal{S}_3 & \mathcal{S}_0-\mathcal{S}_1
\end{pmatrix}\,, \quad \quad && \text{(linear basis)}
\end{alignedat}
\end{equation}
and similarly in the diagonal-antidiagonal and right-left circular bases. Following \cref{eq:definition_of_ai} one can express the Stokes parameters as they are most well-known:
\begin{equation}
\begin{alignedat}{3}
\mathcal{S}_0 = \abs{A_\text{h}}^2 + \abs{A_\text{v}}^2 &= \abs{A_\text{d}}^2 + \abs{A_\text{a}}^2 = \abs{A_\text{R}}^2 + \abs{A_\text{L}}^2, \\
\mathcal{S}_1 &= \abs{A_\text{h}}^2 - \abs{A_\text{v}}^2, \\
\mathcal{S}_2 &= \abs{A_\text{d}}^2 - \abs{A_\text{a}}^2, \\
\mathcal{S}_3 &= \abs{A_\text{R}}^2 - \abs{A_\text{L}}^2.
\end{alignedat}
\end{equation}
One can easily see that $\det\vc{J} \vc{J}^\dagger=\abs{A_\text{h}}^2 \abs{A_\text{v}}^2 - \abs{A_\text{h} A_\text{v}^*}^2=0$ which, following \cref{eq:traceanddeterminant}, leads to $\mathcal{S}_0^2=\mathcal{S}_1^2+\mathcal{S}_2^2+\mathcal{S}_3^2$ (with the equality becoming an inequality for partially-polarised or unpolarised fields). This justifies why every polarisation state, characterised by coordinates $(\mathcal{S}_1,\mathcal{S}_2,\mathcal{S}_3)$, is contained in a Poincaré sphere with radius $\mathcal{S}_0$. In the next section we will show that the $\mathbb{C}^2$ subspace of electromagnetic bispinors gives us exactly the same structure.

\subsection{Energy and the symmetry sphere}
According to \cref{eq:EMenergydensitywithFeFmFpFaFRFL} the electromagnetic energy density is given by the inner product of the electromagnetic bispinor with itself:
\begin{equation}
\begin{alignedat}{2}
W &= \vc{\psi}^\dagger \vdot\vc{\psi} = \norm{\vc{\psi}}^2 \quad \quad && \text{(basis-independent notation)}\\
W_{[\text{EM}]} &= \pmqty{\vc{F}^*_\text{e} & \vc{F}^*_\text{m}} \vdot \pmqty{\vc{F}_\text{e}\\ \vc{F}_\text{m}} = \vc{F}^*_\text{e}\vdot\vc{F}_\text{e} + \vc{F}^*_\text{m}\vdot\vc{F}_\text{m} = W_\text{e} + W_\text{m} \quad \quad && \text{(EM basis)}
\end{alignedat}
\end{equation}
The outer product (in $\mathbb{C}^2$ space) of the electromagnetic bi-spinor with itself produces a tensor:
\begin{equation}
\begin{alignedat}{2}
\tens{W} &= \vc{\psi} \vdot\vc{\psi}^\dagger \quad \quad && \text{(basis-independent notation)}\\
\tens{W}_{[\text{EM}]} &= \pmqty{\vc{F}_\text{e} \\ \vc{F}_\text{m}} \vdot \pmqty{\vc{F}^*_\text{e} & \vc{F}^*_\text{m}} = \begin{pmatrix}
\vc{F}_\text{e} \vdot \vc{F}_\text{e}^*  & \vc{F}_\text{e} \vdot \vc{F}_\text{m}^* \\
\vc{F}_\text{m} \vdot \vc{F}_\text{e}^* & \vc{F}_\text{m} \vdot \vc{F}_\text{m}^*
\end{pmatrix} = \begin{pmatrix}
W_{\text{e}}  & W_{\text{em}} \\
W_{\text{me}} & W_{\text{m}}
\end{pmatrix} \quad \quad && \text{(EM basis)}
\end{alignedat}
\end{equation}

\noindent whose $\mathbb{C}^2$ determinant is always positive from the Cauchy-Schwarz inequality $|\vc{F}_\text{e}|^2\vdot|\vc{F}_\text{m}|^2 \geq |\vc{F}_\text{e}^*\vdot \vc{F}_\text{m}|^2$. This tensor can be expanded into Pauli matrices following \cref{eq:2by2_in_different_bases}, such that it is expressed in the different bases as:
\begin{equation}
\begin{alignedat}{3}
\tens{W}_\text{[RL]}&=\pmqty{W_\text{R}&W_\text{RL}\\W_\text{LR}&W_\text{L}}&&=\frac{W_0}{2}\mqty(\pmat{0})+\frac{W_1}{2}\underbrace{\mqty(\pmat{1})}_{\tens{\sigma}_1}\mathop{+}\frac{W_2}{2}\underbrace{\mqty(\pmat{2})}_{\tens{\sigma}_2}\mathop{+}\frac{W_3}{2}\underbrace{\mqty(\pmat{3})}_{\tens{\sigma}_3}\,,\\
    \tens{W}_\text{[EM]}&=\pmqty{W_\text{e}&W_\text{em}\\W_\text{me}&W_\text{m}} &&=\frac{W_0}{2}\mqty(\pmat{0})+\frac{W_1}{2}\underbrace{\mqty(\pmat{3})}_{\tens{\sigma}_3}\mathop{-}\frac{W_2}{2}\underbrace{\mqty(\pmat{1})}_{\tens{\sigma}_1}\mathop{-}\frac{W_3}{2}\underbrace{\mqty(\pmat{2})}_{\tens{\sigma}_2}\,,\\
    \tens{W}_\text{[PA]}&=\pmqty{W_\text{p}&W_\text{pa}\\W_\text{ap}&W_\text{a}}&&=\frac{W_0}{2}\mqty(\pmat{0})+\frac{W_1}{2}\underbrace{\mqty(\pmat{1})}_{\tens{\sigma}_1}\mathop{-}\frac{W_2}{2}\underbrace{\mqty(\pmat{3})}_{\tens{\sigma}_3}\mathop{+}\frac{W_3}{2}\underbrace{\mqty(\pmat{2})}_{\tens{\sigma}_2}\,,
\end{alignedat}
\end{equation}%
or, written in a basis-independent way:
\begin{equation}
\tens{W}=\frac{1}{2}(W_0\tens{W}_0 + W_1 \tens{W}_1 + W_2 \tens{W}_2 + W_3 \tens{W}_3).
\end{equation}
with the simple definitions:
\begin{equation}
\begin{alignedat}{3}
W_0 =  W_\text{e} + W_\text{m} &= W_\text{p} + W_\text{a} = W_\text{R} + W_\text{L}=W, \\
W_1 &= W_\text{e} - W_\text{m}, \\
W_2 &= W_\text{a} - W_\text{p}, \\
W_3 &= W_\text{R} - W_\text{L}.
\end{alignedat}
\end{equation}
As proven earlier, the determinant of this tensor is always positive and independent of the basis, hence $\det(\tens{W}) = (W_0^2 - W_1^2 - W_2^2 - W_3^2)/4 \geq 0$. This proves that $(W_1,W_2,W_3)$ is a vector in W-space always contained in a Bloch sphere of radius $W_0$, giving rise to the \emph{energy symmetry sphere}: a main result of our work.

Also note that one can obtain the energy-like expressions $W_A$ using the $\tens{W}_A$ operators, in a basis-independent notation (note $\tens{W}_A^\dagger=\tens{W}_A$ is Hermitian hence the following are real numbers) as:
\begin{equation}
\label{eq:relationWAwithtensorWA}
W_A =\vc{\psi}^\dagger\tens{W}_A\vc{\psi} \in \mathbb{R}\,.
\end{equation}
% --------------------------------------------------------------------------------------------
\subsection{Other quadratic quantites}
There is one big difference between Jones vectors and bispinors: the $\mathbb{C}^2$ components of bispinors are themselves $\mathbb{C}^3$-valued vectors. The way we dealt with this in the previous section was by taking a dot product
\begin{equation}
\begin{alignedat}{2}
\tens{W}_{[\text{EM}]}&=(\vc{\psi} \vdot\vc{\psi}^\dagger)_{[\text{EM}]} =  \pmqty{
\vc{F}_\text{e} \vdot \vc{F}_\text{e}^*  & \vc{F}_\text{e} \vdot \vc{F}_\text{m}^* \\
\vc{F}_\text{m} \vdot \vc{F}_\text{e}^* & \vc{F}_\text{m} \vdot \vc{F}_\text{m}^*
} = \frac{1}{2}\pmqty{
W_{\text{0}}+W_{\text{1}}  & -W_{\text{2}}+\ii W_{\text{3}} \\
-W_{\text{2}}-\ii W_{\text{3}} & W_{\text{0}}-W_{\text{1}}
}\,.
\end{alignedat}
\end{equation}
However one might argue that this is just one of many possible products for $\mathbb{C}^3$ vectors. Indeed it is the case that, for example, taking the vector cross-product leads to spin-like quantities
\begin{equation}
\begin{alignedat}{2}
(\vc{\psi} \cp\vc{\psi}^\dagger)_{[\text{EM}]} =  \pmqty{
\vc{F}_\text{e} \cp \vc{F}_\text{e}^*  & \vc{F}_\text{e} \cp \vc{F}_\text{m}^* \\
\vc{F}_\text{m} \cp \vc{F}_\text{e}^* & \vc{F}_\text{m} \cp \vc{F}_\text{m}^*
} = \frac{\omega}{2\ii}\pmqty{
\vc{S}_{\text{0}}+\vc{S}_{\text{1}}  & -\vc{S}_{\text{2}}+\ii \vc{S}_{\text{3}} \\
-\vc{S}_{\text{2}}-\ii \vc{S}_{\text{3}} & \vc{S}_{\text{0}}-\vc{S}_{\text{1}}
}\,.
\end{alignedat}
\end{equation}
Sadly these spin-like quantities are $\mathbb{R}^3$ vectors, therefore, there is no canonical way to define the determinant and hence prove the existence of a spin-sphere in this way. The next product that one can take is the symmetric tensor product which leads to stress-tensor-like quantities
\begin{equation}
\begin{alignedat}{2}
(\vc{\psi} \odot\vc{\psi}^\dagger)_{[\text{EM}]} =  \pmqty{
\vc{F}_\text{e} \odot \vc{F}_\text{e}^*  & \vc{F}_\text{e} \odot \vc{F}_\text{m}^* \\
\vc{F}_\text{m} \odot \vc{F}_\text{e}^* & \vc{F}_\text{m} \odot \vc{F}_\text{m}^*
} = \frac{1}{2}\pmqty{
\tens{T}_{\text{0}}+\tens{T}_{\text{1}}  & -\tens{T}_{\text{2}}+\ii \tens{T}_{\text{3}} \\
-\tens{T}_{\text{2}}-\ii \tens{T}_{\text{3}} & \tens{T}_{\text{0}}-\tens{T}_{\text{1}}
}+\tens{W}\tens{I}_{3}\,,
\end{alignedat}
\end{equation}
where $\tens{I}_{3}=\operatorname{diag}(1,1,1)$ is the identity in $\mathbb{C}^3$ space.
One might wonder why we used the symmetric tensor product rather than the regular one. Usually, the Maxwell stress tensor is defined as the real part of a complex quantity defined with the tensor product. However, the tensor product of any $\mathbb{C}^3$ vector with its conjugate can be split into the symmetric and antisymmetric parts as follows
\begin{equation}
    \vc{F}_i\!\otimes\vc{F}_i^*=\underbrace{\frac{1}{2}\vc{F}_i\odot\vc{F}_i^*}_{\Re(\vc{F}_i\otimes\vc{F}_i^*)}+\underbrace{\frac{1}{2}\vc{F}_i\wedge\vc{F}_i^*}_{\ii\Im(\vc{F}_i\otimes\vc{F}_i^*)}
\end{equation}
where we use symmetric and antisymmetric outer products which are defined as $\vc{F}\odot\vc{F}^*=\vc{F}\otimes\vc{F}^*+\vc{F}^*\!\otimes\vc{F}$, and $\vc{F}\wedge\vc{F}^*=\vc{F}\otimes\vc{F}^*-\vc{F}^*\!\otimes\vc{F}\!$ respectively. The key observation is that a bivector $\vc{F}_i\wedge\vc{F}_i^*$ (directed plane segment with a sense of rotation) is isomorphic to a vector (directed line segments with a sense of translation) $\vc{F}_i\cp\vc{F}_i^*$, this isomorphism is called Hodge duality and it is denoted as $\star(\vc{F}_i\cp\vc{F}_i^*)= \vc{F}_i\wedge\vc{F}_i^*$ (see, e.g. \cite[p.~38]{Lounesto2001}), or in other words any antisymmetric matrix in $\mathbb{C}^3$ is isomorphic to a pseudovector
\begin{equation}
    \star\!\pmqty{0&c&-b\\-c&0&a\\b&-a&0}=\pmqty{a\\b\\c}\qq{or} \star\!\pmqty{a\\b\\c}=\pmqty{0&c&-b\\-c&0&a\\b&-a&0}
\end{equation}
Another useful fact is that $\tr(\vc{F}_i\odot\vc{F}_i^*)=\abs{\vc{F}_i}^2$ so, together, we can write the tensor product as 
\begin{equation*}
    \vc{F}_i\otimes\vc{F}_i^*=\frac{1}{2}[\tens{I}_{3}\abs{\vc{F}_i}^2+(\vc{F}_i\odot\vc{F}_i^*-\tens{I}_{3}\abs{\vc{F}_i}^2)+\star(\vc{F}_i\cp\vc{F}_i^*)]
\end{equation*}
but this is exactly the stress tensor, spin and energy densities of field $\vc{F}_i$ arranged in a single complex tensor 
\begin{equation}\label{eq:outer_decomposition}
    \vc{F}_i\otimes\vc{F}^*_i=\frac{1}{2}(W_i\tens{I}_{3}+\tens{T}_{i}-{\ii\omega}\,{\star\vc{S}_i})\,,
\end{equation}
where $i\in\{\text{e,m,p,a,\textsc{r},\textsc{l}}\}$. Notice also, that the symmetric part is purely real 
\begin{equation}
    \Re(\vc{F}_i\otimes\vc{F}^*_i)=\frac{1}{2}\vc{F}_i\odot\vc{F}_i^*=\frac{1}{2}(W_i\tens{I}_{3}+\tens{T}_{i})\,,
\end{equation}
while the antisymmetric part is purely imaginary
\begin{equation}
    \Im(\vc{F}_i\otimes\vc{F}^*_i)=\frac{1}{2\ii}\vc{F}_i\wedge\vc{F}_i^*=-\frac{\omega}{2}\,{\star\vc{S}_i}\,,
\end{equation}
which justifies us replacing the real part of the regular tensor product with a symmetric product. This is a general behaviour of symmetric and antisymmetric products of a complex vector with its conjugate. We note that the reactive or imaginary stress tensor introduced in \cite{NietoVesperinas2022} can be written for the time-harmonic field simply as 
\begin{equation}\label{eq:TI}
    \tens{T}_I=\Im(\vc{F}_\text{e}\otimes\vc{F}^*_\text{e}+\vc{F}^*_\text{m}\otimes\vc{F}_\text{m})=-\frac{\omega}{2}\star\!(\underbrace{\vc{S}_\text{e}-\vc{S}_\text{m}}_{\vc{S}_1})=-\frac{\omega}{2}\star\!\!\pmqty{S_{1x}\\S_{1y}\\S_{1z}}=\frac{\omega}{2}\pmqty{0&-S_{1z}&S_{1y}\\S_{1z}&0&-S_{1x}\\-S_{1y}&S_{1x}&0}\,.
\end{equation}
As a consequence of \cref{eq:outer_decomposition} the tensor product of the bispinors contains $W_A$, $\tens{T}_A$ and $\vc{S}_A$ and no new quantities
\begin{equation}
\begin{alignedat}{2}
\!\!\!(\vc{\psi} \otimes\vc{\psi}^\dagger)_{[\text{EM}]} = \frac{1}{2}\pmqty{
W_{\text{0}}+W_{\text{1}}  & -W_{\text{2}}+\ii W_{\text{3}} \\
-W_{\text{2}}-\ii W_{\text{3}} & W_{\text{0}}-W_{\text{1}}
}\!\tens{I}_{3}+  \frac{1}{2}\pmqty{
\tens{T}_{\text{0}}+\tens{T}_{\text{1}}  & -\tens{T}_{\text{2}}+\ii \tens{T}_{\text{3}} \\
-\tens{T}_{\text{2}}-\ii \tens{T}_{\text{3}} & \tens{T}_{\text{0}}-\tens{T}_{\text{1}}
}+\frac{\omega\star}{2\ii}\!\pmqty{
\vc{S}_{\text{0}}+\vc{S}_{\text{1}}  & -\vc{S}_{\text{2}}+\ii \vc{S}_{\text{3}} \\
-\vc{S}_{\text{2}}-\ii \vc{S}_{\text{3}} & \vc{S}_{\text{0}}-\vc{S}_{\text{1}}
}.\!\!\!
\end{alignedat}
\end{equation}
The last observable that we introduced in the main text is the canonical linear momentum density, however, simply using $\vdot(\grad)$ leads to a tensor that has both hermitian and non-hermitian parts in the $\mathbb{C}^2$ space
\begin{equation}
\begin{alignedat}{2}
\!\!\![\vc{\psi} \vdot(\grad)\vc{\psi}^\dagger]_{[\text{EM}]} =  \pmqty{
\vc{F}_\text{e} \vdot(\grad) \vc{F}_\text{e}^*  & \vc{F}_\text{e} \vdot(\grad) \vc{F}_\text{m}^* \\
\vc{F}_\text{m} \vdot(\grad) \vc{F}_\text{e}^* & \vc{F}_\text{m} \vdot(\grad) \vc{F}_\text{m}^*
} =\frac{1}{4}\grad\!\pmqty{
W_{\text{0}}+W_{\text{1}}  & -W_{\text{2}}+\ii W_{\text{3}} \\
-W_{\text{2}}-\ii W_{\text{3}} & W_{\text{0}}-W_{\text{1}}
}+ \frac{\ii\omega}{2}\pmqty{
\vc{p}_{\text{0}}+\vc{p}_{\text{1}}  & \vc{p}_{\text{2}}-\ii \vc{p}_{\text{3}} \\
\vc{p}_{\text{2}}+\ii \vc{p}_{\text{3}} & \vc{p}_{\text{0}}-\vc{p}_{\text{1}}
},\!\!\!
\end{alignedat}
\end{equation}
the anti-hermitian part leads to the definition of the momentum-like quantities $\vc{p}_A$, while the hermitian part only contains gradients of already defined quantities $W_A$.\clearpage
% ENERGY STOKES TABLE
% ------------------------
\begin{table}[!h]
\vspace{-.5em}
\footnotesize
\centering
\renewcommand{\arraystretch}{1.3} % Default value: 1
\begin{tabular}{|c|c|c|c|c|c|}
\hline
&  \text{Quantity} &  \text{EM basis} & \text{PA basis} & \text{RL basis}&\text{Description} \\
\hline\hline
\cellcolor{proper}\( W_0 \) &
\cellcolor{proper}\( W \) &\cellcolor{proper}\( \vc{F}_\text{e}^*\!\vdot\!\vc{F}_\text{e}^{\vphantom{*}} + \vc{F}_\text{m}^*\!\vdot\!\vc{F}_\text{m}^{\vphantom{*}} \) & \cellcolor{proper}\( \vc{F}_\text{a}^*\!\vdot\!\vc{F}_\text{a}^{\vphantom{*}} + \vc{F}_\text{p}^*\!\vdot\!\vc{F}_\text{p}^{\vphantom{*}} \) & \cellcolor{proper}\( \vc{F}_\text{R}^*\!\vdot\!\vc{F}_\text{R}^{\vphantom{*}} + \vc{F}_\text{L}^*\!\vdot\!\vc{F}_\text{L}^{\vphantom{*}} \) &{ total energy density}\\
\hline
\cellcolor{altroo}\( W_1 \) & 
\cellcolor{altroo}\( W_\text{e}-W_\text{m} \) &\cellcolor{altroo}\( \vc{F}_\text{e}^*\!\vdot\!\vc{F}_\text{e} - \vc{F}_\text{m}^*\!\vdot\!\vc{F}_\text{m} \) & \( 2\Re(\vc{F}_\text{p}^*\vdot\vc{F}_\text{a}^{\vphantom{*}}) \) & \( 2\Re(\vc{F}_\text{R}^*\vdot\vc{F}_\text{L}^{\vphantom{*}}) \)&{EM energy difference\footnote{Reactive energy density, also called reactive power density $p_\text{react}=-2\omega W_1$ \cite{NietoVesperinas2022, Jackson1998} and time-averaged Lagrangian \cite{Bergman2009}.}} \\
\hline
\cellcolor{crlroo}\( W_2 \) &
\cellcolor{crlroo}\( W_\text{a}-W_\text{p} \) &\(- 2\Re( {\vc{F}_\text{e}^*\vdot\vc{F}_\text{m}} ) \) & \cellcolor{crlroo}\( \vc{F}_\text{a}^*\!\vdot\!\vc{F}_\text{a}^{\vphantom{*}} - \vc{F}_\text{p}^*\!\vdot\!\vc{F}_\text{p}^{\vphantom{*}} \) & \(2\Im(\vc{F}_\text{R}^* \vdot \vc{F}_\text{L}^{\vphantom{*}})\) &$-\omega\cdot${(reactive heli. den.)\footnote{It is sometimes called magnetoelectric energy density \cite{Bliokh2014}. Related to reactive helicity $\mathfrak{S}_\text{react}=-W_2/\omega$, quantity explored in \cite{NietoVesperinas2021,kamenetskii2015microwave}.  }}\\
\hline
\cellcolor{pseudo}\( W_3 \) &
\cellcolor{pseudo}\( W_\text{R}-W_\text{L} \) &\(- 2\Im( {\vc{F}_\text{e}^*\!\vdot\vc{F}_\text{m}} ) \) & \( 2\Im(\vc{F}_\text{p}^*\vdot\vc{F}_\text{a}^{\vphantom{*}}) \) & \cellcolor{pseudo}\( \vc{F}_\text{R}^*\!\vdot\!\vc{F}_\text{R}^{\vphantom{*}} - \vc{F}_\text{L}^*\!\vdot\!\vc{F}_\text{L}^{\vphantom{*}} \) &$\omega\cdot${(helicity density)}\\
\hline\hline
\cellcolor{proper}\( \omega\vc{p}_0 \) &
\cellcolor{proper}\( \omega\vc{p} \) &\cellcolor{proper}\( \Im[\vc{F}_\text{e}^*\!\vdot\!(\grad)\vc{F}_\text{e}^{\vphantom{*}} + \vc{F}_\text{m}^*\!\vdot\!(\grad)\vc{F}_\text{m}^{\vphantom{*}}] \) & \cellcolor{proper}\( \Im[\vc{F}_\text{a}^*\!\vdot\!(\grad)\vc{F}_\text{a}^{\vphantom{*}} + \vc{F}_\text{p}^*\!\vdot\!(\grad)\vc{F}_\text{p}^{\vphantom{*}}] \) & \cellcolor{proper}\( \Im[\vc{F}_\text{R}^*\!\vdot\!(\grad)\vc{F}_\text{R}^{\vphantom{*}} + \vc{F}_\text{L}^*\!\vdot\!(\grad)\vc{F}_\text{L}^{\vphantom{*}}] \) &canonical momentum den.\\
\hline
\cellcolor{altroo}\( \omega\vc{p}_1 \) &
\cellcolor{altroo}\( \omega(\vc{p}_\text{e}-\vc{p}_\text{m}) \) &\cellcolor{altroo}\( \Im[\vc{F}_\text{e}^*\!\vdot\!(\grad)\vc{F}_\text{e}^{\vphantom{*}} - \vc{F}_\text{m}^*\!\vdot\!(\grad)\vc{F}_\text{m}^{\vphantom{*}}]\) & \( \Im[\vc{F}_\text{a}^*\!\vdot\!(\grad)\vc{F}_\text{p}^{\vphantom{*}} - \vc{F}_\text{p}^{\vphantom{*}}\!\vdot\!(\grad)\vc{F}_\text{a}^*] \) & \( \Im[\vc{F}_\text{L}^*\!\vdot\!(\grad)\vc{F}_\text{R}^{\vphantom{*}} - \vc{F}_\text{R}^{\vphantom{*}}\!\vdot\!(\grad)\vc{F}_\text{L}^*] \) &EM mom. difference\\
\hline
\cellcolor{crlroo}\( \omega\vc{p}_2 \) &
\cellcolor{crlroo}\( \omega(\vc{p}_\text{a}-\vc{p}_\text{p}) \) &\(\Im[\vc{F}_\text{e}^{\vphantom{*}}\!\vdot\!(\grad)\vc{F}_\text{m}^*\! - \vc{F}_\text{m}^*\!\vdot\!(\grad)\vc{F}_\text{e}^{\vphantom{*}}] \) & \cellcolor{crlroo}\( \Im[\vc{F}_\text{a}^*\!\vdot\!(\grad)\vc{F}_\text{a}^{\vphantom{*}} - \vc{F}_\text{p}^*\!\vdot\!(\grad)\vc{F}_\text{p}^{\vphantom{*}}] \) & \(\Re[\vc{F}_\text{L}^*\!\vdot\!(\grad)\vc{F}_\text{R}^{\vphantom{*}} - \vc{F}_\text{R}^{\vphantom{*}}\!\vdot\!(\grad)\vc{F}_\text{L}^*]\) &\\
\hline
\cellcolor{pseudo}\( \omega\vc{p}_3 \) &
\cellcolor{pseudo}\( \omega(\vc{p}_\text{R}-\vc{p}_\text{L}) \) &\(\Re[\vc{F}_\text{e}^{\vphantom{*}}\!\vdot\!(\grad)\vc{F}_\text{m}^*\! - \vc{F}_\text{m}^*\!\vdot\!(\grad)\vc{F}_\text{e}^{\vphantom{*}}] \) & \( \Re[\vc{F}_\text{a}^*\!\vdot\!(\grad)\vc{F}_\text{p}^{\vphantom{*}} - \vc{F}_\text{p}^{\vphantom{*}}\!\vdot\!(\grad)\vc{F}_\text{a}^*] \) & \cellcolor{pseudo}\( \Im[\vc{F}_\text{R}^*\!\vdot\!(\grad)\vc{F}_\text{R}^{\vphantom{*}} - \vc{F}_\text{L}^*\!\vdot\!(\grad)\vc{F}_\text{L}^{\vphantom{*}}] \) & chiral momentum den.\footnote{Related to canonical spin $\vc{s}_\text{can}=\vc{p}_3/k$ \cite{Vernon2023} and discussed in e.g. \cite{Bliokh2014}.}\\
\hline\hline
\cellcolor{proper}\( \omega\vc{S}_0 \) &
\cellcolor{proper}\( \omega\vc{S} \) &\cellcolor{proper}\( \Im[\vc{F}_\text{e}^*\!\cp\!\vc{F}_\text{e}^{\vphantom{*}} + \vc{F}_\text{m}^*\!\cp\!\vc{F}_\text{m}^{\vphantom{*}}] \) & \cellcolor{proper}\( \Im[\vc{F}_\text{a}^*\!\cp\!\vc{F}_\text{a}^{\vphantom{*}} + \vc{F}_\text{p}^*\!\cp\!\vc{F}_\text{p}^{\vphantom{*}}] \) & \cellcolor{proper}\( \Im[\vc{F}_\text{R}^*\!\cp\!\vc{F}_\text{R}^{\vphantom{*}} + \vc{F}_\text{L}^*\!\cp\!\vc{F}_\text{L}^{\vphantom{*}}] \) & spin angular mom. den.\\
\hline
\cellcolor{altroo}\( \omega\vc{S}_1 \) &
\cellcolor{altroo}\( \omega(\vc{S}_\text{e}-\vc{S}_\text{m}) \) &\cellcolor{altroo}\( \Im[\vc{F}_\text{e}^*\!\cp\!\vc{F}_\text{e} - \vc{F}_\text{m}^*\!\cp\!\vc{F}_\text{m}] \) & \( 2\Im(\vc{F}_\text{p}^*\cp\vc{F}_\text{a}^{\vphantom{*}}) \) & \( 2\Im(\vc{F}_\text{R}^*\cp\vc{F}_\text{L}^{\vphantom{*}}) \) &{EM spin difference\footnote{Related to the reactive stress tensor $\tens{T}_I=-\omega\star\vc{S}_1/2$ \cite{NietoVesperinas2022} (for more details see \cref{eq:TI}).}}\\
\hline
\cellcolor{crlroo}\( \omega\vc{S}_2 \) &
\cellcolor{crlroo}\( \omega(\vc{S}_\text{a}-\vc{S}_\text{p}) \) &\( 2\Im( {\vc{F}_\text{e}\cp\vc{F}^*_\text{m}} ) \) & \cellcolor{crlroo}\( \Im[\vc{F}_\text{a}^*\!\cp\!\vc{F}_\text{a}^{\vphantom{*}} - \vc{F}_\text{p}^*\!\cp\!\vc{F}_\text{p}^{\vphantom{*}}] \) & \(-2\Re(\vc{F}_\text{R}^* \cp \vc{F}_\text{L}^{\vphantom{*}})\) &$c\cdot\Im${(Poynting vector)\footnote{Imaginary part of complex Poynting vector representing flow of reactive power \cite{kamenetskii2015microwave,Jackson1998}.}}\\
\hline
\cellcolor{pseudo}\( \omega\vc{S}_3 \) &
\cellcolor{pseudo}\( \omega(\vc{S}_\text{R}-\vc{S}_\text{L}) \) &\( 2\Re( {\vc{F}_\text{e}\cp\vc{F}^*_\text{m}} ) \) & \( -2\Re(\vc{F}_\text{p}^*\cp\vc{F}_\text{a}^{\vphantom{*}}) \) & \cellcolor{pseudo}\( \Im[\vc{F}_\text{R}^*\!\cp\!\vc{F}_\text{R}^{\vphantom{*}} - \vc{F}_\text{L}^*\!\cp\!\vc{F}_\text{L}^{\vphantom{*}}] \) &$c\cdot\Re${(Poynting vector)\footnote{Real part of complex Poynting vector representing flow of active power \cite{kamenetskii2015microwave,Jackson1998}.}}\\
\hline\hline
\cellcolor{proper}\( \tens{T}_0 \) & 
\cellcolor{proper}\( \tens{T} \) & \cellcolor{proper}\( \vc{F}_\text{e}^*\!\odot\!\vc{F}_\text{e}^{\vphantom{*}} + \vc{F}_\text{m}^*\!\odot\!\vc{F}_\text{m}^{\vphantom{*}}-W_0\tens{I} \) & \cellcolor{proper}\( \vc{F}_\text{a}^*\!\odot\!\vc{F}_\text{a}^{\vphantom{*}} + \vc{F}_\text{p}^*\!\odot\!\vc{F}_\text{p}^{\vphantom{*}}-W_0\tens{I} \) & \cellcolor{proper}\( \vc{F}_\text{R}^*\!\odot\!\vc{F}_\text{R}^{\vphantom{*}} + \vc{F}_\text{L}^*\!\odot\!\vc{F}_\text{L}^{\vphantom{*}}-W_0\tens{I} \) & Maxwell stress tensor\footnote{Represents flux of canonical momentum density $\vc{p}_0$.}\\
\hline
\cellcolor{altroo}\( \tens{T}_1 \) & \cellcolor{altroo}\( \tens{T}_\text{e}-\tens{T}_\text{m} \) & \cellcolor{altroo}\( \vc{F}_\text{e}^*\!\odot\!\vc{F}_\text{e} - \vc{F}_\text{m}^*\!\odot\!\vc{F}_\text{m}-W_1\tens{I} \) & \( 2\Re(\vc{F}_\text{p}^*\odot\vc{F}_\text{a}^{\vphantom{*}})-W_1\tens{I} \) & \( 2\Re(\vc{F}_\text{R}^*\odot\vc{F}_\text{L}^{\vphantom{*}})-W_1\tens{I} \) &{EM stress difference}\\
\hline
\cellcolor{crlroo}\( \tens{T}_2 \) &
\cellcolor{crlroo}\( \tens{T}_\text{a}-\tens{T}_\text{p} \) &\(- 2\Re( {\vc{F}_\text{e}^*\odot\vc{F}_\text{m}} )-W_2\tens{I} \) & \cellcolor{crlroo}\( \vc{F}_\text{a}^*\!\odot\!\vc{F}_\text{a}^{\vphantom{*}} - \vc{F}_\text{p}^*\!\odot\!\vc{F}_\text{p}^{\vphantom{*}}-W_2\tens{I} \) & \(2\Im(\vc{F}_\text{R}^* \odot \vc{F}_\text{L}^{\vphantom{*}})-W_2\tens{I}\) &\\
\hline
\cellcolor{pseudo}\( \tens{T}_3 \) &
\cellcolor{pseudo}\( \tens{T}_\text{R}-\tens{T}_\text{L} \) &\(- 2\Im( {\vc{F}_\text{e}^*\!\odot\vc{F}_\text{m}} )-W_3\tens{I} \) & \( 2\Im(\vc{F}_\text{p}^*\odot\vc{F}_\text{a}^{\vphantom{*}})-W_3\tens{I} \) & \cellcolor{pseudo}\( \vc{F}_\text{R}^*\!\odot\!\vc{F}_\text{R}^{\vphantom{*}} - \vc{F}_\text{L}^*\!\odot\!\vc{F}_\text{L}^{\vphantom{*}}-W_3\tens{I} \) &chiral stress tensor\footnote{Represents flux of chiral momentum density $\vc{p}_3$.}\\
\hline
\end{tabular}

\caption{Quadratic quantities in different bases, revealing the similarity with Stokes vectors.}
% \vspace{-5em}
\label{tab:energystokes_SM}
\end{table}

\subsection{States on the surface of the energy-symmetry sphere}
The main text Fig.~2 shows six special points on the surface of the symmetry sphere given by $\vc{F}_i = 0$ for $i = \{\text{e},\text{m},\text{p},\text{a},\textsc{r},\textsc{l}\}$. However, in principle, there is nothing special about those specific directions and the specific axes $W_1$, $W_2$, $W_3$ defined in the main text.

In the well-known $\mathbb{C}^2$ space of field polarisations, we typically write the electric field either in the horizontal-vertical basis, $(E_\text{h},E_\text{v})$, the diagonal-antidiagonal basis, $(E_\text{d},E_\text{a})$, or the circular basis, $(E_\text{R},E_\text{L})$, which gives rise to the three axes of the Poincare sphere $\mathcal{S}_1 = |E_\text{h}|^2 - |E_\text{v}|^2$, $\mathcal{S}_2 = |E_\text{d}|^2 - |E_\text{a}|^2$ and $\mathcal{S}_3 = |E_\text{R}|^2 - |E_\text{L}|^2$, but these are not the only three possible bases. In fact, any two orthogonal polarisations can serve as a valid basis (with coefficients $E_\alpha$ and $E_\beta$ each being a linear combination of $E_\text{h}$ and $E_\text{v}$) and correspond to an arbitrary axis $|E_\alpha|^2 - |E_\beta|^2$ on the Poincare sphere with some specific orientation. Similarly, our isomorphic $\mathbb{C}^2$ space of bispinors admits infinitely many basis choices, apart from the EM, PA, RL basis, each with vector-valued coefficients $\vc{F}_\alpha$ and $\vc{F}_\beta$ that are each some linear combination of $\sqrt{\varepsilon}\vc{E}$ and $\sqrt{\mu}\vc{H}$. In that basis one can write $\vc{\psi}_{[\alpha\beta]}=(\vc{F}_\alpha,\vc{F}_\beta)$. This basis will represent an axis going through the origin of the symmetry sphere, with the axis value associated to the quantity $|\vc{F}_\alpha|^2 - |\vc{F}_\beta|^2$. Reversing the argument, every axis crossing the origin of the symmetry sphere, regardless of orientation, will have some associated basis whose vector coefficients are $(\vc{F}_{i\alpha},\vc{F}_{i\beta})$.

This means that every point lying exactly on the surface of the symmetry sphere (where $W_1^2+W_2^2+W_3^2 = W_0^2$) corresponds to a point where a given $\vc{F}_{i\alpha} = 0$. Because $\vc{F}_{i\alpha}$ is always a linear combination of $\sqrt{\varepsilon}\vc{E}$ and $\sqrt{\mu}\vc{H}$, the surface of the field-symmetry sphere necessarily corresponds to electromagnetic states where both quantities are linearly dependent, such that their linear combination can be zero. Hence, on the surface of the symmetry sphere,  $\sqrt{\varepsilon}\vc{E} = \delta \sqrt{\mu}\vc{H}$ with $\delta$ a complex scalar. With some algebra, one can find that for any arbitrary location on the surface of the field-symmetry sphere, with spherical angle coordinates $(\theta_i,\phi_i)$ measured with respect to the arbitrary axis corresponding to $|\vc{F}_{i\alpha}|^2 - |\vc{F}_{i\beta}|^2$ acting as polar axis, the linear dependence of the fields is given specifically by:
\begin{equation}
\label{eq:suppl:maximallyasymmetricfield}
\sin(\theta_i/2) \vc{F}_{i\alpha} = \ee^{i \phi_i} \cos(\theta_i/2) \vc{F}_{i\beta},
\end{equation}
such that the polar angle $\theta_i$ determines the amplitude ratio, while the azimuthal angle $\phi_i$ determines the relative phase between the basis coefficients $\vc{F}_{i\alpha}$ and $\vc{F}_{i\beta}$. This relation is a well-known property of a Bloch sphere, often used in quantum superpositions of two states.

If we choose to express this in terms of $\vc{E}$ and $\vc{H}$, this is equivalent to choosing the EM basis $\vc{\psi}=(\vc{F}_\text{e},\vc{F}_\text{m})$ to describe the fields and define the angles on the sphere $(\theta_1,\phi_1)$, such that the $W_1$ axis plays the role of polar axis. In this case, \cref{eq:suppl:maximallyasymmetricfield} can be written as:
\begin{equation}\label{eq:suppl:maximallyasymmetricfieldW1axis}
\sin(\theta_1/2) \sqrt{\varepsilon}\vc{E} = \ee^{i \phi_1} \cos(\theta_1/2) \sqrt{\mu}\vc{H},
\end{equation}
or, equivalently:
\begin{equation}\label{eq:suppl:maximallyasymmetricfieldW1axis2}
\vc{E} = \underbrace{\frac{\ee^{i \phi_1}}  {\tan(\theta_1/2)}}_{\delta(\theta_1,\phi_1)} \eta \vc{H},
\end{equation}
which provides the linear relationship between the $\vc{E}$ and $\vc{H}$ field corresponding to any point in the surface of the energy symmetry sphere. Six such special points (corresponding to the antipodes for the bases EM, PA, RL, where the sphere intersects the $W_1$, $W_2$ and $W_3$ axes) are shown in Fig. 2 of the main text.

% --------------------------------------------------------------------------------------------
% \clearpage
\subsection{Electromagnetic symmetry sphere for dipolar fields}
Entire three-dimensional electromagnetic fields including near and far field regions can be mapped into the volume of the electromagnetic symmetry sphere, constructing complicated 3D patterns that measure the share of energy density among $W_{1,2,3}$.
In this section we map well-known dipolar fields to the symmetry sphere as well as `synthetic' dipoles inspired by the vectors $\vc{F}_A$ in the main article.
Each example is a superposition of an electric dipole with dipole moment $\vc{p}=(p_x,p_y,p_z)$ and a magnetic dipole with dipole moment $\vc{m}=(m_x,m_y,m_z)$---this way, an $x$-polarised electric dipole is given by $(p_x,p_y,p_z)=(1,0,0)$ and $(m_x,m_y,m_z)=\vc{0}$ both superimposed at the origin. The full electromagnetic fields in free space created by such a magnetoelectric dipole are well known \cite{Jackson1998} and given in SI~units~by:
\begin{equation}
\label{eq:Efielddipole}
\vc{E}(\vc{r}) = \frac{1}{4\pi\varepsilon_0} \left\{ k^2 (\hat{\vc{r}} \times \vc{p}) \times \hat{\vc{r}} \frac{\ee^{ikr}}{r} + \left[ 3(\hat{\vc{r}} \vdot \vc{p})\hat{\vc{r}} - \vc{p} \right] \left( \frac{1}{r^3} - \frac{ik}{r^2} \right) \ee^{ikr} \right\} -\frac{\eta k^2}{4\pi} (\hat{\vc{r}} \times \vc{m}) \left( 1 - \frac{1}{ikr} \right) \frac{\ee^{ikr}}{r}
\end{equation}
\begin{equation}
\label{eq:Hfielddipole}
\vc{H}(\vc{r}) = \frac{ck^2}{4\pi} (\hat{\vc{r}} \times \vc{p}) \left( 1 - \frac{1}{ikr} \right) \frac{\ee^{ikr}}{r} + \frac{1}{4 \pi} \left\{ k^2 (\hat{\vc{r}} \times \vc{m}) \times \hat{\vc{r}} \frac{\ee^{ikr}}{r} + \left[ 3(\hat{\vc{r}} \vdot \vc{m})\hat{\vc{r}} - \vc{m} \right] \left( \frac{1}{r^3} - \frac{ik}{r^2} \right) \ee^{ikr} \right\}
\end{equation}
where $\varepsilon_0$ is the permittivity of free space, $k=\omega/c=2\pi/\lambda$ is the wavenumber of free space, $\eta=\sqrt{\mu_0/\varepsilon_0}$ is the impedance of free space, $\hat{\vc{r}}=\vc{r}/r$ is the unit vector in the radial direction, and $\vc{r}$ is the position vector, with length~$r$.

\begin{figure}[hb!]
    \centering
    \includegraphics[width=\textwidth]{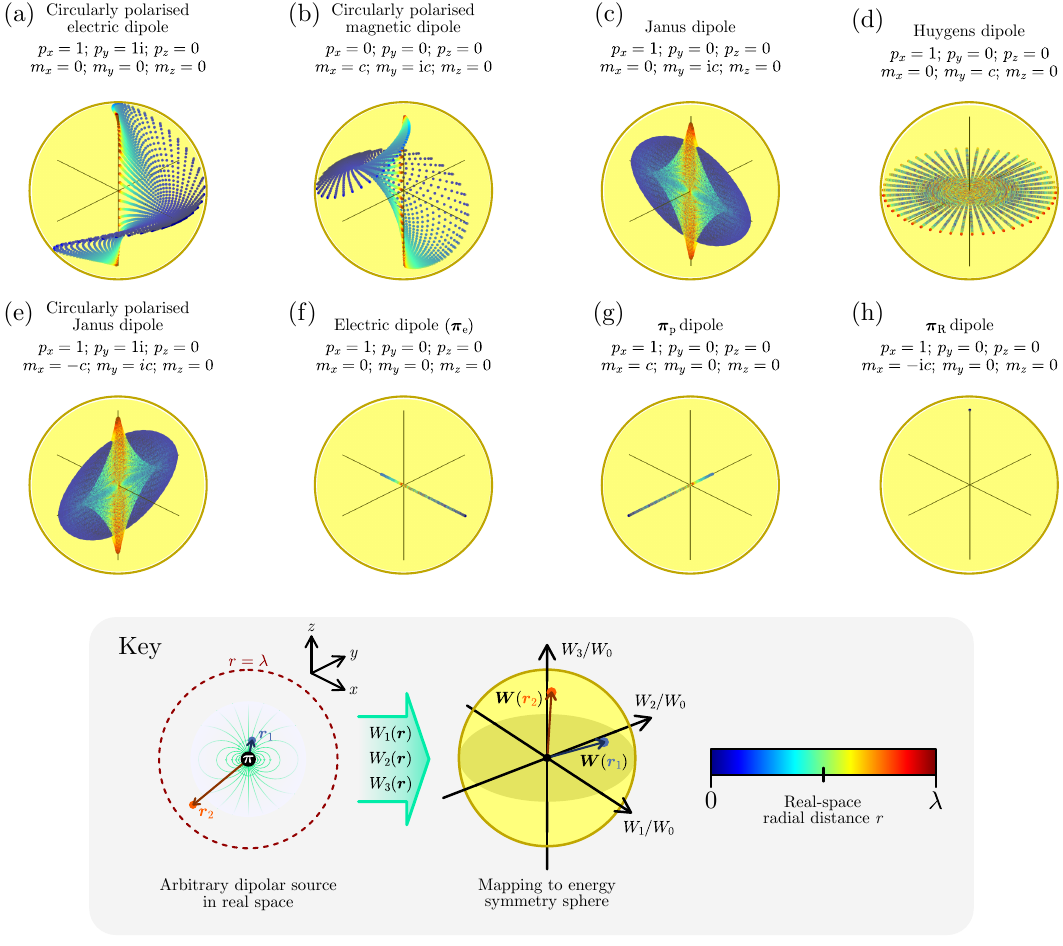}
    \caption{
    The electromagnetic fields produced in a $1\lambda$-radius volume by an assortment of dipoles, mapped to the symmetry sphere with axes normalised by total energy density $W_0$.
    Each marker in the sphere corresponds to the local distribution of $W_0$ among $W_{1,2,3}$ at one of many uniformly spaced positions $\vc{r}_i$ in the dipole's field in real space.
    The colour of the marker corresponds to its real-space radial distance from the dipole, up to a maximum distance $r=\lambda$ (red markers are real-space samples approaching the far field).
    See the key for a visual representation of the mapping from two example positions $\vc{r}_{1,2}$.
    }
    \label{fig:dipole_examples}
\end{figure}

In Fig.~\ref{fig:dipole_examples}, the symmetry sphere mapping is calculated for a selection of dipoles.
To complete the mapping the real-space fields of each dipole (given by \cref{eq:Efielddipole,eq:Hfielddipole}) are calculated in a spherical volume of radius equal to the wavelength $\lambda$ and sampled uniformly in a spatial grid---for each sample the energy densities $W_{1,2,3}$ are calculated and normalised by the local total energy density $W_0$ to place a marker in the symmetry sphere.
Markers are colour coded to represent the radial distance of the real-space field sample from the dipole; red markers are furthest from the dipolar source, approaching the far field, while blue coloured markers correspond to near fields. With this approach, the symmetry sphere for the fields surrounding different electric-magnetic dipoles are shown in \cref{fig:dipole_examples}.

Circular electric and magnetic dipoles (a) and (b) build helical structures (with the same winding sense) within the symmetry sphere while the entire electromagnetic field of the Janus dipole \cite{Picardi2018_2} (c) maps to the $W_2W_3$ plane.
The mapped structures in (a) and (b) can be interchanged visually by a $\hat{D}$ transformation which rotates the sphere $180$ degrees about the $W_3$ axis.
As anticipated in these cases, paraxial far fields collapse on to the $W_3$ axis, indicated by the aggregation of the red markers.
Perhaps confusingly the Huygens dipole field (d), while sitting in the $W_1W_2$ plane, does not completely collapse onto the $W_3$ axis in the far field as shown by the red markers interspersed with near field samples, some situated virtually at the sphere equator.
This occurs because, unlike the other three dipole fields, the Huygens dipole has a dark spot in its far field radiation pattern where transverse field components vanish.
Approaching this dark spot, radial and transverse field components inevitably become comparable resulting in a small far-field non-paraxial region \cite{Vernon2024} where it is possible for $W_{1,2}$ to be large relative to the (albeit strongly suppressed) total energy density $W_0$ (note once more that the sphere axes are normalised by $W_0$).
In (e), a circularly polarised Janus dipole produces a rotated version of the ordinary Janus dipole's symmetry sphere structure, while in (f)-(h) are more `elementary' dipoles which only have one component in the $\mathbb{C}^2$ space of the dipole moment bispinor.
These elementary dipoles are (f): a $\vc{\pi}_\text{e}$ dipole (an electric dipole), (g): a $\vc{\pi}_\text{p}$ dipole and (h): a $\vc{\pi}_\text{R}$ dipole---each of these dipoles has given $\vc{p}$ and $\vc{m}$ dipole moments which eliminate the $\vc{\pi}_\text{m}$, $\vc{\pi}_\text{a}$ and $\vc{\pi}_\text{L}$ components in Eq.~\ref{dipole_bispinor} from (f)-(h) respectively.
The electric dipole (f) radiates a field whose energy density is largely supplied by the electric field more than the magnetic field, hence in the symmetry sphere its field maps onto the positive $W_1$ axis (with a small tail on the negative side of that axis).
Meanwhile a similar situation occurs in (g) as the $\vc{\pi}_\text{p}$ dipole field maps mostly to the negative $W_2$ axis.
In (h) is shown the striking feature of a $\vc{\pi}_\text{R}$ dipole which is that its entire radiated field is right-handed, mapping to a single point at the north pole of the symmetry sphere.

% --------------------------------------------------------------------------------------------
\clearpage
\subsection{Maxwell-like equations with field quadratic quantities}
The dual-symmetric (including magnetic charges) time-harmonic Maxwell's equations in \cref{eq:maxwelldualnormalised}, written in a more standard notation, are:
\begin{equation}
    \label{eq:maxwelldualstandard}
    \begin{split}
        \div\vc{E}&=\rho_\text{e}/\varepsilon\,,\\
        \div\vc{H}&=\rho_\text{m}/\mu\,,\\
        \curl\vc{E}&=+\ii k\eta\vc{H}-\vc{J}_\text{m}\,,\\
        \curl\vc{H}&=-\ii k\vc{E}/\eta+\vc{J}_\text{e}\,.
    \end{split}
\end{equation}
In the main text we mention that certain combination of the spin-like quadratic quantities ($\vc{S}_0, \vc{S}_1, \vc{S}_2, \vc{S}_3$) fulfil Maxwell-like equations. Indeed, from eq.~(7) in the main text, one can check that the complex combinations $\vc{S}_0\pm\ii\vc{S}_3$ satisfy the following equations:
\begin{equation}
    \begin{split}
        \div(\vc{S}_0\pm\ii\vc{S}_3)&=0\,,\\
        \curl(\vc{S}_0\pm\ii\vc{S}_3)&=\pm2\ii k(\vc{S}_0\mp\ii\vc{S}_3)-2(\vc{p}_0\pm\ii\vc{p}_3)\,,
    \end{split}
\end{equation}
making them exactly analogous to time harmonic electric and magnetic fields with twice the frequency of the fields and with currents but no charge densities. Also, the equivalent of duality symmetry of these quantities is simply complex conjugation or parity. One can combine these into a single Helmholtz equation
\begin{equation}\label{eq:helmholtz}
       \!(\laplacian+4k^2)(\vc{S}_0\pm\ii\vc{S}_3)=4k(\vc{p}_3\pm\ii\vc{p}_0)+2\grad\cp(\vc{p}_0\pm\ii\vc{p}_3)\,.\!
\end{equation}
The remaining two spin-like quadratic quantities can also be combined into a complex quantity, however, this time they will be quasistatic fields (taking $k=0$ in \cref{eq:maxwelldualstandard} and resembling electrostatic/magnetostatic fields) with both charge densities and currents 
\begin{equation}
    \begin{split}
        \div(\vc{S}_1\pm\ii\vc{S}_2)&=\frac{2}{c}(W_2\mp\ii W_1)\,,\\
        \curl(\vc{S}_1\pm\ii\vc{S}_2)&=-2(\vc{p}_1\mp\ii\vc{p}_2)\,,
    \end{split}
\end{equation}
These can be combined into a single Poisson equation
\begin{equation}
    \!(\laplacian+0k^2)(\vc{S}_1\pm\ii\vc{S}_2)=\frac{2}{c}\grad(W_2\mp\ii W_1)+2\curl(\vc{p}_1\mp\ii\vc{p}_2)\,.\!\!
\end{equation}
In the same way in which $\vc{S}_0\pm\ii\vc{S}_3$ fulfils the Helmholtz equation with twice the frequency of the fields $\vc{S}_1\pm\ii\vc{S}_2$ fulfils the Poisson equation (Helmholtz equation with zero times the frequency).

% --------------------------------------------------------------------------------------------

\section{Light-matter interaction on different \texorpdfstring{$\mathbb{C}^2$}{C²} bases}

\subsection{The dipole moment bispinor and basis-independent polarisability}
In the same way that electric and magnetic fields may be combined into an electromagnetic bi-spinor $\vc{\psi}$, so can electric $\vc{p}$ and magnetic $\vc{m}$ dipole moments be combined into a dipole moment bispinor $\vc{\pi}$, which can be written in each of the three bases:
\begin{equation}\label{dipole_bispinor}
\begin{alignedat}{4}
    \vc{\pi}_{[\text{EM}]}&=\pmqty{\vc{\pi}_\text{e}\\\vc{\pi}_\text{m}}&&=\frac{1}{2}\pmqty{\vc{p}/\sqrt{\varepsilon}\\\sqrt{\mu}\vc{m}}\,,\\
    \vc{\pi}_{[\text{PA}]}&=\pmqty{\vc{\pi}_\text{p}\\\vc{\pi}_\text{a}}&&=\frac{1}{2\sqrt{2}}\pmqty{\vc{p}/\sqrt{\varepsilon}+\sqrt{\mu}\vc{m}\\\vc{p}/\sqrt{\varepsilon}-\sqrt{\mu}\vc{m}}\,,\\
    \vc{\pi}_{[\text{RL}]}&=\pmqty{\vc{\pi}_\text{R}\\\vc{\pi}_\text{L}}&&=\frac{1}{2\sqrt{2}}\pmqty{\vc{p}/\sqrt{\varepsilon}+\ii\sqrt{\mu}\vc{m}\\\vc{p}/\sqrt{\varepsilon}-\ii\sqrt{\mu}\vc{m}}\,,
\end{alignedat}    
\end{equation}%
In a Rayleigh particle, the linearity of the response ensures that the dipole moments induced in the particle are proportional to the applied fields via a polarisability matrix. In a basis-independent bi-spinor notation, this is written:
\begin{equation}
\label{eq:polarisabilitybasisindependent}
    \vc{\pi} = \tens{A} \vc{\psi}
\end{equation}
For isotropic particles, the polarisability matrix $\tens{A}$ can be represented in any basis as a two-by-two complex-valued matrix, and hence we can apply the results from \cref{eq:2by2_in_different_bases}, as we do below:
\begin{equation}
\label{eq:polarisabilities_in_different_bases}
\begin{alignedat}{3}
    \vc{\pi}_{[\text{EM}]}&=\pmqty{\vc{\pi}_\text{e}\\\vc{\pi}_\text{m}}&&=\pmqty{\alpha_\text{e}&\alpha_\text{em}\\\alpha_\text{me}&\alpha_\text{m}}\pmqty{\vc{F}_\text{e}\\\vc{F}_\text{m}} &&=\bqty{\alpha_0\mqty(\pmat{0})+\alpha_1\underbrace{\mqty(\pmat{3})}_{\sigma_3}-\alpha_2\underbrace{\mqty(\pmat{1})}_{\sigma_1}-\alpha_3\underbrace{\mqty(\pmat{2})}_{\sigma_2}}\pmqty{\vc{F}_\text{e}\\\vc{F}_\text{m}}\,,\\
    \vc{\pi}_{[\text{PA}]}&=\pmqty{\vc{\pi}_\text{p}\\\vc{\pi}_\text{a}}&&=\pmqty{\alpha_\text{p}&\alpha_\text{pa}\\\alpha_\text{ap}&\alpha_\text{a}}\pmqty{\vc{F}_\text{p}\\\vc{F}_\text{a}}&&=\bqty{\alpha_0\mqty(\pmat{0})+\alpha_1\underbrace{\mqty(\pmat{1})}_{\sigma_1}-\alpha_2\underbrace{\mqty(\pmat{3})}_{\sigma_3}+\alpha_3\underbrace{\mqty(\pmat{2})}_{\sigma_2}}\pmqty{\vc{F}_\text{p}\\\vc{F}_\text{a}}\,,\\
    \vc{\pi}_{]\text{RL}]}&=\pmqty{\vc{\pi}_\text{R}\\\vc{\pi}_\text{L}}&&=\pmqty{\alpha_\text{R}&\alpha_\text{RL}\\\alpha_\text{LR}&\alpha_\text{L}}\pmqty{\vc{F}_\text{R}\\\vc{F}_\text{L}}&&=\bqty{\alpha_0\mqty(\pmat{0})+\alpha_1\underbrace{\mqty(\pmat{1})}_{\sigma_1}+\alpha_2\underbrace{\mqty(\pmat{2})}_{\sigma_2}+\alpha_3\underbrace{\mqty(\pmat{3})}_{\sigma_3}}\pmqty{\vc{F}_\text{R}\\\vc{F}_\text{L}}\,,
\end{alignedat}
\end{equation}%
which provides us with four basis-independent measures of polarisability following \cref{eq:definition_of_ai}:
\begin{equation}
\label{eq:definition_of_alphai}
\begin{alignedat}{2}
\alpha_0 = \tfrac{1}{2}(\alpha_\text{R} + \alpha_\text{L}) &= \tfrac{1}{2}(\alpha_\text{e} + \alpha_\text{m}) &&= \tfrac{1}{2}(\alpha_\text{p} + \alpha_\text{a}), \\
\alpha_1 &= \tfrac{1}{2}(\alpha_\text{e} - \alpha_\text{m})  &&, \\
\alpha_2 &= -\tfrac{1}{2}(\alpha_\text{p} - \alpha_\text{a})  &&, \\
\alpha_3 &= \tfrac{1}{2}(\alpha_\text{R} - \alpha_\text{L})   &&.
\end{alignedat}
\end{equation}
Most readers are familiar with the EM basis notation, $\vc{\pi}_\text{EM}$ above. Note that in that representation, the main diagonal is composed exclusively of $\alpha_0$ and $\alpha_1$, providing the sum and difference of electric and magnetic polarisabilities, while the off-diagonal components in $\vc{\pi}_\text{EM}$ are the antisymmetric $\alpha_3$, termed the chiral polarisability, and the symmetric $\alpha_2$, the non-reciprocal polarisability. By changing the basis, we can see that the chiral polarisability $\alpha_3$ is also the difference in the main diagonal elements in the RL basis. This insight is not new, but we believe that \cref{eq:definition_of_alphai} constitutes a very intuitive explanation. Similarly, $\alpha_2$ is the difference in the main diagonal elements in the PA basis.
Because the three matrices in \cref{eq:polarisabilities_in_different_bases} are \emph{similar matrices}, they share the same trace and determinant in $\mathbb{C}^2$ space:
\begin{equation}\label{eq:trace_and_det}
\begin{alignedat}{1}
    \tr{\tens{A}} &= \alpha_\text{e}+\alpha_\text{m} = \alpha_\text{p}+\alpha_\text{a} = \alpha_\text{R}+\alpha_\text{L} = 2\alpha_{0}, \\
    \det{\tens{A}} &= \alpha_\text{e}\alpha_\text{m} - \alpha_\text{em}\alpha_\text{me} = \alpha_\text{p}\alpha_\text{a} - \alpha_\text{pa}\alpha_\text{ap} = \alpha_\text{R}\alpha_\text{L} - \alpha_\text{RL}\alpha_\text{LR}=\alpha_0^2-\alpha_1^2-\alpha_2^2-\alpha_3^2.
\end{alignedat}
\end{equation}

In general the response of a particle can be non-isotropic in the $\mathbb{C}^3$ space. In that case the components $\alpha_i$ with $i\in\{\text{e,m,p,a,\textsc{r},\textsc{l}}\}$ will become three-by-three matrices $\tens\alpha_i$ (same for $\alpha_A$ with $A\in\{0,1,2,3\}$) and the full response $\tens{A}$ can be thought of as a six-by-six matrix. Note however, that same as one can treat the $\mathbb{C}^2$ and $\mathbb{C}^3$ as separate vector spaces the same is true for the matrices acting on them. This means that \cref{eq:polarisabilities_in_different_bases,eq:definition_of_alphai,} will be the same just with $\alpha_i\mapsto\tens\alpha_i$, more care has to be taken in case of \cref{eq:trace_and_det} as one also has to emphasise that $\tr\mapsto\tr_{\mathbb{C}^2}$ and $\det\mapsto\det_{\mathbb{C}^2}$ are taken only across the two-by-two components that act on $\mathbb{C}^2$.

The use of these bases in $\mathbb{C}^2$ space also brings great simplifications to the analytical calculations involving dipoles. For instance, the power extinguished by a dipolar particle has the well known analytical expression:
\begin{equation}
    P_\text{ext} = \frac{\omega}{2}\Im(\vc{E}^*\vdot\vc{p} + \mu \vc{H}^*\vdot\vc{m})\,,
\end{equation}
this can be written in a basis-independent way as a dot product of the electromagnetic and dipolar bi-spinors \cite{Bliokh2014}:
\begin{equation}
    P_\text{ext} = 2\omega\Im(\vc{\psi}^\dagger\vdot\vc{\pi})\,,
\end{equation}
by applying the linear relation between dipolar bispinor and electromagnetic bispinor $\vc{\pi} = \tens{A} \vc{\psi}$ (\cref{eq:polarisabilitybasisindependent}), followed by the basis-independent decomposition $\tens{A} = \alpha_0 \tens{W}_0 + \alpha_1 \tens{W}_1 + \alpha_2 \tens{W}_2 + \alpha_3 \tens{W}_3$ (\cref{eq:2x2decomposition_basisindependent}), we will get 
\begin{equation*}
\begin{split}
    P_\text{ext} &= 
    2\omega\Im(\vc{\psi}^\dagger\vdot\tens{A}\vc{\psi})= 2\omega\Im[\vc{\psi}^\dagger\vdot(\alpha_0 \tens{W}_0 + \alpha_1 \tens{W}_1 + \alpha_2 \tens{W}_2 + \alpha_3 \tens{W}_3)\vc{\psi}]\,,
\end{split}
\end{equation*}
if the particle is isotropic then $\alpha_A$ are scalar and can be pulled through $\vc{\psi}^\dagger$ to obtain
\begin{equation*}
    P_\text{ext}= 2\omega\sum_{A=0}^3\Im[\alpha_A(\vc{\psi}^\dagger\tens{W}_A\vc{\psi})]\,,
\end{equation*}
and using the relation $W_A =\vc{\psi}^\dagger\tens{W}_A\vc{\psi}$ (\cref{eq:relationWAwithtensorWA}), one can derive:
\begin{equation*}
    P_\text{ext}= 2\omega\sum_{A=0}^3\Im(\alpha_A)W_A\,.
\end{equation*}
This leads to a remarkably simple expression for the extinction power, revealing the deep link between the energy-symmetry sphere coordinates $W_i$ and the polarisability-sphere parameters $\alpha_i$:
\begin{equation}
    P_\text{ext}= 2\omega\left[ \Im(\alpha_0)W_0 + \Im(\alpha_1)W_1 + \Im(\alpha_2)W_2 + \Im(\alpha_3)W_3 \right]\,.
\end{equation}
One can also define generalised intensities $cW_A$ and extinction cross sections to write
$\sigma_A^\text{ext}=2k\Im(\alpha_A)$
\begin{equation}
    P_\text{ext}=  \sigma_0^\text{ext}cW_0 + \sigma_1^\text{ext}cW_1 + \sigma_2^\text{ext}cW_2 + \sigma_3^\text{ext}cW_3 \,.
\end{equation}
This carries much insight. For instance, as we showed that paraxial far fields have $W_1=W_2=0$, this reveals that we cannot retrieve $\alpha_1$ and $\alpha_2$ by using far-field dipolar extinction measurements (we cannot distinguish between electric and magnetic polarisability, or retrieve the non-reciprocal polarisability). While $\alpha_0$ and $\alpha_3$ are usually measured, using far-field illumination, as the extinction cross section $\sigma_\text{ext}=2k\Im(\alpha_0)=k\Im(\alpha_\text{e}+\alpha_\text{m})$ and the $g$-factor $g=2\Im(\alpha_3)/\Im(\alpha_0)$ spectra, which can then be used to obtain real parts using Kramers–Kronig relations. Similar simple expressions can be derived for dipolar absorption, force and torque \cite{Bliokh2014}. 

\subsection{Macroscopic Maxwell equations}\label{sec:macmax}

So far we have considered microscopic Maxwell equations and point particles, so we have been limited to homogeneous materials with no structure. If we want to consider material structures, such as finite particles, waveguides, etc., we need to turn to the macroscopic Maxwell equations. 
{Consider $\vc{J}_\text{e}$ and $\vc{J}_\text{m}$ to be only the \emph{free} currents}
\begin{equation}
    \begin{alignedat}{1}
        \curl\vc{E}&=-\frac{\partial\vc{B}}{\partial t}-\vc{J}_\text{m}\,,\\
        \curl\vc{H}&=\frac{\partial\vc{D}}{\partial t}+\vc{J}_\text{e}\,,
    \end{alignedat}
\end{equation}
which in the time-harmonic case ($\partial/\partial t \to -i \omega \to - i k_0 c_0$, with $k_0 = \omega/c_0$) become:
\begin{equation}
    \begin{alignedat}{1}
        \curl\vc{E}&=i k_0 c_0 \vc{B}-\vc{J}_\text{m}\,,\\
        \curl\vc{H}&=- i k_0 c_0 \vc{D}+\vc{J}_\text{e}\,.
    \end{alignedat}
\end{equation}
Now, substituting $c_0 = 1/\sqrt{\varepsilon_0 \mu_0}$, we can write these two equations as a single equation in bi-spinor ($\mathbb{C}^2$) space (and this will be in the EM basis) \cite{muljarov2018resonant}:
\begin{equation}\label{eq:maxwell_material}
\begin{alignedat}{1}
    \pmqty{\curl&0\\0&\curl} \frac{1}{2} \pmqty{\sqrt{\varepsilon_0}\vc{E}\\\sqrt{\mu_0}\vc{H}} &= i k_0 \frac{1}{2} \pmqty{\vc{B}/\sqrt{\mu_0} \\ - \vc{D}/\sqrt{\varepsilon_0}} -\frac{1}{2} \pmqty{\vc{J}_\text{m} \\ - \vc{J}_\text{e}} \,,\\
    \pmqty{\curl&0\\0&\curl} \underbrace{\frac{1}{2} \pmqty{\sqrt{\varepsilon_0}\vc{E}\\\sqrt{\mu_0}\vc{H}}}_{\vc{\psi}_{[\text{EM}]} = \pmqty{\vc{F}_\text{e}\\ \vc{F}_\text{m}}} &=  \underbrace{\pmqty{0&1\\-1&0}}_{\hat{D}_{[\text{EM}]}}\Bigg[ i k_0\underbrace{\frac{1}{2} \pmqty{\vc{D}/\sqrt{\varepsilon_0} \\ \vc{B}/\sqrt{\mu_0}}}_{\vc{\gamma}_{[\text{EM}]} = \pmqty{\vc{G}_\text{e}\\ \vc{G}_\text{m}}}\mathop{-}\underbrace{\frac{1}{2} \pmqty{\vc{J}_\text{e} \\  \vc{J}_\text{m}}}_{\vc{g}_{[\text{EM}]}}\Bigg],
\end{alignedat}
\end{equation}
which motivates us to introduce a bi-spinor with the auxiliary displacement-field $\vc{D}$ and magnetic flux $\vc{B}$, which we define as $\vc{\gamma} = (\vc{D}/2\sqrt{\varepsilon_0}) \uv{e} + (\vc{B}/2\sqrt{\mu_0}) \uv{m} = \vc{G}_\text{e} \uv{e} + \vc{G}_\text{m} \uv{m}$. Note that the prefactors are the \emph{vacuum} permittivity and permeability, and are there just to make the dimensions match. We are \emph{not} assuming a vacuum background, as the information of the material is encoded in the constitutive relations between the fields and auxiliary fields. We also used the duality transformation $\hat{D}$ such that $i \hat{D} = \tens{W}_3$, following \cref{tab:transposed_matrix_basis}, corresponds to the Pauli matrix $-\tens{\sigma}_2 = i \hat{D} = i\spmqty{0&1\\-1&0}$ in the EM basis above. The remaining two equations can be obtained from the microscopic ones \cref{eq:microscopicmaxwell_basisindependent} simply by $\vc{\psi}\mapsto\vc{\gamma}$ and considering $\vc{f}$ to be free charges. The macroscopic Maxwell's equations can thus be written in a $\mathbb{C}^2$-basis-independent form as:
\begin{equation}
\label{eq:macroscopicmaxwell_basisindependent}
\begin{split}
    \div \vc{\gamma} &= \vc{f}\,,\\
    \curl \vc{\psi} &= \hat{D} (\ii k_0\vc{\gamma}-\vc{g})\,.
\end{split}
\end{equation}

\subsection{Constitutive relations}

Now, we will focus on the case of linear and isotropic materials (but including chiral and non-reciprocal properties). The constitutive relations are well known, and we will write them in a compact notation focusing on susceptibilities:
\begin{equation}
    \underbrace{\frac{1}{2} \pmqty{\vc{D}/\sqrt{\varepsilon_0} \\ \vc{B}/\sqrt{\mu_0}}}_{\vc{\gamma}_{[\text{EM}]} = \pmqty{\vc{G}_\text{e}\\ \vc{G}_\text{m}}} = \underbrace{\Bigg(\mqty{ \overbrace{1+\chi_e}^{\varepsilon_r} & \chi_t+\ii \chi_c \\ \chi_t - \ii \chi_c & \underbrace{1+\chi_m}_{\mu_r}}\Bigg)}_{\tens{I} + \tens{\chi}_{[\text{EM}]}} \underbrace{\frac{1}{2} \pmqty{\sqrt{\varepsilon_0}\vc{E}\\\sqrt{\mu_0}\vc{H}}}_{\vc{\psi}_{[\text{EM}]} = \pmqty{\vc{F}_\text{e}\\ \vc{F}_\text{m}}}
\end{equation}
where $\chi_e$ is the electric susceptibility, $\varepsilon_r = 1+\chi_e$ the relative electric permittivity, $\chi_m$ the magnetic susceptibility, $\mu_r = 1+\chi_m$ the relative magnetic permeability, $\chi_c$ is a chiral susceptibility (typically written as the $\kappa$ material parameter in many works), and $\chi_t$ the non-reciprocal susceptibility. This means that the constitutive relations can be written in a basis-independent notation of bi-spinors as:
\begin{equation}
\label{eq:constitutiverelation_basisindependent}
    \vc{\gamma} = (\tens{I} + \tens{\chi}) \vc{\psi} 
\end{equation}
where the tensor $\tens{\chi}$ in $\mathbb{C}^2$ space can, following \cref{eq:2by2_in_different_bases}, be written in different bases:
\begin{equation}
\begin{alignedat}{3}
\tens{\chi}_\text{[RL]}&=\pmqty{\chi_\text{R}&\chi_\text{RL}\\\chi_\text{LR}&\chi_\text{L}}&&=\chi_0\mqty(\pmat{0})+\chi_1\underbrace{\mqty(\pmat{1})}_{\tens{\sigma}_1}\mathop{+}\chi_2\underbrace{\mqty(\pmat{2})}_{\tens{\sigma}_2}\mathop{+}\chi_3\underbrace{\mqty(\pmat{3})}_{\tens{\sigma}_3}\,,\\
    \tens{\chi}_\text{[EM]}&=\pmqty{\chi_\text{e}&\chi_\text{em}\\\chi_\text{me}&\chi_\text{m}} &&=\chi_0\mqty(\pmat{0})+\chi_1\underbrace{\mqty(\pmat{3})}_{\tens{\sigma}_3}\mathop{-}\chi_2\underbrace{\mqty(\pmat{1})}_{\tens{\sigma}_1}\mathop{-}\chi_3\underbrace{\mqty(\pmat{2})}_{\tens{\sigma}_2}\,,\\
    \tens{\chi}_\text{[PA]}&=\pmqty{\chi_\text{p}&\chi_\text{pa}\\\chi_\text{ap}&\chi_\text{a}}&&=\chi_0\mqty(\pmat{0})+\chi_1\underbrace{\mqty(\pmat{1})}_{\tens{\sigma}_1}\mathop{-}\chi_2\underbrace{\mqty(\pmat{3})}_{\tens{\sigma}_3}\mathop{+}\chi_3\underbrace{\mqty(\pmat{2})}_{\tens{\sigma}_2}\,,
\end{alignedat}
\end{equation}%
which provides the definition of the following four basis-independent degrees of freedom of the susceptibility tensor:
\begin{equation}
\begin{alignedat}{2}
\chi_0 &= \tfrac{1}{2}(\varepsilon_r + \mu_r) - 1 = \tfrac{1}{2}(\chi_\text{e} + \chi_\text{m}) = \tfrac{1}{2}(\chi_\text{R} + \chi_\text{L}) = \tfrac{1}{2}(\chi_\text{p} + \chi_\text{a}), \\
\chi_1 &= \tfrac{1}{2}(\varepsilon_r - \mu_r) = \tfrac{1}{2}(\chi_\text{e} - \chi_\text{m}), \\
\chi_2 &= -\chi_t = -\tfrac{1}{2}(\chi_\text{p} - \chi_\text{a}), \\
\chi_3 &= \chi_c = \tfrac{1}{2}(\chi_\text{R} - \chi_\text{L}).
\end{alignedat}
\end{equation}
These four susceptibilities will completely define any isotropic linear medium in a basis-independent way, unlike most formulations that give special importance to the EM basis. Note that each of these susceptibilities $\chi_A(\vc{r})$ may be a function of position, defining any geometry, in the same way that a position-dependent $\varepsilon_r(\vc{r})$ is often used to describe dielectric systems such as particles, slabs, waveguides, etc.

\subsection{Diagonalisation of macroscopic Maxwell equations}

Combining the $\mathbb{C}^2$-basis-independent formulation of the macroscopic Maxwell equation (with no \emph{free} charges and currents) $\curl \vc{\psi} = \hat{D} (\ii k_0\vc{\gamma})$ (\cref{eq:macroscopicmaxwell_basisindependent}) with the basis-independent formulation of the constitutive relation in a linear istropic medium $\vc{\gamma} = (\tens{I} + \tens{\chi}) \vc{\psi}$ (\cref{eq:constitutiverelation_basisindependent}) one may write the resulting macroscopic Maxwell's equation in a linear medium as:
\begin{equation}
\curl \vc{\psi} = \ii k_0 \hat{D} (\tens{I} + \tens{\chi}) \vc{\psi}
\end{equation}
which can be written as a homogeneous equation $\tens{M} \vc{\psi} = 0$:
\begin{equation}
\underbrace{\left[ \ii k_0 \hat{D} (\tens{I} + \tens{\chi}) - \pmqty{\curl&0\\0&\curl} \right]}_{\tens{M}} \vc{\psi} = 0
\end{equation}
This equation can be written, following \cref{eq:2by2_in_different_bases}, in the different bases:
\begin{equation}
\begin{alignedat}{3}
\left[ (k_0\chi_3 - \curl)\mqty(\pmat{0})-\ii k_0 \chi_2 \mqty(\pmat{3}) - \ii k_0 \chi_1 \mqty(\pmat{1}) - k_0(1+\chi_0) \mqty(\pmat{2}) \right] \pmqty{\vc{F}_\text{e}\\\vc{F}_\text{m}} = \vc{0} \,,\\
\left[ (k_0\chi_3 - \curl)\mqty(\pmat{0})-\ii k_0 \chi_2 \mqty(\pmat{1}) - \ii k_0 \chi_1 \mqty(\pmat{3}) + k_0(1+\chi_0) \mqty(\pmat{2}) \right] \pmqty{\vc{F}_\text{p}\\\vc{F}_\text{a}} = \vc{0} \,,\\
\left[ (k_0\chi_3 - \curl)\mqty(\pmat{0})-\ii k_0 \chi_2 \mqty(\pmat{1}) + \ii k_0 \chi_1 \mqty(\pmat{2}) + k_0(1+\chi_0) \mqty(\pmat{3}) \right] \pmqty{\vc{F}_\text{R}\\\vc{F}_\text{L}} = \vc{0} \,.
\end{alignedat}
\end{equation}%
This reveals under which material parameters will Maxwell's equations be diagonalised for each basis EM, RL or PA---by taking the off diagonal terms to zero in each of the three cases. The conditions are the following. 
\newline
For EM uncoupled equations,
\begin{equation}
\begin{alignedat}{1}
    \left\{ \begin{array}{c}
        1+\chi_0 = 0 \\
        \chi_1 = 0
    \end{array} \right\}
    & \xLeftrightarrow{} 
    \left\{ \begin{array}{c}
        \varepsilon_r + \mu_r = 0 \\
        \varepsilon_r - \mu_r = 0
    \end{array} \right\}
    & \xLeftrightarrow{} 
    \left\{ \begin{array}{c}
        \varepsilon_r(\vc{r}) = 0 \\
        \mu_r(\vc{r}) = 0
    \end{array} 
    \quad \text{with} \quad 
     \begin{array}{c}
        \chi_c, \chi_t \\
        \text{free variables}
    \end{array} \right\}
\end{alignedat}
\end{equation}%
For PA uncoupled equations,
\begin{equation}
\begin{alignedat}{1}
    \left\{ \begin{array}{c}
        1 + \chi_0 = 0 \\
        \chi_2 = 0
    \end{array} \right\}
    & \xLeftrightarrow{} 
    \left\{ \begin{array}{c}
        \varepsilon_r + \mu_r = 0 \\
        \chi_t = 0
    \end{array} \right\}
    & \xLeftrightarrow{} 
    \left\{ \begin{array}{c}
        \varepsilon_r(\vc{r}) = -\mu_r(\vc{r}) \\
        \chi_t(\vc{r}) = 0
    \end{array} 
    \quad \text{with} \quad 
     \begin{array}{c}
        \chi_c, \mu_r = -\varepsilon_r \\
        \text{free variables}
    \end{array} \right\}
\end{alignedat}
\end{equation}%
For RL uncoupled equations,
\begin{equation}
\begin{alignedat}{1}
    \left\{ \begin{array}{c}
        \chi_1 = 0 \\
        \chi_2 = 0
    \end{array} \right\}
    & \xLeftrightarrow{} 
    \left\{ \begin{array}{c}
        \varepsilon_r - \mu_r = 0 \\
        \chi_t = 0
    \end{array} \right\}
    & \xLeftrightarrow{} 
    \left\{ \begin{array}{c}
        \varepsilon_r(\vc{r}) = \mu_r(\vc{r}) \\
        \chi_t(\vc{r}) = 0
    \end{array} 
    \quad \text{with} \quad 
     \begin{array}{c}
        \chi_c, \mu_r = \varepsilon_r \\
        \text{free variables}
    \end{array} \right\}
\end{alignedat}
\end{equation}%
The EM uncoupled condition corresponds to $\varepsilon$-and-$\mu$-near-zero (EMNZ) materials \cite{ziolkowski2004propagation,mahmoud2014wave}, well-known to turn Maxwell equations into a quasi-static form and thus uncouple $\vc{E}$ from $\vc{H}$. The RL uncoupled condition corresponds to dual materials, $\varepsilon_r(\vc{r}) = \mu_r(\vc{r})$ (free space being an example), which have been widely discussed as systems that do not perturb the helicity of incoming light, because the right and left-handed components of light are uncoupled \cite{zambrana2013duality}. Finally, the PA uncoupled condition $\varepsilon_r(\vc{r}) = -\mu_r(\vc{r})$ is novel to our knowledge, and it happens for example in a plasma such as a Drude metal with $\mu_r = 1$ and $\varepsilon_r = 1-\omega_p^2/\omega^2$, at an angular frequency $\omega = \omega_{\text{p}}/\sqrt{2}$ where $\omega_\text{p}$ is the plasma frequency. It is worth mentioning that in the interface between a material such as vacuum (fulfilling the RL uncoupled condition) and such plasma (fulfilling the PA uncoupled condition) a surface plasmon mode exists theoretically with infinite wavenumber under that exact condition.

Reversing the argument, for any arbitrary linear material defined by ($\chi_0(\vc{r})$, $\chi_1(\vc{r})$, $\chi_2(\vc{r})$, $\chi_3(\vc{r})$) one could in principle diagonalise the macroscopic Maxwell \cref{eq:macroscopicmaxwell_basisindependent}, which amounts to finding a basis for the electromagnetic bi-spinor in $\mathbb{C}^2$ space, in which the equations are uncoupled - but this would need to be done at each point in space $\vc{r}$, due to the position-dependence of the susceptibilities according to the material geometry, in principle limiting the usefulness of this approach. Only in cases where the basis is the same throughout space (as in the cases described above) would this approach be useful. Inside every block of homogeneous material with constant material parameters, an electromagnetic basis in $\mathbb{C}^2$ space can be found such that its two components are uncoupled. This could represent an interesting approach to solving electromagnetic problems.

% --------------------------------------------------------------------------------------------
\subsection{On the choice of orthonormal basis in \texorpdfstring{$\mathbb{C}^2$}{C²}}
In a general material the time-averaged energy density can be written as follows \cite{Jackson1998}:
\begin{equation}\label{eq:EMenergydensitywithEHDB}
    W = \frac{1}{4}\Re(\vc{D}^\ast\vdot\vc{E} + \vc{B}^\ast\vdot\vc{H}). 
\end{equation}
In a basis-independent way one can write this energy density using the two bispinors $\vc\psi$ and $\vc\gamma$ (see \cref{eq:maxwell_material}) as
\begin{equation}
    W=\Re(\vc\gamma^\dagger\vdot\vc\psi).
\end{equation}
Note that this form should work for any arbitrary material (including chiral, non-reciprocal, anisotropic and even complex), which we will address in future work.
In the case of an arbitrary linear material $\vc\gamma=(\tens{I} + \tens{\chi})\vc\psi$, as a consequence we can understand this as an inner product  
\begin{equation}
    W=\Re[\vc\psi^\dagger(\tens{I} + \tens{\chi})^\dagger\vdot\vc\psi]=\vc\psi^\dagger\vdot \vc\psi+\tfrac{1}{2}(\vc\psi^\dagger \tens{\chi}^\dagger\vdot\vc\psi+\vc\psi^\intercal \tens{\chi}^\intercal\vdot\vc\psi^\ast)=\vc\psi^\dagger\tens{g}\vdot \vc\psi=\Bar{\vc\psi}\vdot \vc\psi\equiv\norm{\vc\psi}^2
\end{equation}
where the hermitian tensor $\tens{g}=\tens{I}+\tfrac{1}{2}(\tens{\chi}+\tens{\chi}^\dagger)$ can be understood as a metric tensor. At the same time the vector $\Bar{\vc\psi}=\vc\psi^\dagger\tens{g}$ plays role of a dual vector for $\vc\psi$ and we denote it by a bar to point out similarity with the Dirac adjoint. The concept of the inner product or metric can be best understood with a simple example, consider two 2D real vectors $\vc{v}$ and $\vc{w}$ together with their inner product $\vc{v}\vdot\vc{w}=\vc{v}^\intercal\tens{g}\vc{w}$. Given a coordinate chart one can express vectors in a basis that is associated with these coordinates, however in general there are two different bases that are associated with a given coordinate system. The first one is called coordinate basis and it is made of vectors that represent the rate of change of the position with respect to the coordinate (we shall denote them $\vc{b}_i$) and the other is the orthonormal basis (we shall denote them $\uv{e}_i=\vc{b}_i/\abs{\vc{b}_i}$) and it is made of unit vectors that point in the same direction as $\vc{b}_i$. In a cartesian basis we have coordinates $(x,y)$ and if the underlying manifold is not cureved then $\vc{b}_x=\uv{e}_x=\uv{x}$ and same for $y$. Vectors can then be written as $\vc{v}=v_x\uv{x}+v_y\uv{y}$ and similarly for $\vc{w}$. However this is not the case if we either choose a curvilinear coordinate  system (for example polar one) or if the underlying space is curved. Consider coordinates $(r,\theta)$ such that $\vc{r}=(x,y)=(r\cos\theta,r\sin\theta)$, then $\vc{b}_r=\pdv*{\vc{r}}{r}$ and $\vc{b}_\theta=\pdv*{\vc{r}}{\theta}$, while $\uv{e}_r=\vc{b}_r=\uv{r}$ and $\uv{e}_\theta=\vc{b}_\theta/r=\uv{\theta}$. One can then choose to express a vector in either of the two bases $\vc{v}=v_r\vc{b}_r+v_\theta\vc{b}_\theta=v_{\hat{r}}\uv{e}_r+v_{\hat{\theta}}\uv{e}_\theta$ and similarly for $\vc{w}$, of course this is just an arbitrary choice of representation and the inner product of the two vectors will be unaffected by this choice
\begin{equation}
    \vc{v}\vdot\vc{w}=\underbrace{\vphantom{\pmqty{w_r\\w_\theta}}(v_r,v_\theta)}_{\vc{v}^\intercal}\underbrace{\pmqty{1&0\\0&r^2}}_{\tens{g}}\underbrace{\pmqty{w_r\\w_\theta}}_{\vc{w}}=\underbrace{\vphantom{\pmqty{w_r\\w_\theta}}(v_{\hat{r}},v_{\hat{\theta}})}_{\vc{v}^\intercal}\underbrace{\pmqty{1&0\\0&1}}_{\tens{g}}\underbrace{\pmqty{w_{\hat{r}}\\w_{\hat{\theta}}}}_{\vc{w}}
\end{equation}
and therefore $\tens{g}$ can be represented by a different matrix depending on the choice of vector basis and in an orthonormal basis it will be just an identity matrix. The same is true for bispinors.
In the case of non-chiral and reciprocal isotropic material which we considered in the initial sections of this work, one can simply choose an orthonormal basis in $\mathbb{C}^2$, i.e. $\uv{i}\vdot\uv{i}=1$ for $\uv{i}$ being one of the following: $\{\uv{e},\uv{m},\uv{p},\uv{a},\uv{r},\uv{l}\}$ defined by writing the abstract bispinor in the EM orthonormal basis as follows
\begin{equation}
    \vc{\psi}(\vc{r})=\tfrac{1}{2} (\sqrt{\varepsilon} \vc{E}(\vc{r}) \otimes\uv{e} +  \sqrt{\mu} \vc{H}(\vc{r})\otimes\uv{m})=\vc{F}_\text{e}(\vc{r}) \otimes\uv{e} +  \vc{F}_\text{m}(\vc{r})\otimes\uv{m}\,,
\end{equation}
where $\varepsilon$ and $\mu$ are the total permittivity and permeability of the linear non-chiral material, and the remaining orthonormal vectors are defined in \cref{eq:C2basisvectors}. However, we note that this arbitrary choice makes $\vc{F}_i$ depend on the material, and the dependence would be very complicated for materials that are chiral and/or non-reciprocal. Hence, starting from \cref{sec:macmax} we instead chose a basis that would be orthonormal in vacuum $\{\uv{e}_0,\uv{m}_0,\uv{p}_0,\uv{a}_0,\uv{r}_0,\uv{l}_0\}$
\begin{equation}
    \vc{\psi}(\vc{r})=\tfrac{1}{2} (\sqrt{\varepsilon_0} \vc{E}(\vc{r}) \otimes\uv{e}_0 +  \sqrt{\mu_0} \vc{H}(\vc{r})\otimes\uv{m}_0)=\vc{F}^0_\text{e}(\vc{r}) \otimes\uv{e}_0 +  \vc{F}^0_\text{m}(\vc{r})\otimes\uv{m}_0\,,
\end{equation}
because in this basis the components of the bispinor $\vc{F}^0_i$ (we dropped the superfluous 0 superscript) do not depend on the material and the full material dependence is then contained in the components of material tensor $(\tens{I} + \tens{\chi})$. In the more general case the material metric in the EM vacuum basis will be
\begin{equation}
    \tens{g}_{[\text{EM}]}=\pmqty{ \uv{e}_0^*\vdot\uv{e}_0 & \uv{e}_0^*\vdot\uv{m}_0 \\ \uv{m}_0^*\vdot\uv{e}_0 & \uv{m}_0^*\vdot\uv{m}_0}=\pmqty{ {\Re\varepsilon_r} & \Re\chi_t+\ii \Re\chi_c \\ \Re\chi_t - \ii \Re\chi_c & {\Re\mu_r}}\,,
\end{equation}
where we used the fact that the material matrix can be written as
\begin{equation}
    \pmqty{ {\varepsilon_r} & \chi_t+\ii \chi_c \\ \chi_t - \ii \chi_c & {\mu_r}}=\underbrace{\pmqty{ {\Re\varepsilon_r} & \Re\chi_t+\ii \Re\chi_c \\ \Re\chi_t - \ii \Re\chi_c & {\Re\mu_r}}}_\text{hermitian part}+\underbrace{\ii \pmqty{ {\Im\varepsilon_r} & \Im\chi_t+\ii \Im\chi_c \\ \Im\chi_t - \ii \Im\chi_c & {\Im\mu_r}}}_\text{anti-hermitian part}
\end{equation}
This means that this vacuum basis will be orthogonal as long as the material is non-chiral and reciprocal and it will be orthonormal as long as $\varepsilon_r=\mu_r=1$. Taking the inner product will indeed produce the desired energy density 
\begin{equation}
\begin{split}
    \norm{\vc\psi}^2&=\tfrac{1}{4}(\sqrt{\varepsilon_0} \vc{E} \otimes\uv{e}_0 +  \sqrt{\mu_0} \vc{H}\otimes\uv{m}_0)^\ast\vdot(\sqrt{\varepsilon_0} \vc{E} \otimes\uv{e}_0 +  \sqrt{\mu_0} \vc{H}\otimes\uv{m}_0)\\
    &=\tfrac{1}{4}[\varepsilon_0\abs{\uv{e}_0}^2\abs{\vc{E}}^2 +  {\mu_0} \abs{\uv{m}_0}^2\abs{\vc{H}}^2+\sqrt{\varepsilon_0\mu_0}(\uv{e}_0^*\vdot\uv{m}_0)\vc{E}^*\vdot\vc{H}+\sqrt{\varepsilon_0\mu_0}(\uv{e}_0\vdot\uv{m}_0^*)\vc{E}\vdot\vc{H}^*]\\
    &=\tfrac{1}{4}\Re[\varepsilon\abs{\vc{E}}^2 +  {\mu} \abs{\vc{H}}^2+2\sqrt{\varepsilon_0\mu_0}\chi_t\Re(\vc{E}^*\vdot\vc{H})+2\sqrt{\varepsilon_0\mu_0}\chi_c\Im(\vc{E}\vdot\vc{H}^*)]\\
    &= \tfrac{1}{4}\Re(\vc{D}^\ast\vdot\vc{E} + \vc{B}^\ast\vdot\vc{H})\,.
\end{split}
\end{equation}
However, note that the abstract bispinor $\vc{\psi}$ and the tensor $(\tens{I} + \tens{\chi})$ are completely unaffected by our choice of basis, it only changes their representations (their $\mathbb{C}^2$ components).

\section{Relation of quadratic quantities to 3D Stokes parameters}
A possible criticism to our main text is the apparently arbitrary choice of the $W$, $\vc{S}$ and $\tens{T}$ quadratic quantities to build a framework of quadratic quantities, and not others. We show here that the seemingly arbitrary choice describes a very general situation. A systematic approach to quadratic quantities would be to consider all possible products of all field components with each other. This `brute force' approach is easily achieved by constructing an outer product in $\mathbb{C}^3$ space $\vc{F}_i^*\!\otimes\vc{F}_i$ between the $\vc{F}_i$ vectors from \cref{eq:phiexpressedinthreebases} with themselves. This is the approach used by the Stokes-Gell-Mann parameters to characterise 3D polarisation when $\vc{F}_i$ is $\sqrt{\varepsilon}\vc{E}$ or $\sqrt{\mu}\vc{H}$. As we will show, this approach can be decomposed into precisely the quadratic quantities in Table 1 in the main text. First, we note that there are 9 Stokes-Gell-Mann parameters, which happen to be the same amount of degrees of freedom as $W_i$, $\vc{S}_i$ and $\tens{T}_i$, that is, $1+3+5$ (note $\tr\tens{T}_i=-W_i$, which is why it only contributes $5$ degrees of freedom). To define 3D Stokes one usually decomposes $\vc{F}_i^*\!\otimes\vc{F}_i$ (where $\vc{F}_i$ can be $\sqrt{\varepsilon}\vc{E}$ or $\sqrt{\mu}\vc{H}$) in terms of Gell-Mann matrices \cite{Carozzi2000,Setala2002}
\begin{equation*}
\frac{1}{3}\pmqty{
\Lambda_0+\Lambda_3+\frac{1}{\sqrt{3}} \Lambda_8 & \Lambda_1-i \Lambda_2 & \Lambda_4-i \Lambda_5 \\
\Lambda_1+i \Lambda_2 & \Lambda_0-\Lambda_3+\frac{1}{\sqrt{3}} \Lambda_8 & \Lambda_6-i \Lambda_7 \\
\Lambda_4+i \Lambda_5 & \Lambda_6+i \Lambda_7 & \Lambda_0-\frac{2}{\sqrt{3}} \Lambda_8
},
\end{equation*}
where the literature is divided on normalisation and also numbering of these parameters.
However, our approach is to further decompose this in terms of physical observables. We previously showed that the outer product $\vc{F}_i\otimes\vc{F}_i^*$ can be written in terms of observables $W_i$, $\vc{S}_i$ and $\tens{T}_i$, see \cref{eq:outer_decomposition}.
Taking a complex conjugate we get 
\begin{equation}
    \vc{F}_i^*\!\otimes\vc{F}_i=\frac{1}{2}(W_i\tens{I}+\tens{T}_{i}+{\ii\omega}\,{\star\vc{S}_i})~,
\end{equation}
which is enough to describe 3D polarisation, and these physically meaningful quadratic quantities contain all the information contained in a brute force approach to quadratics. Also, $\star\vc{S}_i$ has a direct geometrical interpretation; it is a plane normal to $\vc{S}_i$ (the plane of the polarisation ellipse) with an added sense of rotation determined by the orientation of $\vc{S}_i$ (the handedness of the polarisation). To be more explicit, we can write our selection of quadratic quantities, in terms of the corresponding Stokes-Gell-Mann parameters of the $\vc{F}_i^*\!\otimes\vc{F}_i$ tensor, as:
\begin{equation}
    \begin{split}
    W_i=\frac{1}{4}\Lambda_0\,,\quad
        \vc{S}_i=\frac{1}{6\omega}\pmqty{\Lambda_7\\-\Lambda_5\\\Lambda_2},\quad
        \tens{T}_i+W_i\tens{I}=\frac{1}{6}\pmqty{
\Lambda_3+\frac{\Lambda_8}{\sqrt{3}}  & \Lambda_1 & \Lambda_4 \\
\Lambda_1 & -\Lambda_3+\frac{\Lambda_8}{\sqrt{3}}  & \Lambda_6 \\
\Lambda_4 & \Lambda_6 & -\frac{2\Lambda_8}{\sqrt{3}} }
\end{split}
\end{equation}
One can confirm by directly using the definitions that:
\begin{equation}\label{eq:Sperp}
    (\tens{T}_i+W_i\tens{I})\vdot\vc{S}_i=0\,,
\end{equation} 
This means that one can find a vector $\uv{e}_\text{p}(\uv{S}_i)$ and pseudovector $\uv{e}_\text{s}(\uv{S}_i)$ such that $\uv{e}_\text{p}\vdot\uv{S}_i=\uv{e}_\text{s}\vdot\uv{S}_i=\uv{e}_\text{s}\vdot\uv{e}_\text{p}=0$. These can be used to define a tensor $\hat{\tens{e}}_+=\uv{e}_\text{p}\otimes\uv{e}_\text{p}-\uv{e}_\text{s}\otimes\uv{e}_\text{s}$ and a pseudotensor $\hat{\tens{e}}_\times=\uv{e}_\text{s}\otimes\uv{e}_\text{p}+\uv{e}_\text{p}\otimes\uv{e}_\text{s}$ that form a basis of all symmetric tensors perpendicular to $\uv{S}_i$ (satisfying \cref{eq:Sperp}). This basis allows for a surprisingly simple form in terms of the observables, if $\operatorname{sgn}(\vc{p}_i\vdot\vc{S}_i)=\pm1$:
\begin{equation}\label{eq:3stokes}
\begin{split}
    \!\!4W_i=\mathcal{S}_0,\quad4\omega\vc{S}_i=\pm\mathcal{S}_3\uv{S}_i\,,\\
    4(\tens{T}_i+W_i\tens{I})=-\mathcal{S}_1\hat{\tens{e}}_+\mp\mathcal{S}_2\hat{\tens{e}}_\times\,,
\end{split}
\end{equation}
where $\mathcal{S}_A$ are the usual 2D Stokes parameters calculated for the vector $\vc{F}_i$ in the plane to which $\uv{S}_i$ is normal to (plane of polarisation ellipse). Notice that we never assumed a paraxial beam; therefore, \cref{eq:3stokes} is true for an arbitrary field with $\vc{S}_i\neq\vc{0}$. Inverting it we get the four Stokes parameters
\begin{equation}
\begin{aligned}
 \mathcal{S}_0&=-4\tr(\tens{T}_i)=4W_i\,,
 & 
 \mathcal{S}_2&=\mp2\tr(\hat{\tens{e}}_\times\vdot\tens{T}_i)\,,\\
 \mathcal{S}_1&=-2\tr(\hat{\tens{e}}_+\vdot\tens{T}_i)\,,
 & 
 {\mathcal{S}_3}&=\pm4\omega\abs{\vc{S}_i}\,,
 \end{aligned}
\end{equation} 
with two additional real angles sufficient to specify the 3D orientation of $\vc{S}_i$ and hence the orientation of the plane where the above Stokes parameters are defined. This represents a simple approach to describe the three-dimensional polarisation of the $\vc{F}_i$ field based only on the quadratic quantities $W_i$, $\vc{S}_i$ and $\tens{T}_i$. Notice that we only have six degrees of freedom (four Stokes parameters plus two angles for the orientation of vector $\vc{S}_i$), we started with nine, but \cref{eq:Sperp} contains three constraints. 

For example, the linear Stokes parameters for the vector $\vc{F}_i = \vc{F}_e = (1/2)\sqrt{\varepsilon}\vc{E}$ and spin $\vc{S}_i = \vc{S}_e$, assuming that the spin is not aligned with the $z$-axis (in which case they would coincide with the usual definition) and assuming a choice of basis vectors $\uv{e}_\text{s}$ and $\uv{e}_\text{p}$ that matches with the polar and azimuthal unit vectors in spherical coordinates (also called s-polarised and p-polarised) with respect to a spin vector $\vc{S}_i$ taken as radial (and hence satisfying the required orthogonality $\uv{e}_\text{p}\vdot\uv{S}_i=\uv{e}_\text{s}\vdot\uv{S}_i=\uv{e}_\text{s}\vdot\uv{e}_\text{p}=0$) will be uniquely given by components of observables $W_\text{e}$, $\vc{S}_\text{e}$ and $\tens{T}_\text{e}$:
\begin{equation}
    \begin{split}
        \!\!\!\!\mathcal{S}_1={}&4(W_{\text{e}x}+W_{\text{e}y})-4W_{\text{e}z}\frac{4\omega^2 {S_{\text{e}z}^2} }{4(W_{\text{e}x}+W_{\text{e}y})W_{\text{e}z}-T_{\text{e}xz}^2-T_{\text{e}yz}^2+\omega^2(S_{\text{e}y}^2+S_{\text{e}x}^2)}\\
 \!\!\!\!\mathcal{S}_2={}&\pm\frac{4}{\abs{\vc{S}_\text{e}}}\Big\{{S_{\text{e}x}T_{\text{e}yz}}-S_{\text{e}y}{T_{\text{e}xz}}+\frac{S_{\text{e}z}}{{S_{\text{e}x}^2+S_{\text{e}y}^2}}[2S_{\text{e}x}S_{\text{e}y}(W_{\text{e}x}-W_{\text{e}y})
 % \\ &
 -T_{\text{e}xy}({S_{\text{e}x}^2\!-S_{\text{e}y}^2})]\Big\}\,.\!\!
    \end{split}
\end{equation}
\section{Acknowledgements}
SG and FJRF acknowledge support from EIC-Pathfinder-CHIRALFORCE (101046961) which is funded by Innovate UK Horizon Europe Guarantee (UKRI project 10045438). AJV is supported by EPSRC Grant EP/R513064/1. {We thank M Nieto-Vesperinas and N. Levy for useful discussions to improve clarity.}

\hbadness 10000\relax\bibliography{main}